\documentclass[twocolumn,showpacs,aps,floatfix,prd]{revtex4}

\usepackage[normalem]{ulem}

\usepackage{graphicx}
\usepackage{dcolumn}
\usepackage{amsmath}
\usepackage{epsfig}
\usepackage{hhline}
\usepackage{soul}
\usepackage{bm}
\usepackage{relsize}
\usepackage{tabularx,pifont,multirow}

\usepackage{multirow}
\long\def\inst#1{\par\nobreak\kern 4pt\nobreak
     {\itshape #1}\par\vskip 10pt plus 3pt minus 3pt}

\usepackage{longtable}
\usepackage{psfrag}

\RequirePackage{xspace}

\def\Imm       {\ensuremath{\Im m}}
\def\Ree       {\ensuremath{\Re e}}
\def\qqbar {\ensuremath{q\overline q}\xspace}
\def\babar{\mbox{\slshape B\kern-0.1em{\smaller A}\kern-0.1em
    B\kern-0.1em{\smaller A\kern-0.2em R}}}
\def\Abar    {\kern 0.18em\overline{\kern -0.18em A}{}\xspace}
\def\Kbar    {\kern 0.18em\overline{\kern -0.18em K}{}\xspace}
\def\Dbar    {\kern 0.18em\overline{\kern -0.18em D}{}\xspace}
\def\Bbar    {\kern 0.18em\overline{\kern -0.18em B}{}\xspace}
\def\Gbar    {\kern 0.18em\overline{\kern -0.18em \Gamma}{}\xspace}

\def\BB      {\ensuremath{B\Bbar}\xspace} 
\def\Bz      {\ensuremath{B^0}\xspace}
\def\Bzb     {\ensuremath{\Bbar^0}\xspace}
\def\BzBzb   {\ensuremath{\Bz {\kern -0.16em \Bzb}}\xspace}
\def\Bu      {\ensuremath{B^+}\xspace}
\def\Bub     {\ensuremath{B^-}\xspace}

\def\BpBm    {\ensuremath{\Bu {\kern -0.16em \Bub}}\xspace}

\newcommand{\optbar}[1]{\shortstack{{\tiny (\rule[.4ex]{1em}{.1mm})}
  \\ [-.7ex] $#1$}}
\def\BorBbar    {\kern 0.18em\optbar{\kern -0.18em B}{}\xspace}
\def\DorDbar    {\kern 0.18em\optbar{\kern -0.18em D}{}\xspace}
\def\KorKbar    {\kern 0.18em\optbar{\kern -0.18em K}{}\xspace}
\def\CP                {\ensuremath{C\!P}\xspace}
\def\pep2{PEP-II}
\mathchardef\Upsilon="7107
\def\Y#1S{\ensuremath{\Upsilon{(#1S)}}\xspace}

\def\FourS {\Y4S}

\newcommand{\gevcc}{\ensuremath{{\mathrm{\,Ge\kern -0.1em V\!/}c^2}}\xspace}
\newcommand{\gev}{\ensuremath{\mathrm{\,Ge\kern -0.1em V}}\xspace}

\def\pip   {\ensuremath{\pi^+}\xspace}

\def\DeltaE     {\mbox{$\Delta E$}\xspace}

\newcommand{\mev}{\ensuremath{\mathrm{\,Me\kern -0.1em V}}\xspace}
\newcommand{\mevcc}{\ensuremath{{\mathrm{\,Me\kern -0.1em V\!/}c^2}}\xspace}

\newcommand{\BABARPubYear}     {08}
\newcommand{\BABARPubNumber}  {036}

\newcommand{\SLACPubNumber} {13337}

\begin{document}

\begin{flushleft}
\babar-PUB-\BABARPubYear/\BABARPubNumber\\
SLAC-PUB-\SLACPubNumber
\\[10mm]
\end{flushleft}

\title{{\bf \boldmath 
Time-dependent and time-integrated angular analysis
of $B\to\varphi K^0_S\pi^0$ and $\varphi K^\pm\pi^\mp$ }
\bigskip\bigskip
}

%
\author{B.~Aubert}
\author{M.~Bona}
\author{Y.~Karyotakis}
\author{J.~P.~Lees}
\author{V.~Poireau}
\author{E.~Prencipe}
\author{X.~Prudent}
\author{V.~Tisserand}
\affiliation{Laboratoire de Physique des Particules, IN2P3/CNRS et Universit\'e de Savoie, F-74941 Annecy-Le-Vieux, France }
\author{J.~Garra~Tico}
\author{E.~Grauges}
\affiliation{Universitat de Barcelona, Facultat de Fisica, Departament ECM, E-08028 Barcelona, Spain }
% ...
\collaboration{The \babar\ Collaboration}
\noaffiliation

\bigskip\date{\today}

\begin{abstract}
We perform a time-dependent and time-integrated angular analysis of 
the decays $B^0\to\varphi K^{*}(892)^0$,
$\varphi K^{*}_2(1430)^0$, and $\varphi(K\pi)^{*0}_{0}$  
with the final sample of about 465 million $\BB$ pairs recorded with the $\babar$ detector. 
Twenty-four parameters are investigated, 
including the branching fractions, $C\!P$-violation parameters,
and parameters sensitive to final-state interactions.
We use the dependence on the $K\pi$ invariant mass of the interference
between the scalar and vector or tensor components to resolve discrete 
ambiguities of the strong and weak phases.
We use the time-evolution of the $B\to\varphi K^0_S\pi^0$ channel to extract
the $C\!P$-violation phase difference $\Delta\phi_{00}=0.28\pm0.42\pm 0.04$
between the $B$ and $\Bbar$ decay amplitudes. When the $B\to\varphi K^\pm\pi^\mp$
channel is included, the fractions of longitudinal polarization ${f_L}$ 
of the vector-vector and vector-tensor decay modes are measured to be 
$0.494\pm0.034\pm0.013$ and $0.901^{+0.046}_{-0.058}\pm 0.037$, respectively. 
This polarization pattern requires the presence of
a positive-helicity amplitude in the vector-vector decay from 
a currently unknown source.
\end{abstract}


\maketitle

\setcounter{footnote}{0}


\section{INTRODUCTION}
\label{sec:INTRODUCTION}

Charge-Parity ($C\!P$) symmetry violation
has been recognized as one of the fundamental requirements for producing
a matter-dominated universe~\cite{Sakharov} and therefore it
has played an important role in understanding fundamental physics since its
initial discovery in the $K$ meson system in 1964~\cite{Cronin}.
A significant $C\!P$-violating asymmetry has 
been observed in decays of neutral
$B$ mesons to final states containing charmonium, due to interference
between $B^0$-$\Bbar^0$ mixing and direct decay amplitudes~\cite{babarcp}. 
It has been established~\cite{directkaon}
that the $C\!P$-violating decays of the $K_L^0$ meson are due 
to $C\!P$ violation in decay amplitudes, as well as in $K^0$-$\Kbar^0$ mixing,
and this kind of ``direct'' $C\!P$ asymmetry in $B$ decays
has also been observed recently~\cite{btokpi}.
The $C\!P$ asymmetries are generally much larger in $B$ decays than in $K$
decays~\cite{Bander} because they directly probe the least flat
Unitarity Triangle constructed from the Cabibbo-Kobayashi-Maskawa (CKM)
matrix elements $V_{ij}$, which relate weak and flavor quark eigenstates~\cite{Kobayashi}. 
This Triangle reflects the unitarity of the CKM matrix, and two of its angles
are the phase differences of its sides on the complex plane: 
${\alpha}\equiv{\rm arg}\displaystyle(-V^{~}_{td}V_{tb}^*\displaystyle/V^{~}_{ud}V_{ub}^*\displaystyle)$
and ${\beta}\equiv{\rm arg}\displaystyle(-V^{~}_{cd}V^*_{cb}\displaystyle/V^{~}_{td}V^*_{tb}\displaystyle)$.
Due to the large $C\!P$-violating effects, $B$ decays provide an
excellent testing ground of fundamental interactions. 

\begin{figure}[b]
\begin{center}
\centerline{
\setlength{\epsfxsize}{1.00\linewidth}\leavevmode\epsfbox{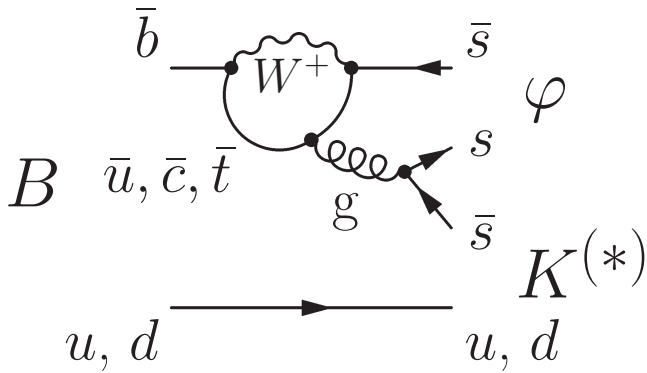}
}
\caption{
Penguin diagram describing the decay $B \to{\varphi K^{(*)}}$.
}
\label{fig:Diagram_phiKst}
\end{center}
\end{figure}

The $C\!P$-violating effects observed to date are self-consistent within
the Standard Model with a single complex phase in the CKM 
mechanism~\cite{Kobayashi}. However, this mechanism alone is believed 
to be insufficient to produce the present matter-dominated universe. 
Therefore, it is important to search for new sources of $C\!P$-violating
interactions. While direct access to new fundamental particles and
interactions may be beyond the energy reach of operating accelerators,
one can look for them in virtual transitions. 
New particles in virtual transitions, including but not limited to 
supersymmetric particles~\cite{supersymmetry},
would provide additional amplitudes with different phases. 
Depending on the model parameters, sizable $C\!P$-violating effects, 
either $B^0$-$\Bbar^0$ mixing-induced or ``direct'', could be
observed in pure penguin modes which involve virtual loops 
as in the example shown in Fig.~\ref{fig:Diagram_phiKst}.
Some of the first observed gluonic 
penguin decays, $\BorBbar\to\eta^\prime{\KorKbar}$~\cite{Behrens:1998dn}
and $\BorBbar\to\varphi{\KorKbar}{}^{(*)}$~\cite{phikst}, remain promising channels in which 
to look for new physics.
The latter type of decay is illustrated in Fig.~\ref{fig:Diagram_phiKst}
and is the focus of this paper.
For example, comparison of the value of $\sin 2\beta$ obtained from 
these modes with that from charmonium modes
such as $\BorBbar\to J/\psi{\KorKbar}{}^{(*)}$~\cite{babarcp, Aubert:2006wv}, 
or measurement of direct $C\!P$ violation, can probe new physics 
participating in penguin loops~\cite{Silvestrini:2007yf}.

The $({V}$-${A})$ nature of the weak interaction 
leads to left-handed fermion couplings in interactions with $W$ bosons, 
such as those shown in Fig.~\ref{fig:Diagram_phiKst}. Combined with helicity 
conservation in strong interactions and spin-flip suppression of relativistic
decay products, this leads to certain expectations of the spin alignment in 
weak $B$ meson decays to light particles with spin, 
such as $\BorBbar\to\varphi{\KorKbar}{}^{*}$~\cite{bvv1}.
However, the large fraction of transverse polarization in the 
$\BorBbar\to\varphi \KorKbar{}^*(892)$ 
decay measured by \babar~\cite{babar:vv} and by Belle~\cite{belle:phikst}
indicates a significant departure from the naive expectation of
predominant longitudinal polarization. This suggests the presence of other contributions
to the decay amplitude, previously neglected, either within or beyond
the Standard Model~\cite{newtheory}.
The presence of a substantial transverse amplitude also allows the 
study of $C\!P$ violation in the angular distribution of  
$\BorBbar\to\varphi{\KorKbar}{}^{*}$ decays, an approach complementary to
either mixing-induced or yield asymmetry studies.
Polarization measurements in $B$ decays are
discussed in a recent review~\cite{bvvreview2006,bib:Amsler2008}.
In Table~\ref{tab:previousresults}, we list $\babar$'s recent
measurements of the branching fraction and 
longitudinal polarization in the decays 
$\BorBbar\to\varphi \KorKbar{}^{(*)}_J$~\cite{babar:phik, babar:phikst, babar:phik1, babar:phikstpl, babar:phiksthigh}.
Measurements in $\BorBbar\to\rho \KorKbar{}^*$ decays have also revealed a large
fraction of transverse polarization~\cite{rhokst}.

\begingroup
\begin{table}[th]
\caption{
$\babar$'s recent measurements of the branching fraction ${\cal B}$ and 
longitudinal polarization fraction $f_L$ in the decays $B\to\varphi K^{(*)}_J$.
The spin $J$ and parity $P$ quantum numbers of the $K^{(*)}_J$
mesons are quoted. The upper limits are shown at the 90$\%$ confidence level. 
For a complete list of all observables in each analysis see the Refs. listed.
Results indicated with ${\dagger}$ are superseded by this analysis.
}
\begin{center}
{
\begin{ruledtabular}
\begin{tabular}{lcccc}
\vspace{-0.3cm}\\
 Mode & $J^P$ & Ref. & ${\cal B}$ (${ 10^{-6}}$) & $f_L$  \\
\vspace{-0.3cm} &&\\
\hline
\vspace{-0.3cm}\\
 $\varphi K^0$  & $0^-$ & \cite{babar:phik}  & $8.4^{+1.5}_{-1.3}\pm0.5$ & 1\\
\vspace{-0.3cm}\\
 $\varphi K^+$  & $0^-$ & \cite{babar:phik}  & $10.0^{+0.9}_{-0.8}\pm0.5$ & 1\\
\vspace{-0.3cm}\\
 $\varphi K_0^{*}(1430)^0$  & $0^+$ & \cite{babar:phikst}$^{\dagger}$  & $4.6\pm0.7\pm0.6$ & 1\\
\vspace{-0.3cm}\\
 $\varphi K_0^{*}(1430)^+$  & $0^+$ & \cite{babar:phik1}   & $7.0\pm1.3\pm0.9$ & 1\\
\vspace{-0.3cm}\\
 $\varphi K^{*}(892)^0$  & $1^-$ & \cite{babar:phikst}$^{\dagger}$   & $9.2\pm0.7\pm0.6$ & $0.51\pm0.04\pm0.02$ \\
\vspace{-0.3cm}\\
 $\varphi K^{*}(892)^+$  & $1^-$ & \cite{babar:phikstpl}   & $11.2\pm1.0\pm0.9$ & $0.49\pm 0.05\pm0.03$  \\
\vspace{-0.3cm}\\
  $\varphi K^{*}(1410)^+$  & $1^-$ & \cite{babar:phik1}  & $<4.8$ &  \\
\vspace{-0.3cm}\\
  $\varphi K^{*}(1680)^0$  & $1^-$ & \cite{babar:phiksthigh} & $<3.5$ &  \\
\vspace{-0.3cm}\\
  $\varphi K_1(1270)^+$  & $1^+$ & \cite{babar:phik1}   & $6.1\pm1.6\pm1.1$ & $0.46^{+0.12~+0.03}_{-0.13~-0.07}$ \\
\vspace{-0.3cm}\\
  $\varphi K_1(1400)^+$  & $1^+$ & \cite{babar:phik1}   & $<3.2$ & \\
\vspace{-0.3cm}\\
 $\varphi K_2^{*}(1430)^0$  & $2^+$ & \cite{babar:phikst}$^{\dagger}$ & $7.8\pm1.1\pm0.6$ & $0.85^{+0.06}_{-0.07}\pm0.04$ \\
\vspace{-0.3cm}\\
 $\varphi K_2^{*}(1430)^+$  & $2^+$ & \cite{babar:phik1}& $8.4\pm1.8\pm0.9$ & $0.80^{+0.09}_{-0.10}\pm 0.03$ \\
\vspace{-0.3cm}\\
 $\varphi K_2(1770)^+$ & $2^-$ & \cite{babar:phik1}   & $<16.0$ & \\
\vspace{-0.3cm}\\
 $\varphi K_2(1820)^+$  & $2^-$ & \cite{babar:phik1}   & $<23.4$ & \\
\vspace{-0.3cm}\\
 $\varphi K_3^{*}(1780)^0$  & $3^-$ & \cite{babar:phiksthigh}  &  $<2.7$ &  \\
\vspace{-0.3cm}\\
  $\varphi K_4^{*}(2045)^0$  & $4^+$ & \cite{babar:phiksthigh}  &  $<15.3$ &  \\
\vspace{-0.3cm}\\
\end{tabular}
\end{ruledtabular}
}
\end{center}
\label{tab:previousresults}
\end{table}
\endgroup

In this analysis, we use the final sample of about 465 million 
$\Upsilon(4S)\to\BB$ pairs recorded with the $\babar$ detector 
at the PEP-II asymmetric-energy $e^+e^-$ storage rings at SLAC. 
We employ all these techniques
for $C\!P$-violation and polarization measurements 
in the study of a single $B$-decay topology $\BorBbar^0\to\varphi(\KorKbar\pi)$. 
Overall, 27 independent parameters sensitive to $C\!P$ violation,
spin alignment, or strong- or weak-interaction phases describe three decay
channels (twelve in either vector-vector or vector-tensor and three in 
vector-scalar decays), which leaves only one overall phase unmeasurable.
The three channels in our amplitude analysis are 
$\BorBbar^0\to\varphi \KorKbar{}^{*}(892)^0$,
$\varphi \KorKbar{}^{*}_2(1430)^0$, and $\varphi(\KorKbar\pi){}^{*0}_0$. 
The latter contribution includes the $\KorKbar{}^{*}_0(1430)^0$ 
resonance together with a nonresonant component, as measured by the 
LASS experiment~\cite{Aston:1987ir}.
While we describe the analysis of these three neutral-$B$ meson
decays, this technique, with the exception of time-dependent measurements,
has also been applied recently to the charged-$B$ meson 
decays~\cite{babar:phik1, babar:phikstpl}.

We use the time-evolution of the $\BorBbar^0\to\varphi K^0_S\pi^0$ channel to extract
the mixing-induced $C\!P$-violating phase difference 
between the $B$ and $\Bbar$ decay amplitudes, which is equivalent
to a measurement of $\sin 2\beta$ to a good approximation.
With the $\BorBbar^0\to\varphi K^\pm\pi^\mp$ channel included, 
the fractions of longitudinal and parity-odd transverse amplitudes
in the vector-vector and vector-tensor decay modes are measured.
We use the dependence on the $K\pi$ invariant mass of the interference
between the scalar and vector, or scalar and tensor components to resolve discrete 
ambiguities of the strong and weak phases.
Using either interference between different channels or $B^0$-$\Bbar^0$ mixing, 
we measure essentially all 27 independent parameters
except for three quantities that characterize the parity-odd transverse 
amplitude in the vector-tensor decay, which is found to be
consistent with zero.


\section{ANALYSIS STRATEGY}
\label{sec:note-overview}

\begin{figure}[t]
\begin{center}
\begin{tabular}{c}
\setlength{\epsfxsize}{1.00\linewidth}\leavevmode\epsfbox{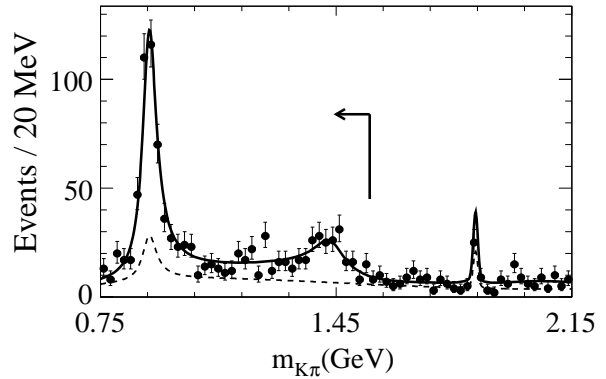}
\end{tabular}
\vspace{0.0cm}
\caption{
Invariant ${K\pi}$ mass distribution from the $B\to\varphi K^{\pm}\pi^{\mp}$ 
analysis from Refs.~\cite{babar:phikst, babar:phiksthigh}.
The solid (dashed) line is a projection of the signal-plus-background (background only)
fit result. The narrow charm background peak at 1.865 GeV comes from 
$\Dbar^0$ decays to $K\pi$ and is not associated with  
$\varphi K^{\pm}\pi^{\mp}$ production.
The arrow indicates the mass range considered in this analysis.
}
\label{fig:dalitz}
\end{center}
\end{figure}

Earlier studies of $\BorBbar^0\to\varphi K^\pm\pi^\mp$ decays 
by the $\babar$ collaboration~\cite{babar:phikst, babar:phiksthigh}
indicate the presence of three significant $K\pi$ partial waves:
$(K\pi)_0^{*0}$ (spin $J=0$, including the resonance $K^{*}_0(1430)^0$),
$K^{*}(892)^0$ ($J=1$), and
$K^{*}_2(1430)^0$ ($J=2$).
These correspond to the following decays, with the number of
independent amplitudes characterizing different spin projections
given in parentheses:
$\BorBbar^0\to\varphi (\KorKbar\pi)_0^{*0}$ (one),
$\varphi \KorKbar^{*}(892)^0$ (three), and
$\varphi \KorKbar^{*}_2(1430)^0$ (three).
No significant contribution from other final states
has been found with $K\!\pi$ invariant mass $m_{K\!\pi}$ up to 
2.15 GeV~\cite{babar:phiksthigh, babar:phik1}.
See Fig.~\ref{fig:dalitz} for an 
illustration of the $\BorBbar^0\to\varphi K^{\pm}\pi^{\mp}$ contributions.
Therefore, we limit our analysis to the mass
range $m_{K\pi}<1.55$ GeV without any significant loss of 
$\BorBbar^0\to\varphi K^{\pm}\pi^{\mp}$ signal through charmless $K\pi$ resonant or
nonresonant production.

There has been no extensive study of the $\BorBbar^0\to\varphi K_S^0\pi^0$ decay,
except for the study of $\BorBbar^0\to\varphi \KorKbar^{*}(892)^0$~\cite{babar:vv}.
However, due to isospin symmetry of the $K^0\pi^0$ and $K^\pm\pi^\mp$ systems,
the same amplitude composition is expected in the 
$\varphi K^\pm\pi^\mp$ and $\varphi K_S^0\pi^0$ final states.
We do not expect any charmless resonance structure
in the $\varphi K^\pm$ or $\varphi\pi^\mp$ combinations, while we 
veto the charm resonance states, such as $D^\pm_{(s)}\to\varphi\pi^\pm$.

It is instructive to do a simple counting of the amplitude parameters
in $\BorBbar^0\to\varphi \KorKbar\pi$ decays with the three $K\pi$ spin contributions
discussed above. With seven independent
$A_{J\lambda}$ complex amplitudes for $B$ decays and seven $\Abar_{J\lambda}$
amplitudes for $\Bbar$ decays, we could construct 28 independent real
parameters. Here $J$ refers to the spin of the $K\pi$ system and $\lambda$
to the spin projection of the $\varphi$ meson onto the direction 
opposite to the $B$ meson flight direction in the $\varphi$ rest frame.
However, one overall phase is not measurable and we are left with
27 real measurable parameters. Among these parameters, 26 parameters have been
or can be measured in the decay $\BorBbar^0\to\varphi K^\pm\pi^\mp$~\cite{babar:phikst}.
Those are branching fractions, polarization parameters, strong phases,
and $C\!P$ asymmetries.
Some of the phases are extracted from the interference
effects between different modes. However, due to limited statistics
some of the $C\!P$ asymmetries were not measured in earlier analyses and
we now extend those measurements.

\begingroup
\begin{table*}[th]
\caption{\label{tab:parameters}
Definitions of 27 real parameters measurable with the $B^0\to\varphi K\pi$ decays.
Three resonance final states with spin $J=0,1,2$ are considered in the $K\pi$ spectrum.
The branching fraction ${\cal B}$ is calculated as a ratio of the average
partial decay widths for $B^0$ ($\Gamma$) and $\Bbar^0$ ($\Gbar$) and the
total width $\Gamma_{\rm total}$
where we neglect any difference in the $B^0$ and $\Bbar^0$ widths.
This definition allows for differences between the $B^0$ and $\Bbar^0$ 
decay amplitudes, $A_{J\lambda}$ and $\Abar_{J\lambda}$, as discussed in the text.
}
\begin{center}
{
\begin{ruledtabular}
\setlength{\extrarowheight}{1.5pt}
\begin{tabular}{ccccc}
\vspace{-3mm} & & & & \\
   parameter
 & definition   & $\varphi K^{*}_0(1430)$ & $\varphi K^*(892)$ & $\varphi K^{*}_2(1430)$ \cr
 &              & $J=0$                   & $J=1$              & $J=2$                   \cr
\vspace{-3mm} & & &  \\
\hline
\vspace{-3mm} & & & \\
  ${\cal B}_J$ 
  & $\frac{1}{2}(\Gbar_J+\Gamma_J)/\Gamma_{\rm total}$ 
  & ${\cal B}_0$ & ${\cal B}_1$  & ${\cal B}_2$  \\
\vspace{-3mm} & & & \\
  ${f_{LJ}}$  
  & $\frac{1}{2}({|\Abar_{J0}|^2/\Sigma|\Abar_{J\lambda}|^2}+{|A_{J0}|^2/\Sigma|A_{J\lambda}|^2})$
  & 1  & ${f_{L1}}$ & ${f_{L2}}$ \\
\vspace{-3mm} & & & \\
  ${f_{\perp J}}$ 
  & $\frac{1}{2}({|\Abar_{J\perp}|^2/\Sigma|\Abar_{J\lambda}|^2}+{|A_{J\perp}|^2/\Sigma|A_{J\lambda}|^2})$ 
  & none  &  ${f_{\perp 1}}$  &  ${f_{\perp 2}}$  \\
\vspace{-3mm} & & & \\
  ${\phi_{\parallel J}}$ 
  & $\frac{1}{2}({\rm arg}(\Abar_{J\parallel}/\Abar_{J0})+{\rm arg}(A_{J\parallel}/A_{J0}))$  
  & none  &  ${\phi_{\parallel 1}}$  &  ${\phi_{\parallel 2}}$  \\
\vspace{-3mm} & & & \\
  ${\phi_{\perp J}}$  
  & $\frac{1}{2}({\rm arg}(\Abar_{J\perp}/\Abar_{J0})+{\rm arg}(A_{J\perp}/A_{J0})-\pi)$ 
  & none  &  ${\phi_{\perp 1}}$  &  ${\phi_{\perp 2}}$  \\
\vspace{-3mm} & & & \\
  ${\delta_{0J}}$  
  & $\frac{1}{2}({\rm arg}(\Abar_{00}/\Abar_{J0})+{\rm arg}(A_{00}/A_{J0}))$
  & 0 &  ${\delta_{01}}$  &  ${\delta_{02}}$  \\
\vspace{0mm} &  & &\\
  ${\cal A}_{C\!PJ}$ 
 & $(\Gbar_J-\Gamma_J)/(\Gbar_J+\Gamma_J)$ 
 &  ${\cal A}_{C\!P0}$ &  ${\cal A}_{C\!P1}$ &  ${\cal A}_{C\!P2}$ \\
\vspace{-3mm} & & & \\
  ${\cal A}_{C\!PJ}^0$ 
  & $({|\Abar_{J0}|^2/\Sigma|\Abar_{J\lambda}|^2}-{|A_{J0}|^2/\Sigma|A_{J\lambda}|^2})/
     ({|\Abar_{J0}|^2/\Sigma|\Abar_{J\lambda}|^2}+{|A_{J0}|^2/\Sigma|A_{J\lambda}|^2})$
  & 0  & ${\cal A}_{C\!P1}^0$    & ${\cal A}_{C\!P2}^0$  \\
\vspace{-3mm} &  & &\\
  ${\cal A}_{C\!PJ}^{\perp}$ 
  & $({|\Abar_{J\perp}|^2/\Sigma|\Abar_{J\lambda}|^2}-{|A_{J\perp}|^2/\Sigma|A_{J\lambda}|^2})/
     ({|\Abar_{J\perp}|^2/\Sigma|\Abar_{J\lambda}|^2}+{|A_{J\perp}|^2/\Sigma|A_{J\lambda}|^2})$
  & none & ${\cal A}_{C\!P1}^{\perp}$  & ${\cal A}_{C\!P2}^{\perp}$ \\
\vspace{-3mm} &  & &\\
  $\Delta \phi_{\parallel J}$  
  & $\frac{1}{2}({\rm arg}(\Abar_{J\parallel}/\Abar_{J0})-{\rm arg}(A_{J\parallel}/A_{J0}))$  
  & none & $\Delta \phi_{\parallel 1}$  & $\Delta \phi_{\parallel 2}$ \\
\vspace{-3mm} & & & \\
  $\Delta \phi_{\perp J}$  
  & $\frac{1}{2}({\rm arg}(\Abar_{J\perp}/\Abar_{J0})-{\rm arg}(A_{J\perp}/A_{J0})-\pi)$ 
  & none & $\Delta \phi_{\perp 1}$  & $\Delta \phi_{\perp 2}$ \\
\vspace{-3mm} & & & \\
  ${\Delta\delta_{0J}}$  
  & $\frac{1}{2}({\rm arg}(\Abar_{00}/\Abar_{J0})-{\rm arg}(A_{00}/A_{J0}))$
  & 0 & ${\Delta\delta_{01}}$ &  ${\Delta\delta_{02}}$ \\
\vspace{0mm} & & & \\
  ${\Delta\phi_{00}}$  
  & $\frac{1}{2}{\rm arg}(A_{00}/\Abar_{00})$ 
  & ${\Delta\phi_{00}}$ & none & none \cr
\vspace{-3mm} & & & \\
\end{tabular}
\end{ruledtabular}
}
\end{center}
\end{table*}
\endgroup

Finally, one parameter, which relates the phases of the
$B$ and $\Bbar$ decay amplitudes, can be measured  
using only the interference between decays with and without
$B^0$-$\Bbar^0$ mixing, to final states which can be
decomposed as $C\!P$ eigenstates, such as $\varphi K_S^0\pi^0$.
In Table~\ref{tab:parameters} all 27 real parameters measurable 
with $\BorBbar^0\to\varphi \KorKbar\pi$ decays are summarized.
These parameters are expressed in terms of the
$A_{J\lambda}$ and $\Abar_{J\lambda}$ amplitudes for
$B^0\to\varphi K^+\pi^-$ or $\varphi K^0\pi^0$ and 
$\Bbar^0\to\varphi K^-\pi^+$ or $\varphi\Kbar^0\pi^0$ decays.
We also refer to a transformed set of amplitudes 
$A_{J0}$ and $A_{J\pm1}=(A_{J\parallel}\pm A_{J\perp})/\sqrt{2}$.
The parameters in Table~\ref{tab:parameters} are expressed as 
six $C\!P$-averaged and six $C\!P$-violating parameters for
the vector-vector and vector-tensor decays. 
The $\pi$ in the definitions of $\phi_{\perp J}$ and $\Delta \phi_{\perp J}$
accounts for the sign flip $A_{\perp J}=-\Abar_{\perp J}$ 
if $C\!P$ is conserved.
The parameterization in Table~\ref{tab:parameters} is motivated by the negligible
$C\!P$ violation expected in these decays.
Therefore, the polarization parameters specific to either $B$ (superscript ``$-$'')  
or $\Bbar$ (superscript ``$+$'') are the $C\!P$-averaged parameters
with small $C\!P$-violating corrections which are either 
multiplicative (for rates) or additive (for phases):
\begin{eqnarray}
\label{eq:brpm}
 {\cal B}_J^\pm&=&{\cal B}_J\cdot(1\pm{\cal A}_{C\!PJ})/2 \\
\label{eq:flpm}
 f_{LJ}^\pm &=& f_{LJ}\cdot(1\pm{\cal A}_{C\!PJ}^0) \\
\label{eq:fppm}
 f_{\perp J}^\pm &=& f_{\perp J}\cdot(1\pm{\cal A}_{C\!PJ}^{\perp}) \\
\label{eq:phiparpm}
 \phi_{\parallel J}^\pm &=& \phi_{\parallel J}\pm\Delta \phi_{\parallel J} \\
\label{eq:phiperppm}
 \phi_{\perp J}^\pm &=& \phi_{\perp J}\pm\Delta\phi_{\perp J}+\frac{\pi}{2}\pm\frac{\pi}{2}\\
\label{eq:d0pm}
 \delta_{0J}^\pm &=& \delta_{0J}\pm\Delta\delta_{0J}.  
\end{eqnarray}

In this Section we discuss further the method for the measurement 
of the relative phase, along with all the other parameters.
First we review the angular distributions, follow with a discussion of 
the $K\pi$ invariant mass distributions critical to
separating different partial waves,
then introduce interference effects between amplitudes
from different decays, and 
finally discuss time-dependent distributions.


\subsection{Angular distributions}
\label{sec:note-angular}

\begin{figure}[b]
\begin{center}
\centerline{
\setlength{\epsfxsize}{1.00\linewidth}\leavevmode\epsfbox{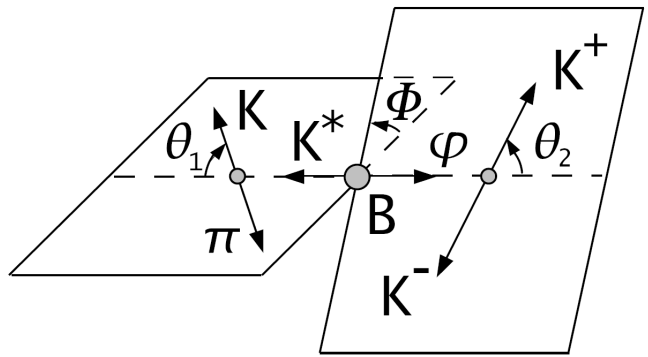}
}
\caption{
Definition of decay angles given in the rest frames of the decaying parents.
}
\label{fig:decay}
\end{center}
\end{figure}

We discuss here the angular distribution of the decay products in the 
chain $B\to\varphi K^*\to(K^+K^-)(K\pi)$ integrated over time.
First we look at the decay of a $B$ meson only and leave 
the $\Bbar$ for later discussion, which involves $C\!P$ violation.
Angular momentum conservation in the decay of a spinless $B$ meson
leads to three possible spin projections of the $\varphi$ meson onto  
its direction of flight, each corresponding to a complex amplitude
$A_{J\lambda}$ with $\lambda=0$ or $\pm 1$. The three $\lambda$
values are allowed with the $K^*$ spin states $J\ge1$, but
only $\lambda=0$ contributes with a spin-zero $K^*$. 
The angular distributions can be expressed
as functions of ${\cal H}_i=\cos\theta_i$ and $\Phi$. Here $\theta_i$ is the angle between the direction
of the $K$ meson from the $K^*\to K\pi$ ($\theta_1$) or $\varphi\to K\Kbar$
($\theta_2$) and the direction opposite to the $B$ in the $K^*$ or $\varphi$
rest frame, and $\Phi$ is the angle between the decay planes of the two
systems, as shown in Fig.~\ref{fig:decay}.
The differential decay width is 
\begin{eqnarray}
\label{eq:helicityfull}
{d^3\Gamma \over d{\cal H}_1 \,d{\cal H}_2\,d\Phi} \propto
\left|~\sum_{}
A_{J\lambda} Y_{J}^{\lambda}({\cal H}_1,\Phi) Y_{1}^{-\!\lambda}(-{\cal H}_2,0)~\right|^2,
\end{eqnarray}
where $Y_{J}^{\lambda}$ are the spherical harmonic functions, with $J=2$ for
$K_2^{*}(1430)$, $J=1$ for $K^{*}(892)$, and $J=0$ for $(K\pi)_0^{*}$,
including $K_0^{*}(1430)$.
We do not consider higher values of $J$ because no significant contribution
from those states is expected.
Only resonances with spin-parity combination $P=(-1)^J$ are possible
in the decay $K^{*}\to K\pi$ due to parity conservation in these
strong-interaction decays.

If we ignore interference between modes with different spins $J$ of
the $K\pi$ system in Eq.~(\ref{eq:helicityfull}), then for each decay
mode we have three complex amplitudes $A_{J\lambda}$ which
appear in the angular distribution.
We discuss interference between different modes later in this Section.
The differential decay rate for each decay mode
involves six real quantities $\alpha_{iJ}^-$,
including terms that account for interference between amplitudes of common $J$.
\begin{eqnarray}
 {d^3\Gamma_J \over \Gamma_J\,d{\cal H}_1 \,d{\cal H}_2\,d\Phi} =
  \sum_{i}\alpha_{iJ}^-\,{f}_{iJ}({\cal H}_1,\,{\cal H}_2,\,\Phi)\,,
\label{eq:fangular}
\end{eqnarray}
where the functions ${f}_{iJ}\,({\cal H}_1,\,{\cal H}_2,\,\Phi)$
are given in Table~\ref{tab:ffunctions}.
The $\alpha_{iJ}^-$ parameters are defined as:
\begin{widetext}
\begin{eqnarray}
\label{eq:alpha1}
\alpha_{1J}^-
&=&{|A_{J0}|^2 \over \Sigma|A_{J\lambda}|^2} = f_{LJ}^-\\
\label{eq:alpha2}
\alpha_{2J}^-
&=&{|A_{J\parallel}|^2 + |A_{J\perp}|^2 \over \Sigma|A_{J\lambda}|^2}
={|A_{J+1}|^2 + |A_{J-1}|^2 \over \Sigma|A_{J\lambda}|^2}
=(1-f_{LJ}^-)\\
\label{eq:alpha3}
\alpha_{3J}^-
&=&{ |A_{J\parallel}|^2 - |A_{J\perp}|^2  \over \Sigma|A_{J\lambda}|^2}
={2}\cdot{{\Ree}(A_{J+1}A^*_{J-1}) \over \Sigma|A_{J\lambda}|^2}
=(1-f_{LJ}^--2\cdot f_{\perp J}^-)\\
\label{eq:alpha4}
\alpha_{4J}^-
&=&{{\Imm}(A_{J\perp}A^*_{J\parallel})\over \Sigma|A_{J\lambda}|^2}
={{\Imm}(A_{J+1}A^*_{J-1}) \over \Sigma|A_{J\lambda}|^2}
=\sqrt{f_{\perp J}^-\cdot(1-f_{LJ}^--f_{\perp J}^-)}\cdot\sin(\phi_{\perp J}^--\phi_{\parallel J}^-)\\
\label{eq:alpha5}
\alpha_{5J}^-
&=&{{\Ree}(A_{J\parallel}A^*_{J0})  \over \Sigma|A_{J\lambda}|^2}
={ {\Ree}(A_{J+1}A^*_{J0} + A_{J-1}A^*_{J0}) \over \sqrt{2}\cdot\Sigma|A_{J\lambda}|^2}
=\sqrt{f_{LJ}^-\cdot(1-f_{LJ}^--f_{\perp J}^-)}\cdot\cos(\phi_{\parallel J}^-)\\
\label{eq:alpha6}
\alpha_{6J}^-
&=&{{\Imm}(A_{J\perp}A^*_{J0})  \over \Sigma|A_{J\lambda}|^2}
={ {\Imm}(A_{J+1}A^*_{J0} - A_{J-1}A^*_{J0}) \over \sqrt{2}\cdot\Sigma|A_{J\lambda}|^2}
=\sqrt{f_{\perp J}^-\cdot f_{LJ}^-}\cdot\sin(\phi_{\perp J}^-).
\\
\nonumber
\end{eqnarray}
\end{widetext}

The above terms are specific to the $B^0$ decays and are denoted with
the superscript ``$-$'', as introduced in Eqs.~(\ref{eq:brpm}--\ref{eq:d0pm}).
The angular distributions for the $\Bbar^0$ decays are described by the
same Eq.~(\ref{eq:fangular}), but with $\alpha_{iJ}^-$ replaced by
$\alpha_{iJ}^+$, and with definitions given by Eqs.~(\ref{eq:alpha1}--\ref{eq:alpha6}), 
replacing $A$ by $\Abar$ and superscript ``$-$'' by ``$+$''.

\begingroup
\begin{table*}[th]
\caption{\label{tab:ffunctions}
Parameterization of the angular distribution in Eq.~(\ref{eq:fangular})
in the $B^0\to\varphi(K\pi)_J$ decays where
three resonance final states with spin $J=0,1,2$ are considered.
The common constant is quoted for each decay mode and is omitted 
from each individual function below.
The three helicity angle parameters $({\cal H}_1,\,{\cal H}_2,\,\Phi)$ 
are discussed in the text.
}
\begin{center}
{
\begin{ruledtabular}
\setlength{\extrarowheight}{1.5pt}
\begin{tabular}{cccc}
\vspace{-3mm} & & & \\
              & $\varphi K^{*}_0(1430)^0$ & $\varphi K^*(892)^0$ & $\varphi K^{*}_2(1430)^0$ \cr
              & $J=0$                   & $J=1$              & $J=2$                   \cr
\vspace{-3mm} & & &  \\
\hline
\vspace{-3mm} & & & \\
common constant & ${3}/{4\pi}$ & ${9}/{8\pi}$ & ${15}/{32\pi}$ \\
\vspace{-3mm} & & & \\
${f}_{1J}\,({\cal H}_1,\,{\cal H}_2,\,\Phi)$
 & ${\cal H}_2^2$ 
 & ${\cal H}_1^2{\cal H}_2^2$   
 & $(3{\cal H}_1^2-1)^2{\cal H}_2^2$   \\
\vspace{-3mm} & & & \\
${f}_{2J}\,({\cal H}_1,\,{\cal H}_2,\,\Phi)$
 & 0  
 & $\frac{1}{4}(1-{\cal H}_1^2)(1-{\cal H}_2^2)$    
 & $3{\cal H}_1^2(1-{\cal H}_1^2)(1-{\cal H}_2^2)$    \\
\vspace{-3mm} & & & \\
${f}_{3J}\,({\cal H}_1,\,{\cal H}_2,\,\Phi)$
 & 0  
 & $\frac{1}{4}(1-{\cal H}_1^2)(1-{\cal H}_2^2)\cos2\Phi$    
 & $3{\cal H}_1^2(1-{\cal H}_1^2)(1-{\cal H}_2^2)\cos2\Phi$    \\
\vspace{-3mm} & & & \\
${f}_{4J}\,({\cal H}_1,\,{\cal H}_2,\,\Phi)$
 & 0  
 & $-\frac{1}{2}(1-{\cal H}_1^2)(1-{\cal H}_2^2)\sin2\Phi$    
 & $-6{\cal H}_1^2(1-{\cal H}_1^2)(1-{\cal H}_2^2)\sin2\Phi$    \\
\vspace{-3mm} & & & \\
${f}_{5J}\,({\cal H}_1,\,{\cal H}_2,\,\Phi)$
 & 0  
 & ${\sqrt{2}}{\cal H}_1\sqrt{1-{\cal H}_2^2}\,{\cal H}_2\sqrt{1-{\cal H}_2^2}\,\cos\Phi$    
 & $\sqrt{6}{\cal H}_1\sqrt{1-{\cal H}_1^2}\,(3{\cal H}_1^2-1)\,{\cal H}_2\sqrt{1-{\cal H}_2^2}\,\cos\Phi$    \\
\vspace{-3mm} & & & \\
${f}_{6J}\,({\cal H}_1,\,{\cal H}_2,\,\Phi)$
 & 0  
 & $-{\sqrt{2}}{\cal H}_1\sqrt{1-{\cal H}_1^2}\,{\cal H}_2\sqrt{1-{\cal H}_2^2}\,\sin\Phi$    
 & $-\sqrt{6}{\cal H}_1\sqrt{1-{\cal H}_1^2}\,(3{\cal H}_1^2-1)\,{\cal H}_2\sqrt{1-{\cal H}_2^2}\,\sin\Phi$    \\
\vspace{-3mm} & & & \\
\end{tabular}
\end{ruledtabular}
}
\end{center}
\end{table*}
\endgroup


\subsection{Mass distributions}
\label{sec:note-mass}

The differential decay width given in Eq.~(\ref{eq:helicityfull})
is parameterized as a function of helicity angles. However,
it also depends on the invariant mass $m_{K\pi}$ of the $K\pi$ resonance, and
the amplitudes should be considered as functions of $m_{K\pi}$.
Without considering interference between different modes,
as shown in Eq.~(\ref{eq:fangular}), this mass dependence decouples
from the angular dependence. Nonetheless, this dependence is
important for separating different $K\pi$ states. The interference 
effects will be considered in the next subsection.
A relativistic spin-$J$ Breit-Wigner (B-W) complex amplitude $R_J$
can be used to parameterize the resonance masses
with $J=1$ and $J=2$~\cite{bib:Amsler2008}:
\begin{equation}
R_{J}({m})= 
{{m}_J\Gamma_J({m})\over{
({m}_J^2-{m}^2)-i{m}_J\Gamma_J({m})}} = \sin\delta_{J}e^{i\delta_{J}},
\label{eq:mass-bw1}
\end{equation}
where we use the following convention:
\begin{equation}
\cot\delta_{J} = {{{m}^{2}_{J}-{m}^{2}}\over {{m}_{J}\Gamma_{J}({m})}}\,.
\label{eq:mass-bw2}
\end{equation}
The mass-dependent widths are given by:
\begin{eqnarray}
\Gamma_{1}(m) &=& \Gamma_1 {{{m}_1}\over{{m}}}{{1+r^2q_1^2}\over{1+r^2q^2}}
\bigg({{q}\over{q_1}}\bigg)^3,\\[5pt]
\label{eq:mass-bw3}
\Gamma_{2}(m) &=& \Gamma_2 {{{m}_2}\over{{m}}}
{{9+3r^2q_2^2+r^4q_2^4}\over{3+3r^2q^2+r^4q^4}}
\bigg({{q}\over{q_2}}\bigg)^5,
\label{eq:inter-13}
\end{eqnarray}
where $\Gamma_J$ is the resonance width, 
${m}_J$ is the resonance mass,
$q$ is the momentum of a daughter particle in the resonance system
after its two-body decay ($q_J$ is evaluated at $m={m}_J$), 
and $r$ is the interaction radius.

The parameterization of the scalar $(K\pi)^{*0}_0$ mass 
distribution requires more attention. Studies of $K\pi$ scattering
were performed by the LASS experiment~\cite{Aston:1987ir}.
It was found that the scattering is elastic up to about 1.5~GeV
and can be parameterized with the amplitude:
\begin{equation}
R_0(m) = \sin\delta_{0}e^{i\delta_{0}},
\label{eq:mass-lass1}
\end{equation}
where 
\begin{equation}
\delta_{0} = \Delta R + \Delta B, 
\label{eq:mass-lass2}
\end{equation}
$\Delta R$ represents a resonant $K^{*}_0(1430)^0$ contribution and 
$\Delta B$ represents a nonresonant contribution.
The mass dependence of $\Delta B$ is described by means of an effective
range parameterization of the usual type: 
\begin{equation}
\cot\Delta B = {1\over {aq}} + {1\over 2}bq,
\label{eq:mass-lass3}
\end{equation}
where $a$ is the scattering length and $b$ is the effective range. 
The mass dependence of $\Delta R$ is described by means of a B-W
parameterization of a form similar to Eq.~(\ref{eq:mass-bw2}):
\begin{equation}
\cot\Delta R = {{{m}^{2}_{0}-{m}^{2}}\over {{m}_{0}\Gamma_{0}({m})}}, 
\label{eq:mass-lass4}
\end{equation}
where $m_0$ is the resonance mass, and $\Gamma_{0}({m})$ is defined as
\begin{equation}
\Gamma_{0}(m) = \Gamma_0 {{{m}_0}\over{{m}}}\bigg({{{q}}\over{{q_0}}}\bigg).
\label{eq:mass-bw0}
\end{equation}

The invariant amplitude $M_J(m)$ is proportional to $R_J(m)$:
\begin{equation}
M_J(m) \propto {m\over q}R_J(m)
\label{eq:mass-lass6}
\end{equation}
and can be expressed, for example for $J=0$, as
\begin{eqnarray}
M_{0}(m) \propto  {m\over q\cot\Delta{B}-iq}   
\nonumber \\
+~e^{2i\Delta{B}} {\Gamma_0 m_0^2/q_0 \over (m_0^2-m^2)-im_0\Gamma_0(m) }\,.
\label{eq:mass-lass7}
\end{eqnarray}

\begin{figure*}[t]
\centerline{
\setlength{\epsfxsize}{0.425\linewidth}\leavevmode\epsfbox{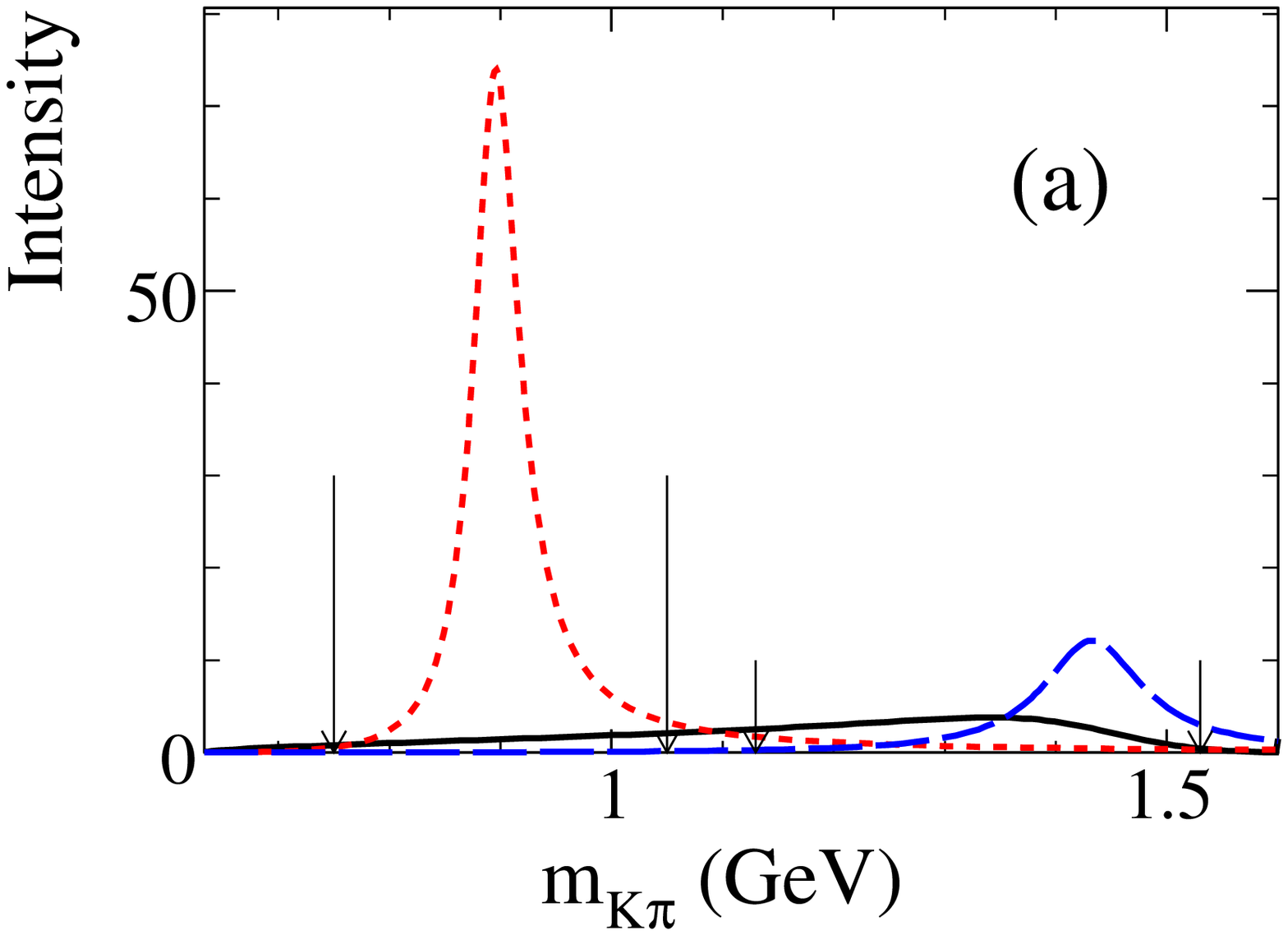} 
\setlength{\epsfxsize}{0.425\linewidth}\leavevmode\epsfbox{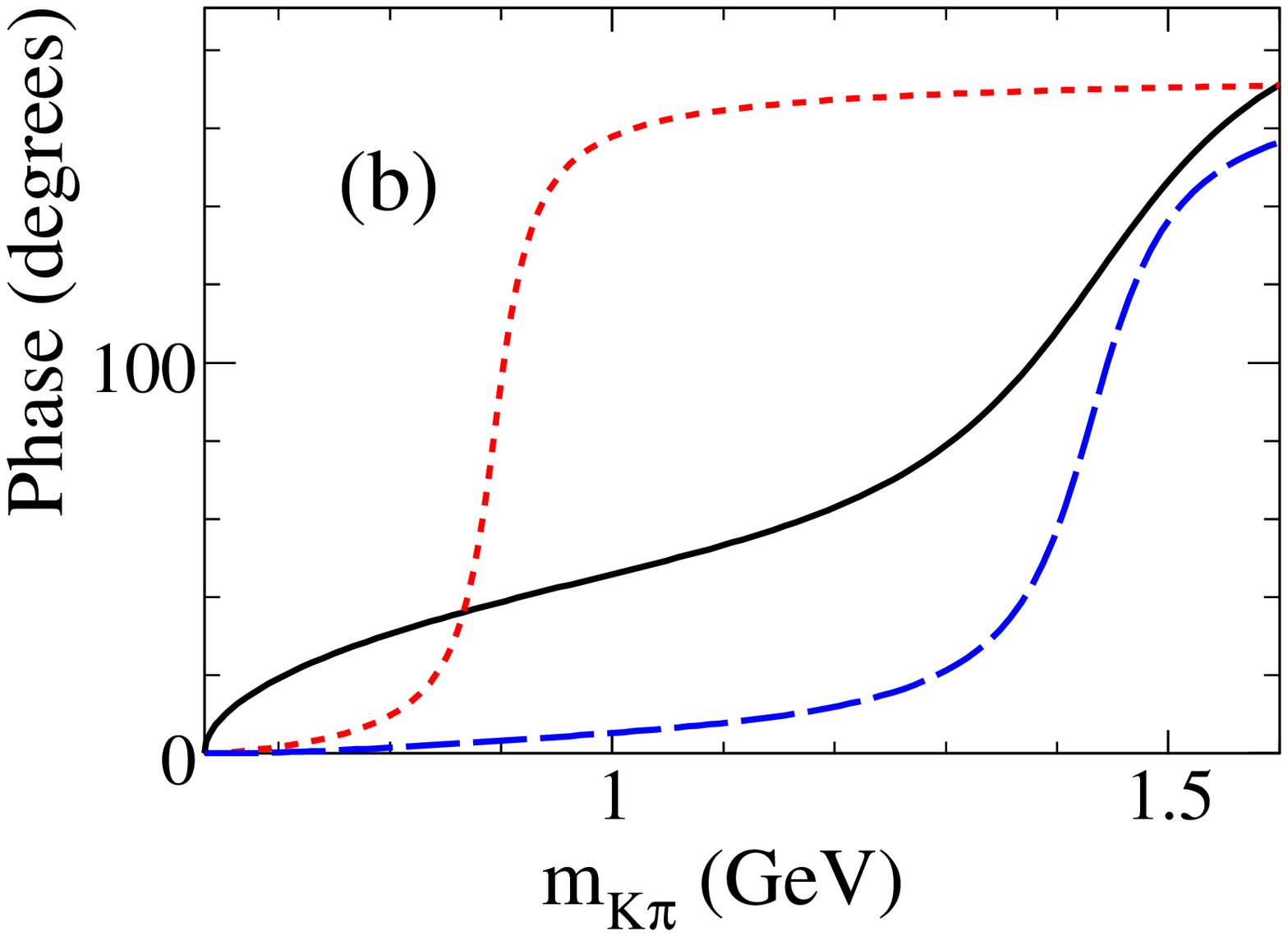}
}
\caption{\label{fig:lass-shape}
Intensity $|M_{J}(m_{K\pi})|^2$ (a) and phase arg($M_{J}(m_{K\pi})$) (b)
of the invariant amplitudes for $J=0$ (solid), $J=1$ (dashed), and 
$J=2$ (long-dashed) ${K\pi}$ contributions as a function
of the invariant ${K\pi}$ mass $m_{K\pi}$.
The taller two arrows indicate the low $m_{K\pi}$ region,
while the shorter two arrows indicate the high $m_{K\pi}$ region. 
The relative intensity of the amplitudes is taken from Fig.~\ref{fig:dalitz},
while the absolute intensity is shown in arbitrary units.
}
\end{figure*}

The resulting  $(K\pi)^{*0}_0$ invariant mass distribution
is shown in Fig.~\ref{fig:lass-shape}, along with the phase and
distributions for the other resonances.
The mass parameters describing the three spin states in the $m$ 
distribution are shown in Table~\ref{tab:mass}. 
Measurements of the LASS experiment are used for the
parameters of the $J=0$ contribution and for the 
interaction radius~\cite{Aston:1987ir, wmdLASS}. 
The values of $m_0$, $\Gamma_0$, $a$, and $b$ used in this analysis
are different from those quoted in Ref.~\cite{Aston:1987ir} due to better handling
of the fit to the LASS data~\cite{wmdLASS}. The two sets of values are consistent
within errors and lead to similar results. 

\begingroup
\begin{table}[t]
\caption{\label{tab:mass}
Parameterization of the $K\pi$ invariant mass distribution
in the $B^0\to\varphi(K\pi)_J$ decays where
three resonance final states with spin $J=0,1,2$ are considered.
The resonance mass $m_J$, width $\Gamma_J$~\cite{bib:Amsler2008, Aston:1987ir, wmdLASS}, 
interaction radius $r$, scattering length $a$,
and effective range $b$ are considered~\cite{Aston:1987ir, wmdLASS}. 
Combined errors are quoted, except for $(K\pi)^{*0}_0$ where
the systematic errors are quoted last while the central values 
and statistical errors have been updated~\cite{wmdLASS}
with respect to Ref.~\cite{Aston:1987ir}.
}
\begin{center}
{
\begin{ruledtabular}
\setlength{\extrarowheight}{1.5pt}
\begin{tabular}{cccc}
\vspace{-3mm} & & & \\
              & $(K\pi)^{*0}_0$ & $K^*(892)^0$ & $K^{*}_2(1430)^0$ \cr
              & $J=0$                   & $J=1$              & $J=2$                   \cr
\vspace{-3mm} & & &  \\
\hline
\vspace{-3mm} & & & \\
 $m_J$ (MeV)        & $1435\pm 5\pm 5$   & $896.00\pm 0.25$  & $1432.4\pm 1.3$  \\
 $\Gamma_J$ (MeV)   & $279\pm 6\pm21$    & $50.3\pm 0.6$  & $109\pm 5$  \\
 $r$  (GeV$^{-1}$)         &-- &$3.4\pm 0.7$ &$2.7\pm 1.3$ \\
 $a$  (GeV$^{-1}$)         & $1.95\pm 0.09\pm0.06$  &-- &-- \\
 $b$  (GeV$^{-1}$)         & $1.76\pm 0.36\pm0.67$  &-- &-- \\
\vspace{-3mm} & & & \\
\end{tabular}
\end{ruledtabular}
}
\end{center}
\end{table}
\endgroup

To account for the three-body kinematics in the analysis of $B^0\to{\varphi K\pi}$ 
decays, we multiply the amplitude squared $|M_J(m)|^2$ by the 
phase-space factor $F(m)$:
\begin{equation}
F(m)= 2\times m\times\left[m^2_{\rm max}(m)-m^2_{\rm min}(m)\right],
\label{eq:dalitzfactor}
\end{equation}
where $m^2_{\rm max}$ and $m^2_{\rm min}$ are the maximum and minimum values
of the Dalitz plot range of $m^2_{\varphi K}$ ($m_{\varphi K}$ is the $\varphi K$ invariant mass) at any given value of $m_{K\!\pi}$,
see kinematics section of Ref.~\cite{bib:Amsler2008}.
Due to slow dependence of the factor in Eq.~(\ref{eq:dalitzfactor}) on
$m$ in any small range of $m$, the difference of this
approach from the quasi-two-body approximation is small.


\vspace*{10pt}
\subsection{Interference effects}
\label{sec:note-interference}

The differential decay width discussed in Eq.~(\ref{eq:helicityfull})
involves interference terms between resonances with different spins $J$.
These interference terms have unique angular and mass dependences 
which cannot be factorized in the full distribution. We can parameterize the mass and angular amplitude for
each spin state $J$ as follows:

\begin{widetext}
\begin{eqnarray}
\label{eq:inter-01}
A_0(m_{K\!\pi},\theta_1,\theta_2,\Phi)&=&  Y_{0}^0({\cal H}_1,\Phi) Y_{1}^0(-{\cal H}_2,0) M_0(m_{K\!\pi})A_{00} \\
\vspace{-5mm}\nonumber\\
\label{eq:inter-02}
A_1(m_{K\!\pi},\theta_1,\theta_2,\Phi)&=&  \sum_{\lambda=0,\pm1} Y_{1}^\lambda({\cal H}_1,\Phi) Y_{1}^{-\lambda}(-{\cal H}_2,0) M_1(m_{K\!\pi})A_{1\lambda} \\
\label{eq:inter-03}
A_2(m_{K\!\pi},\theta_1,\theta_2,\Phi)&=&  \sum_{\lambda=0,\pm1} Y_{2}^\lambda({\cal H}_1,\Phi) Y_{1}^{-\lambda}(-{\cal H}_2,0) M_2(m_{K\!\pi})A_{2\lambda}.  
\end{eqnarray}
\end{widetext}

\begingroup
\begin{table*}[t]
\caption{\label{tab:f7functions}
Parameterization of the angular distribution in Eq.~(\ref{eq:fang-interf}).
Interference between either  $J=0$ and $J=1$, or $J=0$ and $J=2$, 
contributions in the $B^0\to\varphi(K\pi)_J$ decays
is considered. 
The common constant is quoted for each decay mode and is omitted
from each individual function below.
The three helicity angle parameters $({\cal H}_1,\,{\cal H}_2,\,\Phi)$
are discussed in the text.
}
\begin{center}
{
\begin{ruledtabular}
\setlength{\extrarowheight}{1.5pt}
\begin{tabular}{ccc}
\vspace{-3mm} & & \\
              & $\varphi K^*(892)/\varphi(K\pi)^{*}_0$ & $\varphi K^{*}_2(1430)/\varphi(K\pi)^{*}_0$ \cr
              & $J=1$               & $J=2$                 \cr
\vspace{-3mm} & & \\
\hline
\vspace{-3mm} & & \\
common constant & ${3\sqrt{3}}/{4\pi}$ & $3\sqrt{5}/{8\pi}$ \\
\vspace{-3mm} & & \\
${f}_{7J}\,({\cal H}_1,\,{\cal H}_2,\,\Phi)$
 & ${\cal H}_1{\cal H}_2^2$
 & $(3{\cal H}_1^2-1){\cal H}_2^2$ \\
\vspace{-3mm} & & \\
${f}_{8J}\,({\cal H}_1,\,{\cal H}_2,\,\Phi)$
 & $\frac{1}{\sqrt{2}}\sqrt{1-{\cal H}_1^2}\sqrt{1-{\cal H}_2^2}\,{\cal H}_2\cos\Phi$
 & ${\sqrt{6}}\sqrt{1-{\cal H}_1^2}\,{\cal H}_1\sqrt{1-{\cal H}_2^2}\,{\cal H}_2\cos\Phi$ \\
\vspace{-3mm} & & \\
${f}_{9J}\,({\cal H}_1,\,{\cal H}_2,\,\Phi)$
 & $-\frac{1}{\sqrt{2}}\sqrt{1-{\cal H}_1^2}\sqrt{1-{\cal H}_2^2}\,{\cal H}_2\sin\Phi$
 & $-{\sqrt{6}}\sqrt{1-{\cal H}_1^2}\,{\cal H}_1\sqrt{1-{\cal H}_2^2}\,{\cal H}_2\sin\Phi$ \\
\vspace{-3mm} & & \\
\end{tabular}
\end{ruledtabular}
}
\end{center}
\end{table*}
\endgroup

The interference will appear in the angular-mass distributions as
$2\Ree (A_i(m_{K\!\pi},\theta_1,\theta_2,\Phi)A_j^*(m_{K\!\pi},\theta_1,\theta_2,\Phi))$.
As we can see from Fig.~\ref{fig:lass-shape}, the overlap between the
$P$- and $D$-wave $K\pi$ contributions is negligibly small, and we will consider
only the interference between the $J=0$ and $J=1$, or $J=0$ and $J=2$ amplitudes.
The resulting two interference terms, properly normalized, are defined for $J=1$ and $J=2$:
\begin{eqnarray}
\frac{2\Ree(A_JA_0^{\ast})}{ \sqrt{\Sigma|A_{J\lambda}|^2}|A_{00}|}
 = \sum_{i=7}^{9}\alpha_{iJ}^-(m_{K\!\pi})\,{f}_{iJ}({\cal H}_1,\,{\cal H}_2,\,\Phi)\,,
\label{eq:fang-interf}
\end{eqnarray}
where the angular dependence is defined in Table~\ref{tab:f7functions}, 
and  $\alpha_{iJ}^-(m_{K\!\pi})$ are defined for  $i=7,8,9$ as:
\begin{widetext}
\begin{eqnarray}
\label{eq:alpha7}
\alpha_{7J}^- (m_{K\!\pi}) &=& 
\sqrt{f_{LJ}^-}~\Ree(M_J(m_{K\!\pi})M^{\ast}_0(m_{K\!\pi})e^{-i\delta_{0J}^-})\\
\label{eq:alpha8}
\alpha_{8J}^- (m_{K\!\pi}) &=& 
\sqrt{1-f_{LJ}^--f_{\perp J}^-}~\Ree(M_J(m_{K\!\pi})M^{\ast}_0(m_{K\!\pi})e^{i\phi_{\parallel J}^-}e^{-i\delta_{0J}^-})\\
\label{eq:alpha9}
\alpha_{9J}^- (m_{K\!\pi}) &=& 
\sqrt{f_{\perp J}^-}~\Imm(M_J(m_{K\!\pi})M^{\ast}_0(m_{K\!\pi})e^{i\phi_{\perp J}^-}e^{-i\delta_{0J}^-}).
\end{eqnarray}
\end{widetext}
The above terms are specific to the $B^0$ decays and are denoted with
superscript ``$-$'', as introduced in Eqs.~(\ref{eq:brpm}--\ref{eq:d0pm}).
The interference distributions for the $\Bbar^0$ decays are described by the
same Eq.~(\ref{eq:fang-interf}), but replacing $A$ by $\Abar$
and $\alpha_{iJ}^-$ by $\alpha_{iJ}^+$ 
and with definitions given by Eqs.~(\ref{eq:alpha7}--\ref{eq:alpha9}), 
replacing superscript ``$-$'' by ``$+$''.

The main difference now is that the $\alpha_{iJ}^-(m_{K\!\pi})$ parameters, as defined 
for $i=7,8,9$, have a different dependence on mass to those defined for $i=1$-$6$ 
in Eqs.~(\ref{eq:alpha1}--\ref{eq:alpha6}). This dependence now includes the phase of
the resonance amplitude as a function of mass. This dependence becomes crucial
in resolving the phase ambiguities.

As can be seen from Eq.~(\ref{eq:fangular}) and Eqs.~(\ref{eq:alpha1}--\ref{eq:alpha6}),
for any given set of values 
$(\phi_{\parallel J}, \phi_{\perp J}, \Delta\phi_{\parallel J}, \Delta\phi_{\perp J})$ a 
simple transformation of phases, for example 
$(2\pi-\phi_{\parallel J},\pi-\phi_{\perp J}, -\Delta\phi_{\parallel J}, -\Delta\phi_{\perp J})$,
gives rise to another set of values that satisfy the above equations in an identical
manner. This results in a four-fold ambiguity (two-fold for each of $B^0$ and $\Bbar^0$ decays).
At any given value of $m_{K\!\pi}$ the distributions, including the interference terms
in Eqs.~(\ref{eq:alpha8}) and (\ref{eq:alpha9}), are still invariant under the above transformations
if we flip the sign of the phase,  
arg$(M_J(m_{K\!\pi})M^{\ast}_0(m_{K\!\pi})e^{-i\delta_{0J}^\pm})$.
At a given value of $m_{K\!\pi}$ this phase has to be determined from the
data and we cannot resolve the ambiguity. 
However, the mass dependence of this phase is unique, given that the parameters $\delta_{0J}^\pm$
are constant. Therefore, the two ambiguous solutions
for each  $B^0$ and $\Bbar^0$ decay can be fully resolved from the 
$m_{K\!\pi}$ dependence of the angular distributions in 
Eq.~(\ref{eq:fang-interf}).

This technique of resolving the two ambiguous solutions 
in $B\to VV$ decays has been 
introduced in the analysis of $B^0\to J/\psi K^{*0}$ decays~\cite{jpsikpi}
and has been used in $\babar$'s earlier analysis of both
$B\to\varphi K^{*0}$ and $\varphi K^{*\pm}$ decays~\cite{babar:phikst, babar:phik1}.
It is based on Wigner's causality principle~\cite{wigner},
where the phase of a resonant amplitude increases with increasing invariant mass,
see Eq.~(\ref{eq:mass-bw1}).
As a result, both the $P$-wave and $D$-wave resonance phase shifts increase
rapidly in the vicinity of the resonance, while the corresponding $S$-wave increases only
gradually, as seen in Fig.~\ref{fig:lass-shape}.


\subsection{Time-dependent distributions}
\label{sec:note-time}

Measurement of the time-dependent $\CP$ asymmetry ${\cal A}(\Delta t)$
in the decay of a neutral $B$ meson to a $\CP$ eigenstate, 
dominated by the tree-level $b\to c$ amplitude
or by the penguin $b\to s$ amplitude,
such as ${B^0\rightarrow(c\bar{c})K^0_S}$ or ${B^0\rightarrow(s\bar{s})K^0_S}$,
where $(c\bar{c})$ and $(s\bar{s})$
are charmonium or quarkonium states respectively, 
gives an approximation, $\beta_{\rm eff}$, to the CKM Unitarity Triangle angle
$\beta$~\cite{babarbook}. The $\CP$ asymmetry is defined by 
\begin{eqnarray}
{\cal A}(\Delta t) = \frac{N(\Delta t, B^0_{\rm tag})-N(\Delta t, \Bbar^0_{\rm tag})}{N(\Delta t, B^0_{\rm tag})+N(\Delta t, \Bbar^0_{\rm tag})} 
~~~~~~~~~~~~~~~~~~~~
\nonumber \\
 = S\sin(\Delta{m_B}\Delta t) - C\cos(\Delta{m_B}\Delta t)\, ,~~ \label{eq:timedep1}
\label{eq:adeltat}
\end{eqnarray}
%
and
%
\begin{eqnarray}
-\sin({ 2\beta_{\rm eff}})=
{\Imm}\left(\frac{q}{p} {\frac{\Abar}{A} }\right)\displaystyle\bigg/
\left|\frac{q}{p} {\frac{\Abar}{A} }\right|
~~~~~~~~~~~~~~
\nonumber \\
=\eta_{\scriptscriptstyle C\!P}\times{ S }\displaystyle/\sqrt{1-{ C }^2}\, ,~~~~~~~~~~~~~~
\label{eq:timedep}
\end{eqnarray}
where $N(\Delta t, B^0_{\rm tag})$ or 
$N(\Delta t, \Bbar^0_{\rm tag})$ is the number of events observed to
decay at time $\Delta t$, in which the flavor of the $B$ meson opposite to 
that of the decaying $B$ at $\Delta t=0$ (referred to as the flavor ``tag'') 
is known to be $B^0$ or $\Bbar^0$ respectively,
$\eta_{\scriptscriptstyle C\!P}=\pm1$ is the $C\!P$ eigenvalue
of the final state; amplitudes ${A}$ and $\Abar$ describe
the direct decays of $B^0$ and $\Bbar^0$ respectively to the final state;
and $\Delta{m_B}$ is the mixing frequency due to the difference in masses 
between the $B$ meson eigenstates.
We use a convention with ${\Abar}=\eta_{\scriptscriptstyle C\!P}\times{A}$
in the absence of $C\!P$ violation.
The above asymmetry follows from the time evolution of each flavor:
\begin{eqnarray}
N(\Delta t, B^0_{\rm tag})\propto\frac{e^{-\left|\Delta t\right|/\tau_B}}{4\tau_B} 
~~~~~~~~~~~~~~~~~~~~~~~~~~~~~~~~
\nonumber
\\
\times \left( 1 + S\sin(\Delta m_B\Delta t)-C\cos(\Delta m_B\Delta t)\right)
\label{eq:deltatb}
\\
N(\Delta t, \Bbar^0_{\rm tag})\propto\frac{e^{-\left|\Delta t\right|/\tau_B}}{4\tau_B} 
~~~~~~~~~~~~~~~~~~~~~~~~~~~~~~~~
\nonumber
\\
\times \left( 1 - S\sin(\Delta m_B\Delta t)+C\cos(\Delta m_B\Delta t)\right).
\label{eq:deltatbar}
\end{eqnarray}

The $B^0$-$\Bbar^0$ mixing parameters ${q}$ and ${p}$ can be expressed 
to a good approximation using the Wolfenstein phase convention within the Standard Model~\cite{bib:Amsler2008}:
\begin{eqnarray}
\label{eq:qtopphase}
{\rm arg}\left(\frac{q}{p}\right) = -2\beta  \\
\label{eq:qtopratio}
\left|\frac{q}{p}\right| = 1. 
\end{eqnarray}

The value of $\sin2\beta$ from the charmonium $b\to c$
decays, which defines the phase of the mixing diagram in the
Wolfenstein parameterization, is well measured~\cite{babarcp,bib:Amsler2008}:
\begin{eqnarray}
\sin2\beta = 0.681\pm 0.025 
\nonumber \\
{\rm or} ~~2\beta = (0.75 \pm 0.03)~{\rm rad}
\label{eq:sin2beta}
\end{eqnarray}
where the phase ambiguity of $\beta$ in the range $[0,\pi]$ has been
resolved using vector-vector charmonium $B$ decays~\cite{jpsikpi} and 
the decay $B^0\to K^+ K^- K^0$~\cite{BtoKKK}. 
Should there be a  
new physics contribution to the mixing diagram, its effect is absorbed into the 
definition of $\beta$ in Eq.~(\ref{eq:qtopphase}), which should be 
valid for the analysis discussed in this paper.
New physics effects in the $b\to c$ 
 amplitude are unlikely to be significant as this transition is not
 suppressed in the Standard Model. 
Therefore the comparison of $\sin2\beta$ in Eq.~(\ref{eq:sin2beta}) 
with $\sin2\beta_{\rm eff}$  measured in $b\to s$ transitions would be a test of new physics
in the penguin $B$ decays.

There is an alternative sign convention for the choice of the 
direct-${C\!P}$ violation parameter defined by: 
\begin{eqnarray}
C = -{\cal A}_{C\!P} = {|A|^2 - |\Abar|^2 \over |A|^2 + |\Abar|^2}.
\label{eq:c}
\end{eqnarray}

Given the approximation in Eq.~(\ref{eq:qtopratio})
and our phase convention in Eq.~(\ref{eq:qtopphase}), 
the value of $S$ can be expressed as:
\begin{eqnarray}
S = 
\sqrt{1-{\cal A}_{C\!P}^2} \times
\sin\left({ -2\beta + {\rm arg}\left( {\Abar \over {A}} \right) }\right) \, .
\label{eq:s}
\end{eqnarray}

Therefore, we have $S = -\eta_{\scriptscriptstyle C\!P}\times\sin2\beta$
when ${\Abar}=\eta_{\scriptscriptstyle C\!P}\times{A}$.
When we measure $S$ in Eq.~(\ref{eq:s}) with $b\to s$ decays,
we can safely assume that the value of $\beta$ has been measured in
charmonium decays, as given in Eq.~(\ref{eq:sin2beta}).
Therefore, we are ultimately interested in the measurement of
${\rm arg}( {\Abar / {A}} )$.
Any large deviation from arg($\eta_{\scriptscriptstyle C\!P}$) 
would be a signal of new physics.

In the study of the time-evolution in Eq.~(\ref{eq:timedep1}) we can use the
decay $B^0\to\varphi (K_S^0\pi^0)^{*0}_0$ with an $S$-wave $K\pi$ contribution.
This final state is a ${C\!P}$-eigenstate with 
$\eta_{\scriptscriptstyle C\!P} = +1$ as we discuss below.
However, 
the situation is more complicated in the general case of 
$B^0\to\varphi K_S^0\pi^0$ decays
where the final state is no longer a $C\!P$-eigenstate.
The amplitude for this decay is a superposition of  $C\!P$ eigenstates.

In Ref.~\cite{Dunietz1991} it was shown that the $C\!P$ quantum numbers
are independent of the $K_S^0\pi^0$ system, that is independent of $J$, 
for the decay $B^0\to(c\bar{c})K_S^0\pi^0$. The same analysis applies to $B^0\to(s\bar{s})K_S^0\pi^0$.
The $C\!P$ parity is defined
only by the $(s\bar{s})$ spin alignment $\lambda$
(an alternative analysis that introduces the 
eigenstate of the transversity $\tau$ is 
sometimes used to separate the $C\!P$ eigenstates~\cite{Dunietz1991}).
For example, all longitudinal decays $B^0\to\varphi K_S^0\pi^0$,
corresponding to $\lambda=0$, are $C\!P$-even, including the 
decay $B^0\to\varphi (K_S^0\pi^0)^{*0}_0$.
Overall, we conclude that in the decay $B^0\to\varphi(K_S^0\pi^0)_J$
we have three amplitudes with definite $C\!P$:
\begin{eqnarray}
A_{J0} & &  ~~~~~~\eta_{\scriptscriptstyle C\!P}=+1 \\
A_{J\parallel}& = ~ (A_{J+}+A_{J-})/\sqrt{2} & ~~~~~~\eta_{\scriptscriptstyle C\!P}=+1 \\
A_{J\perp}& = ~ (A_{J+}-A_{J-})/\sqrt{2} & ~~~~~~\eta_{\scriptscriptstyle C\!P}=-1.
\label{eq:cpphikst}
\end{eqnarray}
Similarly, $\eta_{\scriptscriptstyle C\!P}=-1$ for the $B$ decay to the final 
state $f_0 K_S^0\pi^0$, which will be considered as a background
decay in our analysis.
We do not discuss the $C\!P$ properties of the interference terms with the
product of amplitudes of different $C\!P$; these terms are integrated 
over in our analysis of the time evolution. These terms could be 
considered in a future experiment with higher statistics.

We can express the time-evolution coefficient in the
decay $B^0\to\varphi(K_S^0\pi^0)^{*0}_0$,  as
\begin{eqnarray}
S_{00} = 
-\sqrt{1-{\cal A}_{00}^2} \times \sin({ 2\beta + 2\Delta\phi_{00} })\, ,
\label{eq:sphikstzero1430}
\end{eqnarray}
where ${\cal A}_{00}={\cal A}_{C\!P0}$.
For the decays $B^0\to\varphi(K_S^0\pi^0)$
with $K^{*}(892)^0$ ($J=1$) or  $K^{*}_2(1430)^0$ ($J=2$)  
intermediate resonances there are three time-evolution terms, 
one for each amplitude, if we ignore the interference terms:
\begin{widetext}
\begin{eqnarray}
S_{J0} & = & -\sqrt{1-{\cal A}_{J0}^2} 
         \times \sin({ 2\beta + 2\Delta\delta_{0J}  + 2\Delta\phi_{00}  }) 
\label{eq:sphikst892a} \\
S_{J\parallel} & = & -\sqrt{1-{\cal A}_{J\parallel}^2} 
         \times \sin({ 2\beta + 2\Delta\delta_{0J} - 2\Delta\phi_{\parallel J} + 2\Delta\phi_{00}  })
\label{eq:sphikst892b} \\
S_{J\perp} & = & +\sqrt{1-{\cal A}_{J\perp}^2} 
         \times \sin({ 2\beta + 2\Delta\delta_{0J} - 2\Delta\phi_{\perp J} + 2\Delta\phi_{00}  }). 
\label{eq:sphikst892c}
\end{eqnarray}
The three corresponding direct-$C\!P$ violation terms ${\cal A}_{J0}$, ${\cal A}_{J\parallel}$,
and ${\cal A}_{J\perp}$ can be obtained from the direct-$C\!P$ violation and polarization
parameters measured in the $\BorBbar^0\to\varphi(K^{\pm}\pi^{\mp})$ decays:
\begin{eqnarray}
{\cal A}_{J0} & = & \frac{{\cal A}_{C\!PJ}+{\cal A}_{C\!PJ}^0}{1+{\cal A}_{C\!PJ}\times{\cal A}_{C\!PJ}^0} 
\label{eq:acpphikst892a} \\
{\cal A}_{J\perp} & = & \frac{{\cal A}_{C\!PJ}+{\cal A}_{C\!PJ}^\perp}{1+{\cal A}_{C\!PJ}\times{\cal A}_{C\!PJ}^\perp} 
\label{eq:acpphikst892b} \\
{\cal A}_{J\parallel} & = & 
\frac{
{\cal A}_{C\!PJ}-f_{LJ}\times({\cal A}_{C\!PJ}+{\cal A}_{C\!PJ}^0)-f_{\perp J}\times({\cal A}_{C\!PJ}+{\cal A}_{C\!PJ}^\perp)
}
{
1-f_{LJ}\times(1+{\cal A}_{C\!PJ}\times{\cal A}_{C\!PJ}^0)-f_{\perp J}\times(1+{\cal A}_{C\!PJ}\times{\cal A}_{C\!PJ}^\perp)
}.
\label{eq:acpphikst892c}
\end{eqnarray}
\end{widetext}

As can be seen from Eqs.~(\ref{eq:sphikstzero1430}--\ref{eq:acpphikst892c}),
there is only one parameter $\Delta\phi_{00}$ that is not measurable in the
$\BorBbar^0\to\varphi(K^{\pm}\pi^{\mp})$ decays. Therefore, the above parameterization allows
us to measure $\Delta\phi_{00}$ from the time evolution of the $B^0\to\varphi(K_S^0\pi^0)$
decays while all other parameters are measured in the mode
$\BorBbar^0\to\varphi(K^{\pm}\pi^{\mp})$, which has a significantly larger reconstructed yield.

The angular distributions in Eq.~(\ref{eq:helicityfull}) can be simplified
after integrating over the angle $\Phi$.  The resulting angular distribution 
will not have interference terms between different amplitudes for a given $J$.
This makes the time evolution parameterization relatively simple
with just two terms: longitudinal ($f_{LJ}$) and transverse ($1-f_{LJ}$) polarization.
The longitudinal time-evolution is parameterized by the $S_{J0}$ 
coefficient, and the transverse time-evolution is parameterized by the expression
\begin{eqnarray}
S_{JT}=\frac{f_{\perp J}\times(1+{\cal A}_{C\!PJ}\times{\cal A}_{C\!PJ}^{\perp})}
	{1-{f_{LJ}\times(1+{\cal A}_{C\!PJ}\times{\cal A}_{C\!PJ}^0)}}\times{\cal S}_{J\perp}\nonumber\\
	+\left(1-\frac{f_{\perp J}\times(1+{\cal A}_{C\!PJ}\times{\cal A}_{C\!PJ}^{\perp})}
	{1-{f_{LJ}\times(1+{\cal A}_{C\!PJ}\times{\cal A}_{C\!PJ}^0)}}\right)\times{\cal S}_{J\parallel}\,.
\label{eq:sjt}
\end{eqnarray}
In a similar manner,
the longitudinal direct ${C\!P}$-violation term is parameterized by the $C_{J0}=-{\cal A}_{J0}$ 
coefficient, and the transverse direct ${C\!P}$-violation term is parameterized by 

\begin{eqnarray}
C_{JT}=-\frac{f_{\perp J}\times(1+{\cal A}_{C\!PJ}\times{\cal A}_{C\!PJ}^{\perp})}
	{1-{f_{LJ}\times(1+{\cal A}_{C\!PJ}\times{\cal A}_{C\!PJ}^0)}}\times{\cal A}_{J\perp}\nonumber\\
	-\left(1-\frac{f_{\perp J}\times(1+{\cal A}_{C\!PJ}\times{\cal A}_{C\!PJ}^{\perp})}
	{1-{f_{LJ}\times(1+{\cal A}_{C\!PJ}\times{\cal A}_{C\!PJ}^0)}}\right)\times{\cal A}_{J\parallel}\,.
\label{eq:cjt}
\end{eqnarray}

As an example, let us consider the $J=1$ case.
Since $A_{1\perp}$ and $A_{1\parallel}$ have opposite $C\!P$-parity
and it has been measured that 
$f_{\perp1}\simeq f_{\parallel1} \equiv (1-f_{\perp1}-f_{L1})$~\cite{babar:vv,babar:phikst,belle:phikst},
in the Standard Model we expect to a good approximation
$S_{1T}\simeq0$, $S_{10}=\sin2\beta$, and $C_{1T}=C_{10}=0$.


\section{EVENT RECONSTRUCTION}
\label{sec:note-reco}

We use a sample of $(465.0\pm 5.1)$ million  $e^+e^-\to\FourS\to\BB$ events
collected with the \babar\ detector~\cite{babar} at the PEP-II $e^+e^-$
asymmetric-energy storage rings. 
The center-of-mass system of the $\FourS$ resonance is boosted
providing roughly 250 $\mu$m average separation between the two
$B$ meson decay vertices.
The $e^+e^-$ center-of-mass energy $\sqrt{s}$ is equal to $10.58$ GeV,
corresponding to the $\FourS$ resonance.

We fully reconstruct the $\BorBbar^0\to\varphi(1020)\KorKbar^{*0}\to(K^+K^-)(\KorKbar\pi)$ 
candidates with two $(K\pi)$ final states, $K^0_S\pi^0$ and $K^\pm\pi^\mp$.
The neutral pseudoscalar mesons are reconstructed
in the final states $K^0_S\to\pi^+\pi^-$ and $\pi^0\to\gamma\gamma$.
The dominant background in our analysis comes from  
$e^+e^-\to \qqbar$ production ($q=u,d,s,c$). 
A data sample equivalent in luminosity to 
$12\%$ of the on-$\FourS$-resonance sample has been collected with $\sqrt{s}$ 40 MeV 
below the $\FourS$ resonance (off-resonance data) for studies of this background.
A detailed GEANT4-based Monte Carlo (MC) simulation~\cite{Agostinelli:2002hh}
of the detector has been used to model all processes. This simulation has been
extensively tested and tuned with high-statistics validation samples.

Momenta of charged particles are measured
in a tracking system consisting of a silicon vertex tracker (SVT) with five
double-sided layers and a 40-layer drift chamber (DCH), both within the 1.5-T
magnetic field of a solenoid.
Identification of charged particles (PID) is provided
by measurements of the energy loss in the tracking devices ($dE/dx$) and by
a ring-imaging Cherenkov detector (DIRC).
Photons are detected by a CsI(Tl) electromagnetic calorimeter (EMC).
We use minimal information from the muon identification system (IFR)
to make a loose veto of the charged muon tracks.

We require all charged-particle tracks (except for those from the
$K^0_S\to\pi^+\pi^-$ decay) used in reconstructing the $B$ candidate to
originate from within 1.5 cm in the $x$-$y$ plane and 10 cm
in the $z$ direction from the nominal beam spot.  
We veto leptons from our charged-particle track samples 
by demanding that tracks have DIRC, EMC,
and IFR signatures that are inconsistent with either electrons or muons.
Further pion and kaon PID requirements are
based on a likelihood selection developed from $dE/dx$ and
Cherenkov angle information from the tracking detectors and DIRC, respectively.
The typical efficiency of PID requirements is greater than 95\% for
charged tracks in our final states. 
Photons are reconstructed from energy deposits in the electromagnetic
calorimeter that are not associated with a charged track.  We require 
that all photon candidates have an energy greater than 30 MeV in the EMC.

The invariant mass of the candidate
$K^0_S$ is required
to lie within the range $|m_{\pi^+\pi^-} - m_{K\!^0}| < 12$ MeV.  We perform a
vertex-constrained fit to require that the two
tracks originate from a common vertex, and require that the lifetime
significance of the $K^0_S$ be $\tau/\sigma_{\tau}>5$, where
$\tau$ and $\sigma_{\tau}$ are the $K^0_S$ lifetime and its uncertainty 
determined from the vertex-constrained fit.
For the $K^0_S$ candidates, we also require the cosine of the angle
between the flight direction from the interaction point
and momentum direction to be greater than 0.995.
The efficiency of $K_S^0$ selection requirements is about 90$\%$.

We select neutral-pion candidates from two photon clusters with the
requirement that the $\gamma\gamma$ 
invariant mass satisfy $120 < m_{\gamma\gamma} < 150$ MeV.  The mass of a $\pi^0$ 
candidate meeting this criterion is then constrained to the nominal
value~\cite{bib:Amsler2008} and, when
combined with other tracks or neutrals to form a $B$ candidate, to
originate from the $B$ candidate vertex.  This
procedure improves the mass and energy resolution of the parent particle.
The purity of $K_S^0$ and $\pi^0$ selection is 92$\%$ (88$\%$)
and 90$\%$ (68$\%$), respectively, in the signal sample (combinatorial
background) based on MC simulation studies.

\begin{figure}[t]
\centerline{
\setlength{\epsfxsize}{0.50\linewidth}\leavevmode\epsfbox{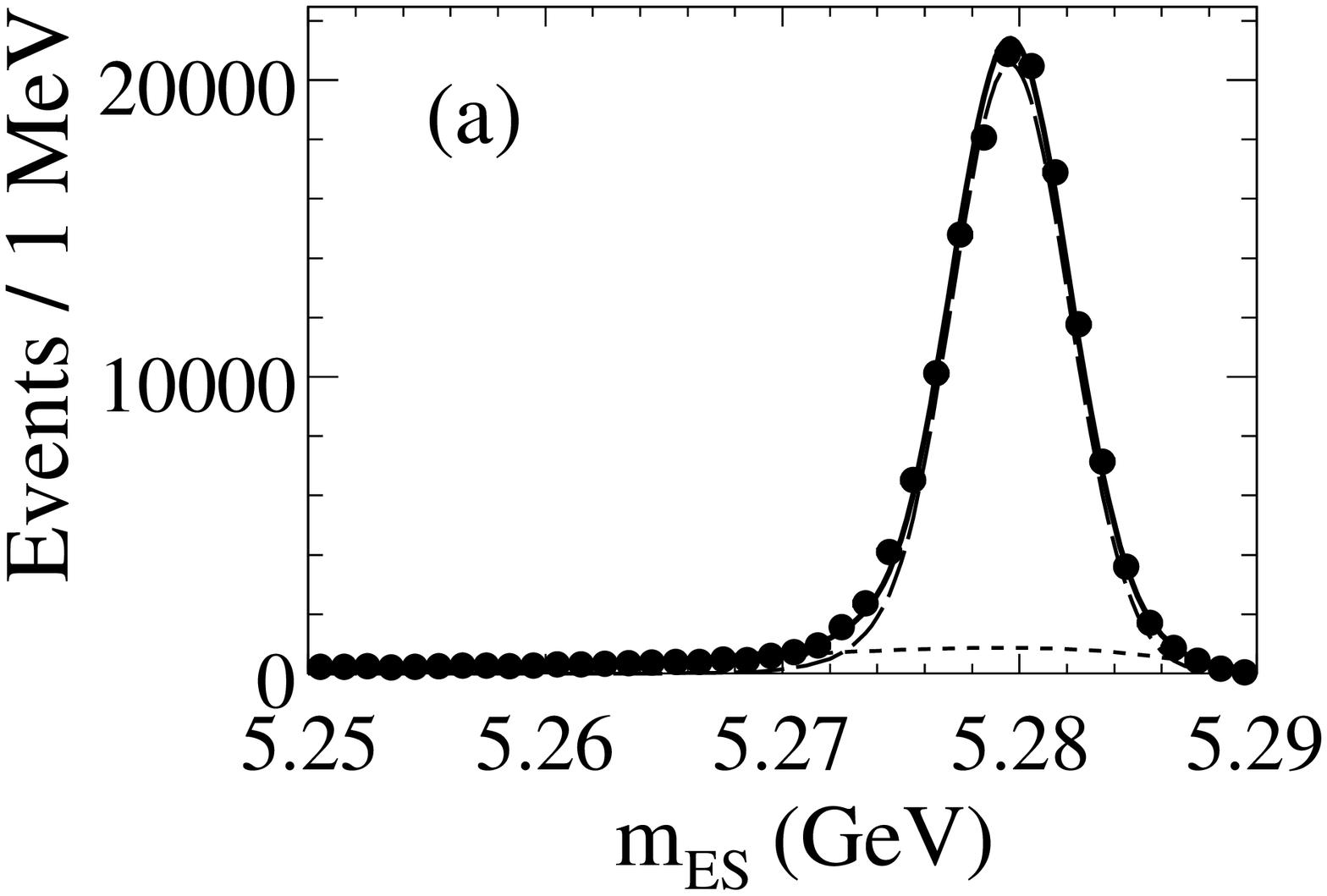}
\setlength{\epsfxsize}{0.50\linewidth}\leavevmode\epsfbox{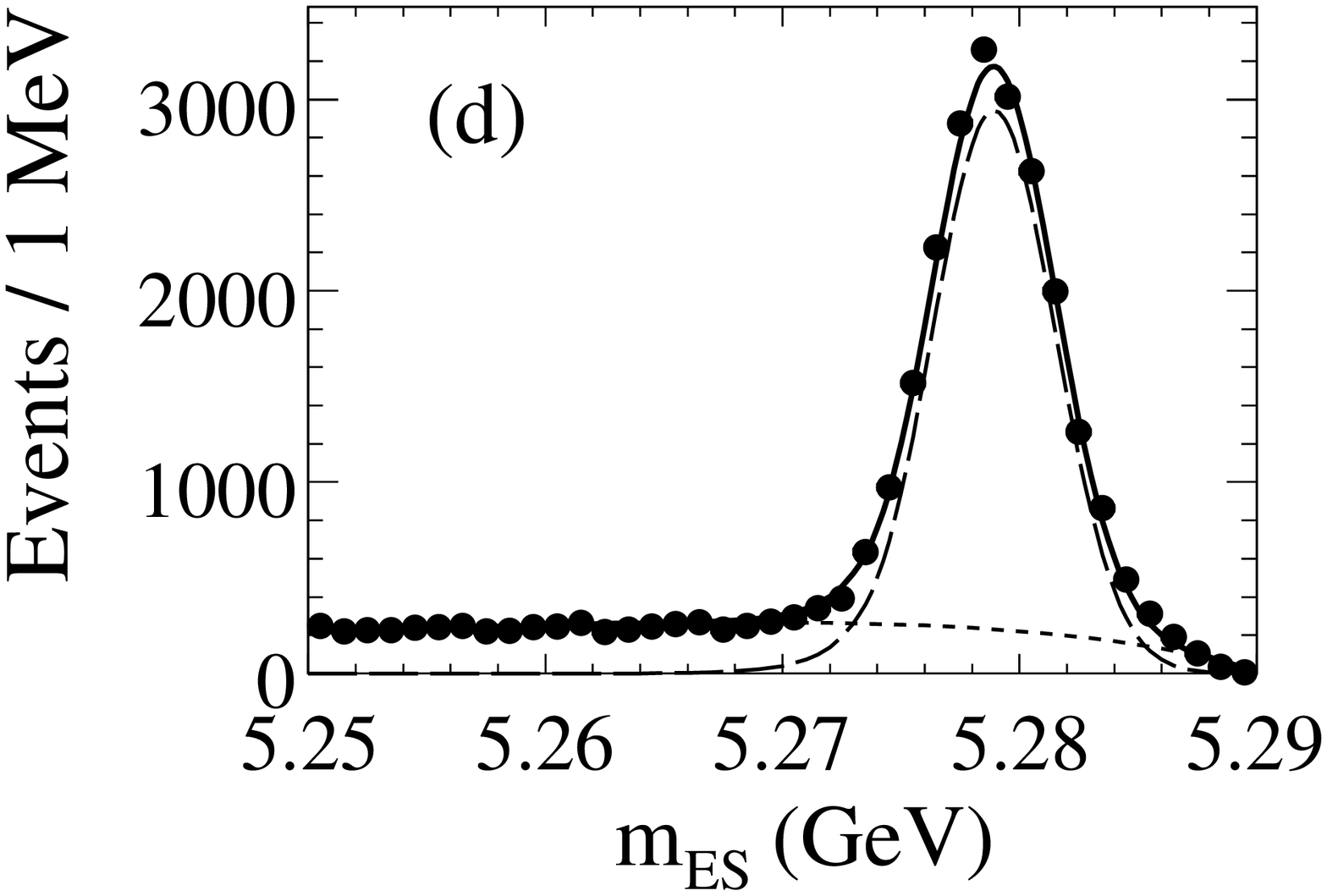}
}
\centerline{
\setlength{\epsfxsize}{0.50\linewidth}\leavevmode\epsfbox{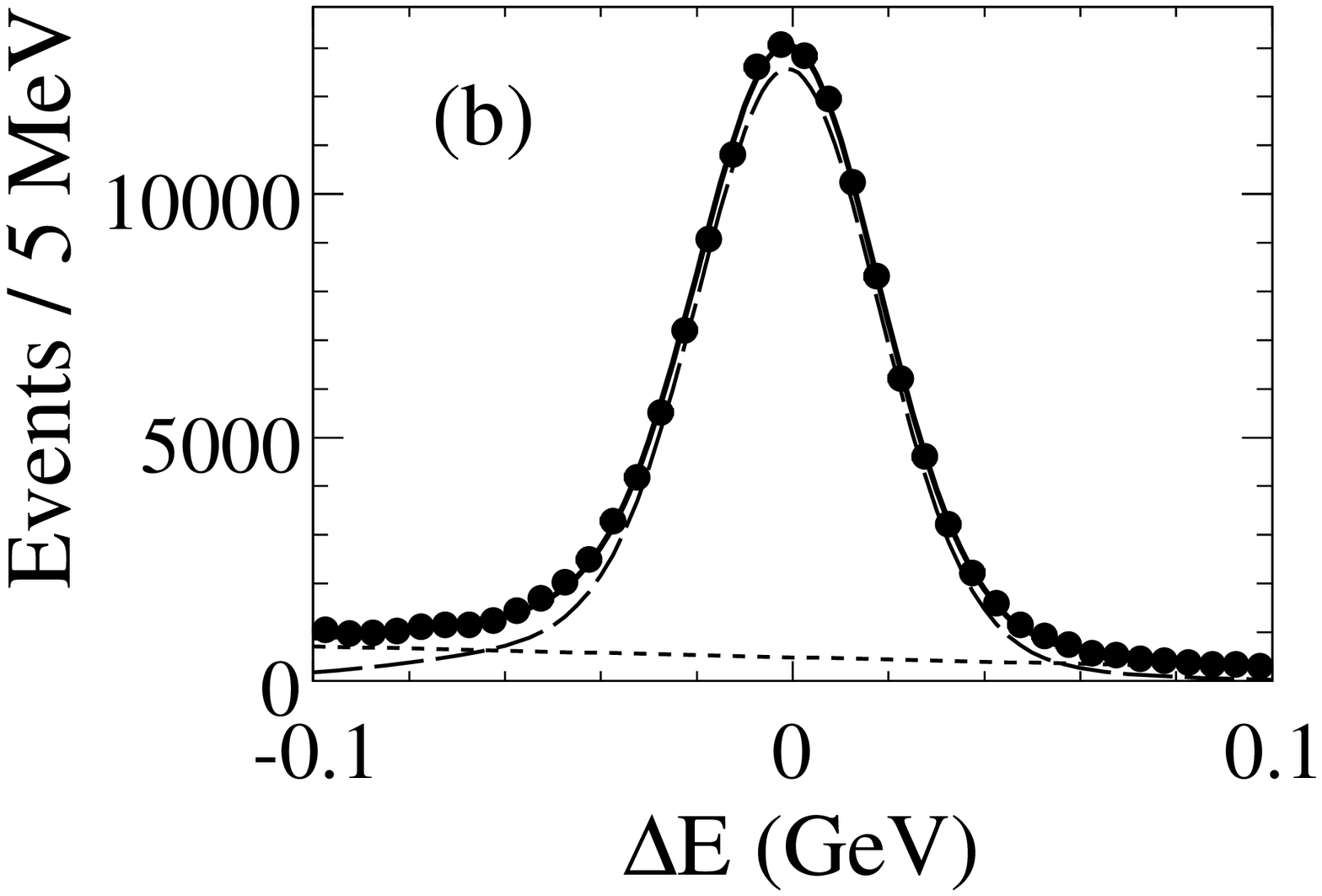}
\setlength{\epsfxsize}{0.50\linewidth}\leavevmode\epsfbox{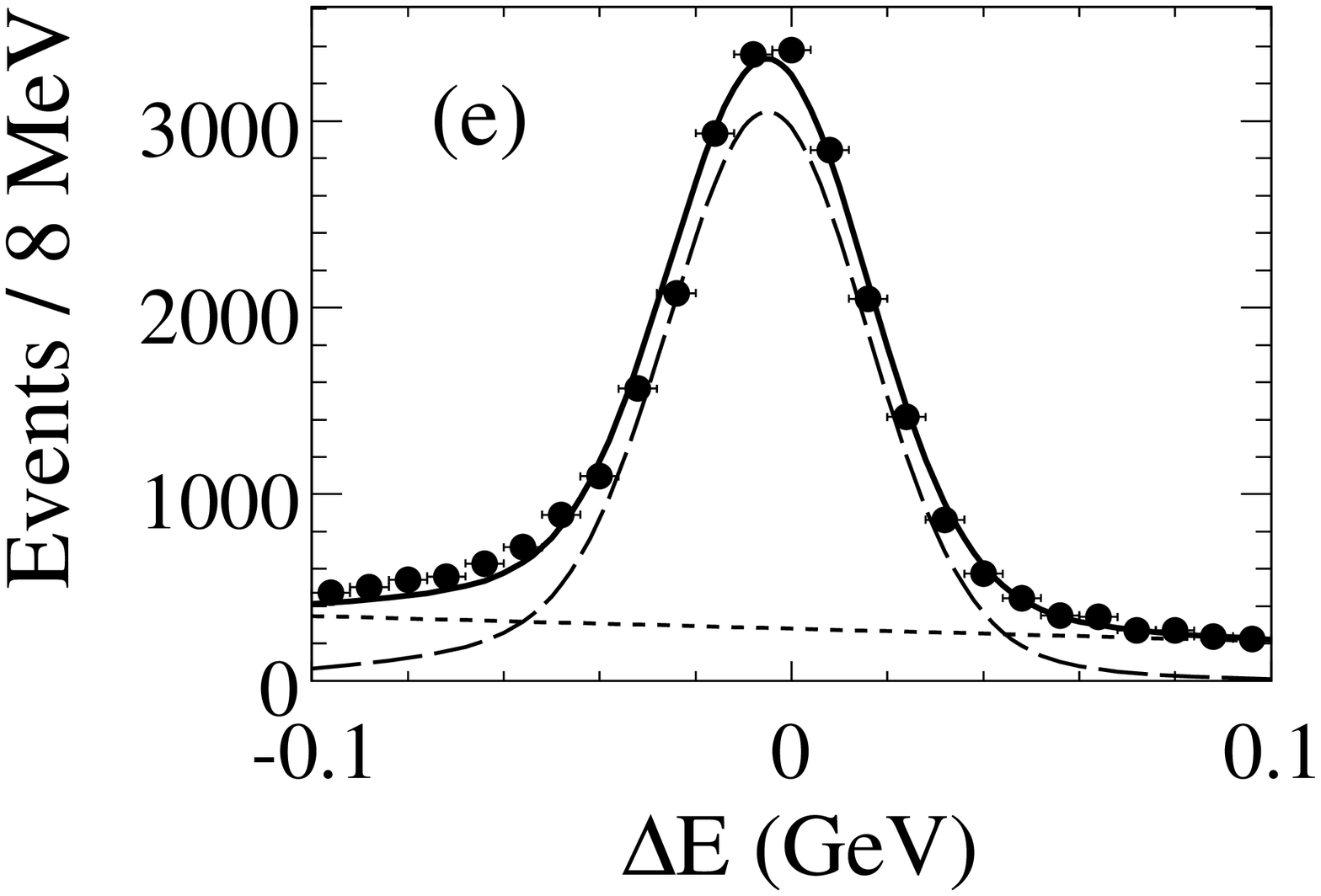}
}
\centerline{
\setlength{\epsfxsize}{0.50\linewidth}\leavevmode\epsfbox{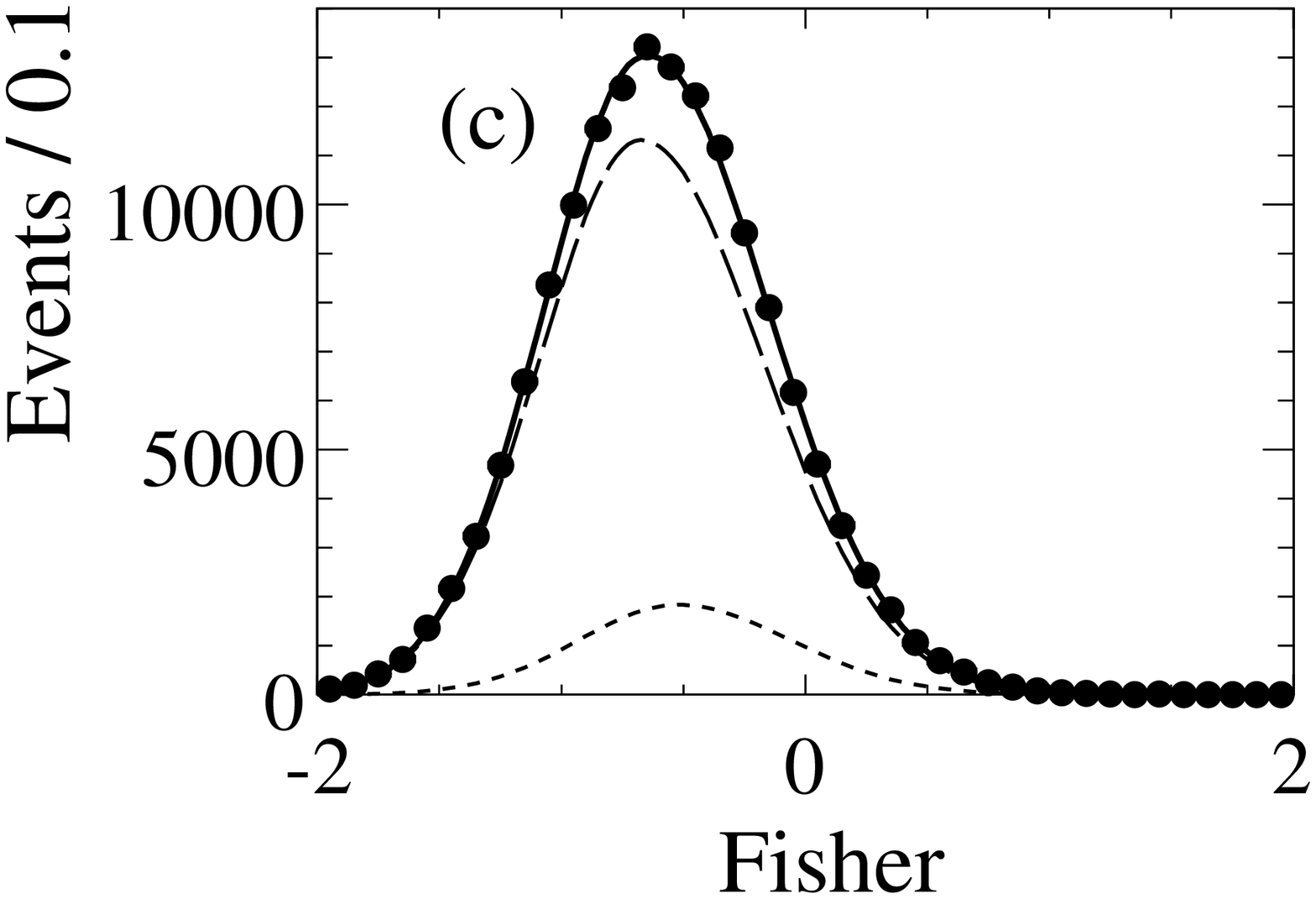}
\setlength{\epsfxsize}{0.50\linewidth}\leavevmode\epsfbox{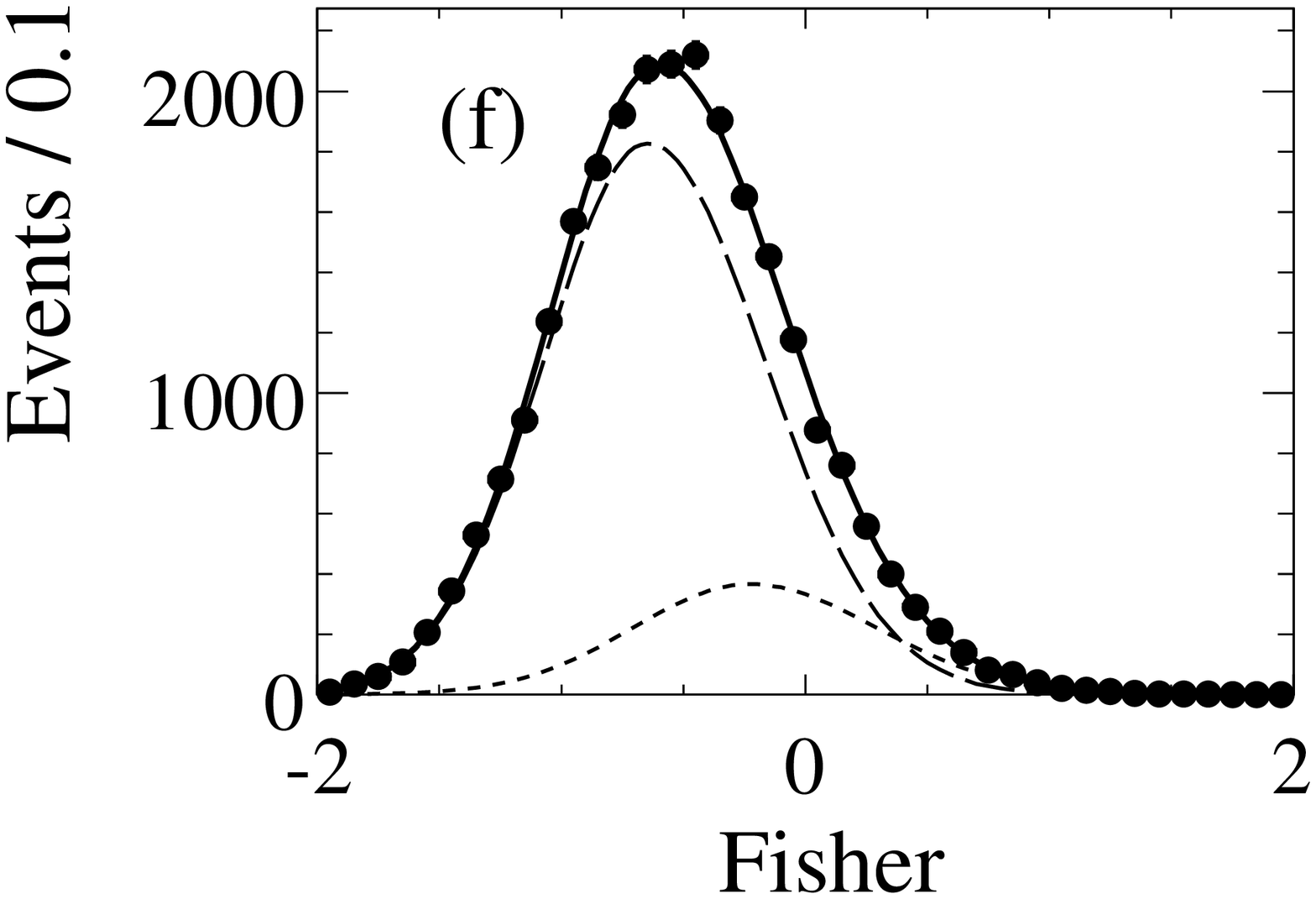}
}
\caption{\label{fig:control}
Validation of kinematic variables with the high statistics sample
of ${B^0\to {D}^-\pi^{+}} \to ({K^+\pi^-\pi^-})({\pi^+})$ decays.
Projections onto the variables $m_{\rm{ES}}$, $\Delta E$, and ${\cal F}$ are 
shown from top to bottom for MC (a), (b), (c), and data (d), (e), (f).
The dots with error bars represent the MC simulation (left) or data (right).
The long dashed lines represent the signal and the solid lines show 
signal-plus-background parameterization. A small fraction
of combinatorial background is present (dashed lines) 
due to combinations of other $B$ decays in the MC and
also $\qqbar$ continuum in the data.
}
\end{figure}

We identify $B$ meson candidates using two main 
kinematic variables, 
beam energy-substituted mass $m_{\rm ES}$:
\begin{equation}
m_{\rm{ES}} = [{ (s/2 + \mathbf{p}_{\Upsilon} \cdot
\mathbf{p}_B)^2 / E_{\Upsilon}^2 - \mathbf{p}_B^{\,2} }]^{1/2},
\end{equation}
and the energy difference $\Delta E$:
\begin{equation}
\Delta{E}=(E_{\Upsilon}E_B-\mathbf{p}_{\Upsilon}\cdot\mathbf{p}_B-s/2)/\sqrt{s},
\end{equation}
where $(E_B,\mathbf{p}_B)$ is the four-momentum of the $B$ candidate,
and $(E_{\Upsilon},\mathbf{p}_{\Upsilon})$ is the $e^+e^-$ initial-state four-momentum,
both in the laboratory frame.
Both variables are illustrated in Fig.~\ref{fig:control}.
The distribution of $\Delta{E}$ is expected to peak at zero and 
$m_{\rm{ES}}$ at the $B$ mass
around 5.280 GeV.
The $\Delta{E}$ resolution is dominated by the decay product energy and momentum
measurements and is typically 34 and 20~MeV for the subchannels
with and without a $\pi^0$, respectively.
The typical $m_{\rm{ES}}$ resolution is 2.6 MeV and is dominated by the
beam energy uncertainties.
We require $m_{\rm{ES}}>5.25$ GeV and $|\Delta{E}|<$ 100~MeV to retain
sidebands for later fitting of parameters to describe the backgrounds.

The requirements on the invariant masses of the resonances are
$0.99 < m_{K\!\Kbar} < 1.05$ GeV and $0.75 < m_{K^{0}_S\pi^{0}} < 1.55$ GeV, 
$0.75 < m_{K\!^{\pm}\pi^{\mp}} < 1.05$ GeV, 
or $1.13 < m_{K\!^{\pm}\pi^{\mp}} < 1.53$ GeV for the $\varphi$ and $K^{*0}$, 
respectively. 
Here we separate the $K^{\pm}\pi^{\mp}$ invariant mass into two ranges
for later fitting. 
The two ranges simplify the fit configuration and allow us to test 
the nonresonant $B\to\varphi K\pi$ contribution independently, 
therefore providing an independent crosscheck of the parameterization.

To reject the dominant $e^+e^-\to \qqbar$ 
background, we use variables calculated in the center-of-mass frame. 
We require $|\cos\theta_T| < 0.8$, where $\theta_T$ is the angle between the $B$-candidate
thrust axis and that of the rest of the event.
The angle $\theta_T$ is the most powerful of the event shape variables we employ.  
The distribution of $|\cos\theta_T|$ is
sharply peaked near 1 for combinations drawn from jet-like $\qqbar$
pairs and is nearly uniform for the isotropic $B$-meson decays.
Further use of the event topology is made via the construction of a Fisher
discriminant ${\cal F}$, which is subsequently used as a discriminating variable in
the likelihood fit. 

Our Fisher discriminant is an optimized
linear combination of the remaining event shape information, excluding
$\cos\theta_T$. The variables entering the Fisher discriminant are the
angles with respect to the beam axis of the $B$ momentum and $B$ thrust axis, 
and the zeroth and second angular moments $L_{0,2}$ of the energy flow about 
the $B$ thrust axis, all calculated in the $\Upsilon(4S)$ center-of-mass frame.
The moments are defined by
\begin{eqnarray}
\label{eq:fisher}
L_j = \sum_i p_i\times\left|\cos\theta_i\right|^j\, ,
\end{eqnarray}
where $\theta_i$ is the angle with respect to the $B$ thrust axis of
track or neutral cluster $i$, $p_i$ is its momentum, and the sum
excludes the $B$ candidate.  The coefficients used to combine these
variables are chosen to maximize the separation (difference of means
divided by quadrature sum of errors) between the signal and continuum
background distributions of $ L_j$, and are determined from studies of signal
MC and off-peak data.  We have studied the optimization of ${\cal F}$\ for a variety
of signal modes, and find that the optimal sets of coefficients are
nearly identical for all. Because the information contained in ${\cal F}$\ is
correlated with $|\cos\theta_T|$, the separation
between signal and background is dependent on the $|\cos\theta_T|$
requirement made prior to the formation of ${\cal F}$. 
The ${\cal F}$ variable is illustrated in Fig.~\ref{fig:control}.

In order to establish that the MC simulation reproduces 
the kinematic observables in the data, such as $m_{\rm{ES}}$,  $\Delta E$, and ${\cal F}$,
we use high-statistics $B^0$-meson decays with similar kinematics and topology.
For example, in Fig.~\ref{fig:control} we illustrate reconstructed
${B^0\to {D}^-\pi^{+}} \to ({K^+\pi^-\pi^-})({\pi^+})$ decays.
There is good agreement between data and MC. The deviations in 
the means of the distributions are about 0.7~MeV for $m_{\rm{ES}}$, 5~MeV for $\Delta E$, and negligible for ${\cal F}$.
We take these corrections into account when we study the $B^0\to\varphi K\pi$ decays.

The $B$ background contribution is generally found to be small due to
selection on the narrow $\varphi$ resonance, PID requirements on the kaons, 
and good momentum resolution, important in particular for $\Delta E$. 
We remove $\BorBbar^0\to\varphi K^{\pm}\pi^{\mp}$ signal candidates that 
have decay products with invariant masses within 12 MeV of the nominal 
mass values for $D_s^\pm$ or $D^\pm$. 
This removes the background from the 
$B\to D_s^\pm K$, $D^\pm K$, $D_s^\pm \pi$, and $D^\pm \pi$ decays.
To reduce combinatorial background in the 
$B^0\to\varphi K^0_S\pi^0$ analysis with low-momentum
$\pi^0$ candidates, we require ${\cal H}_1<0.8$. 
Certain types of $B$ background, such as potential $B^0\to f_0(980)K\pi$,
 cannot be distinguished on an
event-by-event basis, and so cannot be removed by vetoes. We incorporate these contributions into the fit.
The remaining $B$ background events were found to be 
random combinations of tracks and can be treated as combinatorial 
background, similar to random tracks from $\qqbar$ production.

When more than one candidate is reconstructed,
which happens in $5\%$ of $\varphi K^\pm\pi^\mp$ and $10\%$ of 
$\varphi K^0_S\pi^0$ events,
we select one candidate per event based on the lowest value of the $\chi^2$ 
of the four-track vertex for $\BorBbar^0\to\varphi K^{\pm}\pi^{\mp}$ or
of the fitted $\BorBbar^0\to\varphi K^0_S\pi^0$ decay tree.

In the self-tagging $B$-decay mode $\BorBbar^0\to\varphi K^{\pm}\pi^{\mp}$,
we define the $b$-quark flavor sign $Q$ to be opposite
to the charge of the kaon candidate.
For each reconstructed $\BorBbar^0\to\varphi K^0_S\pi^0$ signal candidate ($B_{\rm sig}$) 
we use the 
remaining tracks in the event to determine the decay vertex position and flavor 
of the other $B$ decay, $B_{\rm tag}$.  
A neural network based on kinematic and particle
identification information assigns each event to one of seven mutually
exclusive tagging categories ($c_{\rm tag}$)~\cite{Aubert:2007hm}, including a
category for events in which a tag flavor is not determined. 
The $B$-flavor-tagging
algorithm is trained to identify primary leptons, kaons, soft pions,
and high-momentum charged particles from the other $B$
and correlate this information to the $B$ flavor.
The performance of this algorithm is evaluated using a data sample
($B_{\rm flav}$ sample) of fully-reconstructed $\Bz \to D^{(*)-} \pip
/ \rho^+/a^+_1$ decays. The effective tagging efficiency is measured to
be $(31.2\pm0.3)\%$.

We determine the proper time difference, $\Delta t$,  between $B_{\rm sig}$ and
$B_{\rm tag}$ from the spatial separation between their decay vertices. 
The $B_{\rm tag}$ vertex is reconstructed from the
remaining charged tracks in the event, and its uncertainty dominates
the $\Delta t$ resolution $\sigma_{\Delta t}$. The average proper time
resolution is $\langle\sigma_{\Delta t}\rangle \approx 0.7$~ps. Only
events that satisfy $|\Delta t|<15$~ps and
$0.1 < \sigma_{\Delta t} < 2.5$~ps are retained.

Overall, selection requirements discussed here have been optimized 
to retain large signals and wide sidebands of observables for
later fitting, as discussed in the next section. Statistical 
precision of the measurements was the main optimization factor,
while individual systematic uncertainties were kept small compared 
to statistical errors.
After applying all selection criteria
and using the measured branching fractions summarized 
in Table~\ref{tab:previousresults}, 
we expect to observe about 
177 (15) $\BorBbar^0\to\varphi (\KorKbar\pi)^{*0}_0$, 
473 (34) $\BorBbar^0\to\varphi \KorKbar^*(892)^0$, and
156 (9) $\BorBbar^0\to\varphi \KorKbar^*_2(1430)^0$
events in the $\varphi K^\pm\pi^\mp$ ($\varphi K^0_S\pi^0$) channel.
The larger reconstruction efficiency and secondary branching fractions
in the $\varphi K^\pm\pi^\mp$ channel result in the dominance of this
decay mode in the signal parameter measurements, except for the
measurement of the $\Delta\phi_{00}$ parameter, which is possible only
with the $\varphi K^0_S\pi^0$ channel.


\section{ANALYSIS METHOD}
\label{sec:note-analysis}

We use an unbinned, extended maximum-likelihood (ML) fit~\cite{babar:vv}
to extract the 27 parameters  defined in Table~\ref{tab:parameters}, which  
describe three decay 
channels (12 in either $B^0\to\varphi K^{*}(892)^0$ or 
$\varphi K^{*}_2(1430)^0$, and three in
$\varphi(K\pi)^{*0}_0$ decays).
We perform a joint fit to the data for three modes: $\varphi K^{0}_S\pi^0$;
$\varphi K^{\pm}\pi^{\mp}$ in the lower $K\pi$ mass range (0.75 - 1.05 GeV);
and $\varphi K^{\pm}\pi^{\mp}$ in the higher $K\pi$ mass range (1.13 - 1.53 GeV). 
To simplify treatment of the likelihood function, we 
separate the two ranges of the $K^\pm\pi^\mp$ invariant mass.
However, in the joint fit
the likelihood function ${\cal L}$ is written as a product
of three independent likelihood functions, 
one for each of the above three modes, as discussed below.

Due to the relatively low statistics, we simplify the angular analysis in the 
$B^0\to\varphi K^0_S\pi^0$ decay mode. We integrate the angular distributions 
in Eq.~(\ref{eq:helicityfull}) over the angle $\Phi$. Therefore only the longitudinal 
polarization fractions $f_{LJ}$, the yields, and the time-evolution are measured 
with the $B^0\to\varphi K^0_S\pi^0$ decays. 
We constrain the relative signal yields for the same
spin-$J$ contributions in the two subchannels 
$\varphi (K^{\pm}\pi^{\mp})_J$ and $\varphi (K^0_S\pi^0)_J$ 
taking into account the isospin relationship, 
daughter branching fractions, 
and reconstruction efficiency corrections. 
The isospin relationship requires that the 
$K^{*0}\to K^{+}\pi^{-}$ and $K^0\pi^0$ fractions
are 2/3 and 1/3 of the total $K\pi$ decay rate, and we ignore
any isospin violation as being negligible for the measurements
in this analysis.
All other signal parameters in Table~\ref{tab:parameters} are constrained
to be the same when they appear in both channels.

\subsection{Likelihood function}

The likelihood function for $B^0\to\varphi K^0_S\pi^0$ is written as:
\begin{equation}
{{\cal L}} =  \prod_c \exp\left( -N_c\right) \prod_i^{N_c} \left( \sum_{j}n_j
f_{j}^{c}  {\cal P}_{j}^{c}(\vec{x}_{i};~\vec{\zeta};~\vec{\xi})  \right)\,,
\end{equation}
where $n_j$ is the unconstrained (except if noted otherwise)
number of events for each event type $j$, $f_{j}^{c}$ is the fraction of events
of component $j$ for each tagging category $c$,
$N_c = \sum_j f_{j}^{c}n_j $  is the number of events found by the fit for 
tagging category $c$, and
${\cal P}_{j}^{c}(\vec{x}_{i};~\vec{\zeta};~\vec{\xi})$ is the
probability density function (PDF).

The data model has five event types $j$:
the signal $B^0\to\varphi(K\pi)_J$ with ${J=0,1,2}$, 
a possible background from $B^0\to f_0(980)K^{*0}$,
and combinatorial background.
The combinatorial background PDF is found to account
well for both the dominant light $\qqbar$ events 
and the random tracks from $B$ decays.
Each event candidate $i$ is characterized by a set of 10 observables
$\vec{x}_{i}=\{m_{\rm{ES}},~\Delta E,~{\cal F},~m_{K\pi},~m_{\!K\Kbar},~
{\cal H}_1,~{\cal H}_2,~c_{\rm tag},~
\Delta t,~\sigma_{\Delta t} \}_i$,
the kinematic observables
$m_{\rm{ES}}$, $\Delta E$, ${\cal F}$, the $K^*$ and $\varphi$ invariant masses
$m_{K\pi}$ and $m_{\!K\Kbar}$, 
the helicity angles ${\cal H}_1$, ${\cal H}_2$, 
the flavor tag $c_{\rm tag}$, the proper time difference 
$\Delta t$ and its event-by-event error $\sigma_{\Delta t}$. 
The PDFs are split into the seven tagging categories.
The polarization parameters quoted in Table~\ref{tab:parameters}
are denoted by $\vec{\zeta}$, and the remaining parameters by $\vec{\xi}$.
Most of the $\vec{\zeta}$ parameters, except for $\Delta\phi_{00}$,
appear also in the likelihood function for the $\BorBbar^0\to\varphi K^{\pm}\pi^{\mp}$ decays.

The likelihood function for $\BorBbar^0\to\varphi K^{\pm}\pi^{\mp}$ decays, in 
either the lower or the higher $K\pi$ mass range, is
\begin{equation}
{\cal L} = \exp\left(-\sum_j n_{j}\right)\,\prod_i{\left(
\sum_{j,k}n_{j}^{k}\,{\cal P}_{j}^{k}({\vec{\rm x}_i};~{\mu^k},~{\vec{\zeta}},~{\vec{\xi}})
\right)}
\end{equation}
where the index $j$ represents the three event types used in our data model:
the signal $B^0\to\varphi(K\pi)^0$ ($j = 1$) which combines the two
dominant modes in a given mass range 
($\varphi K^{*}(892)^0$ and $\varphi(K\pi)^{*0}_0$ in the lower or 
$\varphi K^{*}_2(1430)^0$ and $\varphi(K\pi)^{*0}_0$ in the higher mass range),
a possible background from $B^0\to f_0(980)K^{*0}$ ($j = 2$),
and combinatorial background ($j = 3$).
The superscript $k$ corresponds to the value of the flavor sign $Q=\pm1$
and allows for a $C\!P$-violating difference between
the $B^0$ and $\Bbar^0$ decay amplitudes ($A$ and $\Abar$).

In the signal event type, the yield and asymmetry, $n_{\rm sig}$ and ${\cal A}_{C\!P}$,
 of the
$\BorBbar^0\to\varphi (K^\pm\pi^\mp)_J$ mode
with $J=1$ in the lower mass range or $J=2$ in the higher mass range,
and those of the $\BorBbar^0\to\varphi(K^\pm\pi^\mp)_0^{*0}$ mode are parameterized
by applying the fraction  $\mu^k$ of the $\varphi (K\pi)_J$ yield to $n_1^k$.
Hence, $n_{\rm sig}=n_1^+\times\mu^+ + n_1^-\times\mu^-$,
${\cal A}_{C\!P}=(n_1^+\times\mu^+ - n_1^-\times\mu^-)/n_{\rm sig}$,
and the $\varphi(K^\pm\pi^\mp)_0^{*0}$ yield is $n_1^+\times(1-\mu^+) + n_1^-\times(1-\mu^-)$.
This treatment is necessary to include interference between the
two decay modes as we discuss below,
while we ignore interference in the $B^0\to\varphi K^0_S\pi^0$
channel due to low statistics.
The PDF is formed from the following set of observables
{$\vec{\rm x}_i$}~=\{$m_{\rm{ES}}$, $\Delta E$, ${\cal F}$, $m_{K\pi}$, 
$m_{K\Kbar}$, ${\cal H}_1$, ${\cal H}_2$, $\Phi$, $Q$\}, 
and the dependence on $\mu^k$ and polarization parameters {$\vec{\zeta}$}
~$\equiv\{f_{LJ}$, $f_{\perp J}$, $\phi_{\parallel J}$,
$\phi_{\perp J}$, $\delta_{0J}$, ${\cal A}_{C\!PJ}^0$, ${\cal A}_{C\!PJ}^{\perp}$,
$\Delta \phi_{\parallel J}$, $\Delta \phi_{\perp J}$, $\Delta\delta_{0J}$\}
is relevant only for the signal PDF ${\cal P}_{1}^k$.

The remaining PDF parameters {$\vec{\xi}$}, in both the 
$\BorBbar^0\to\varphi K^0_S\pi^0$ and  $\varphi K^{\pm}\pi^{\mp}$ channels,
are left free to vary in the fit for the combinatorial
background and are fixed to the values extracted from
MC simulation and calibration $B\to\Dbar\pi$ decays 
for the other event types.
We minimize the $-2\ln{\cal L}$ function using MINUIT~\cite{minuit}
in the ROOT framework~\cite{root}.
The statistical error on a parameter is given by its change 
when the quantity $-2\ln{\cal L}$ increases by
one unit.  The statistical significance is taken as the square root of the
difference between the value of $-2\ln{\cal L}$ for zero signal and the
value at its minimum. 
We have tested this procedure with simulated samples and found
good agreement with the statistical expectations.

\subsection{PDF parameterization}

The PDF
${\cal P}_{j}^{k}$({$\vec{\rm x}_i$};~$\mu^k$,~{$\vec{\zeta}$};~{$\vec{\xi}$})
for a given $\BorBbar^0\to\varphi K^\pm\pi^\mp$ or $\varphi K^0_S\pi^0$ candidate $i$ is taken to be 
a joint PDF for the helicity angles,
resonance masses, and $Q$, and the product of
the PDFs for each of the remaining variables.
The assumption of negligible correlations in the selected data sample 
among the discriminating variables,
except for resonance masses and helicity angles where relevant,
 has been validated by evaluating the correlation coefficients. This assumption was further tested 
with the MC simulation.

For the parameterization of the signal PDFs we use double-Gaussian functions for signal
$\Delta E$ and $m_{\rm{ES}}$.
For the background we use low-degree polynomials as required by the data or,
in the case of $m_{\rm{ES}}$,
an empirical threshold ARGUS function~\cite{argus-func}:
\begin{eqnarray}
f(x) \propto x\sqrt{1-x^2}~{\rm exp}[-\xi_1(1-x^2)] \ ,
\label{eq:bmass}
\end{eqnarray}
where $x \equiv m_{\rm{ES}}/E_{\rm beam}$ and $\xi_1$ is a parameter
that is determined from the fit with a typical value of about 25.

For both signal and background, the Fisher distribution ${\cal F}$ is
described well by a Gaussian function with different widths to the
left and right of the mean.  For the combinatorial background
distribution, we also include a second Gaussian function with a larger
width to account for a small tail in the signal ${\cal F}$ region.  This additional
component of the PDF is important, because it prevents the background
probability from becoming too small in the region where
signal lies. 

A relativistic spin-$J$ B-W amplitude
parameterization is used for the resonance
masses~\cite{bib:Amsler2008,f0mass},
except for the $(K\pi)^{*0}_0$ $m_{K\pi}$ amplitude, which is 
parameterized with the LASS function~\cite{Aston:1987ir, wmdLASS}.
The latter includes the $K_0^{*}(1430)^0$ resonance
together with a nonresonant component.
The detailed treatment of the invariant mass distribution
is discussed in Section~\ref{sec:note-mass}.
We found that no additional correction to the $K\pi$ invariant
mass parameterization is necessary because resolution effects
of only a few MeV are negligibly small compared with the resonance widths. 
On the other hand, we convolve resolution effects into the $m_{K\Kbar}$ parameterization.

\begin{figure*}[t]
\centerline{
\setlength{\epsfxsize}{0.3\linewidth}\leavevmode\epsfbox{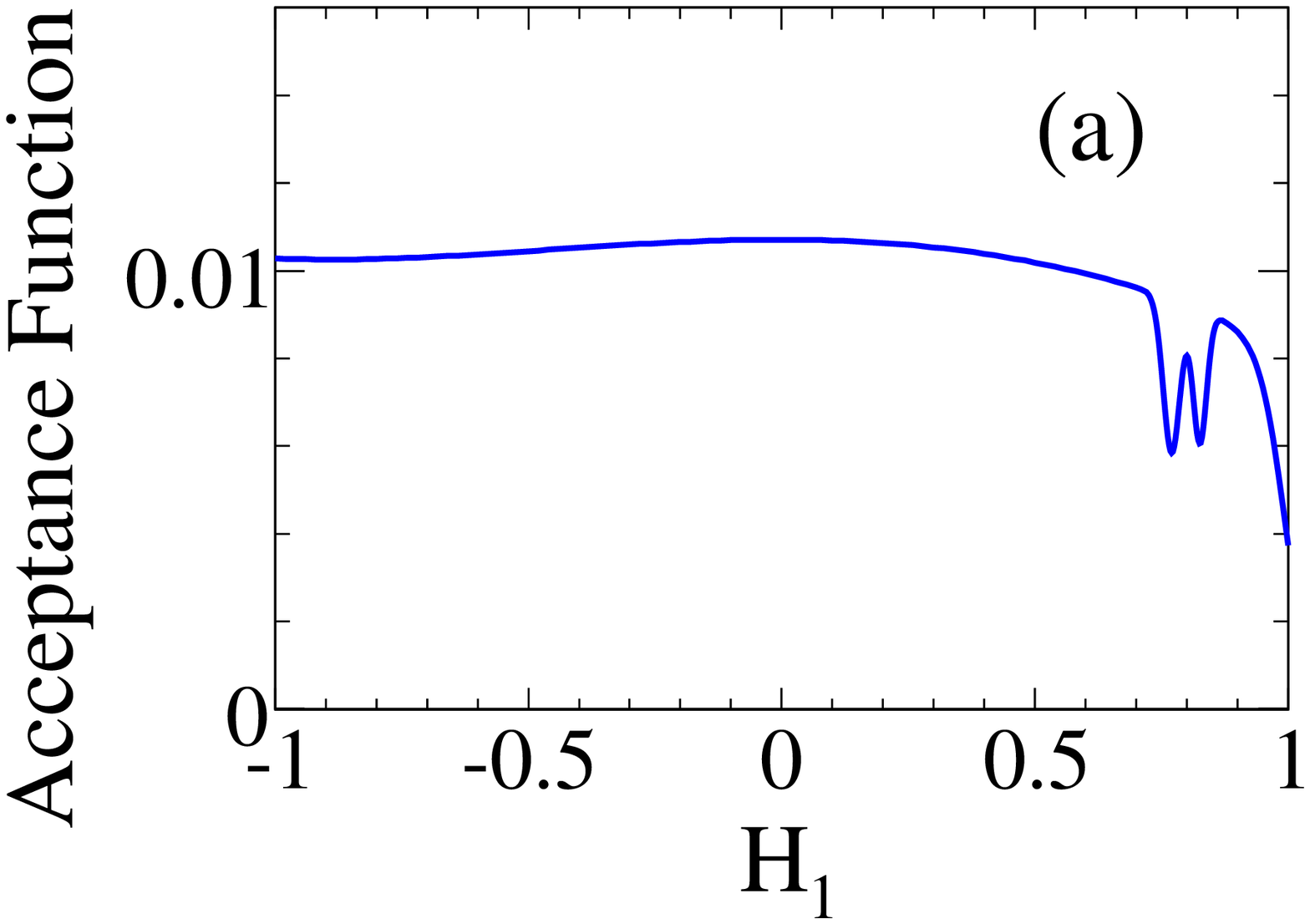}
\setlength{\epsfxsize}{0.3\linewidth}\leavevmode\epsfbox{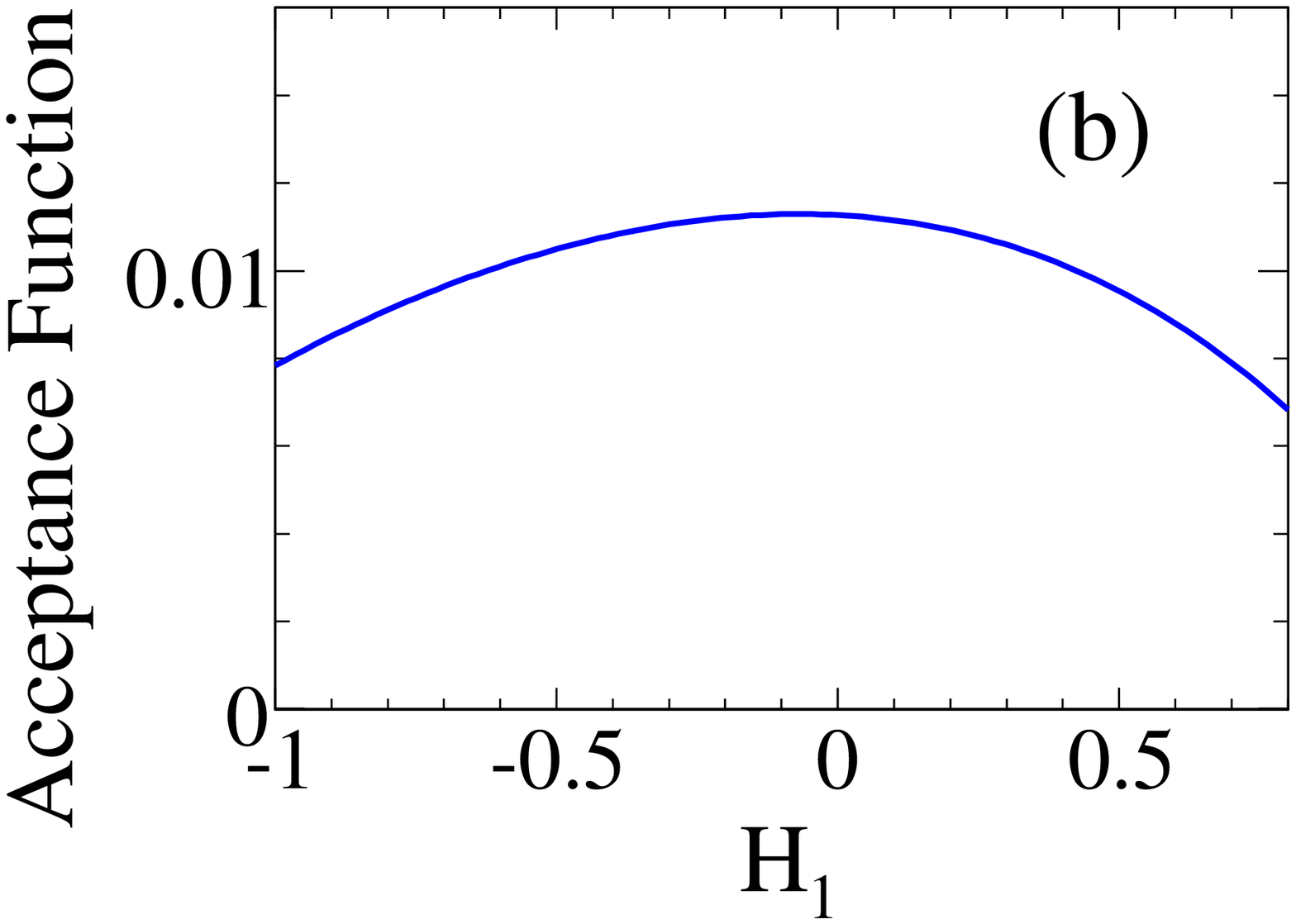}
\setlength{\epsfxsize}{0.3\linewidth}\leavevmode\epsfbox{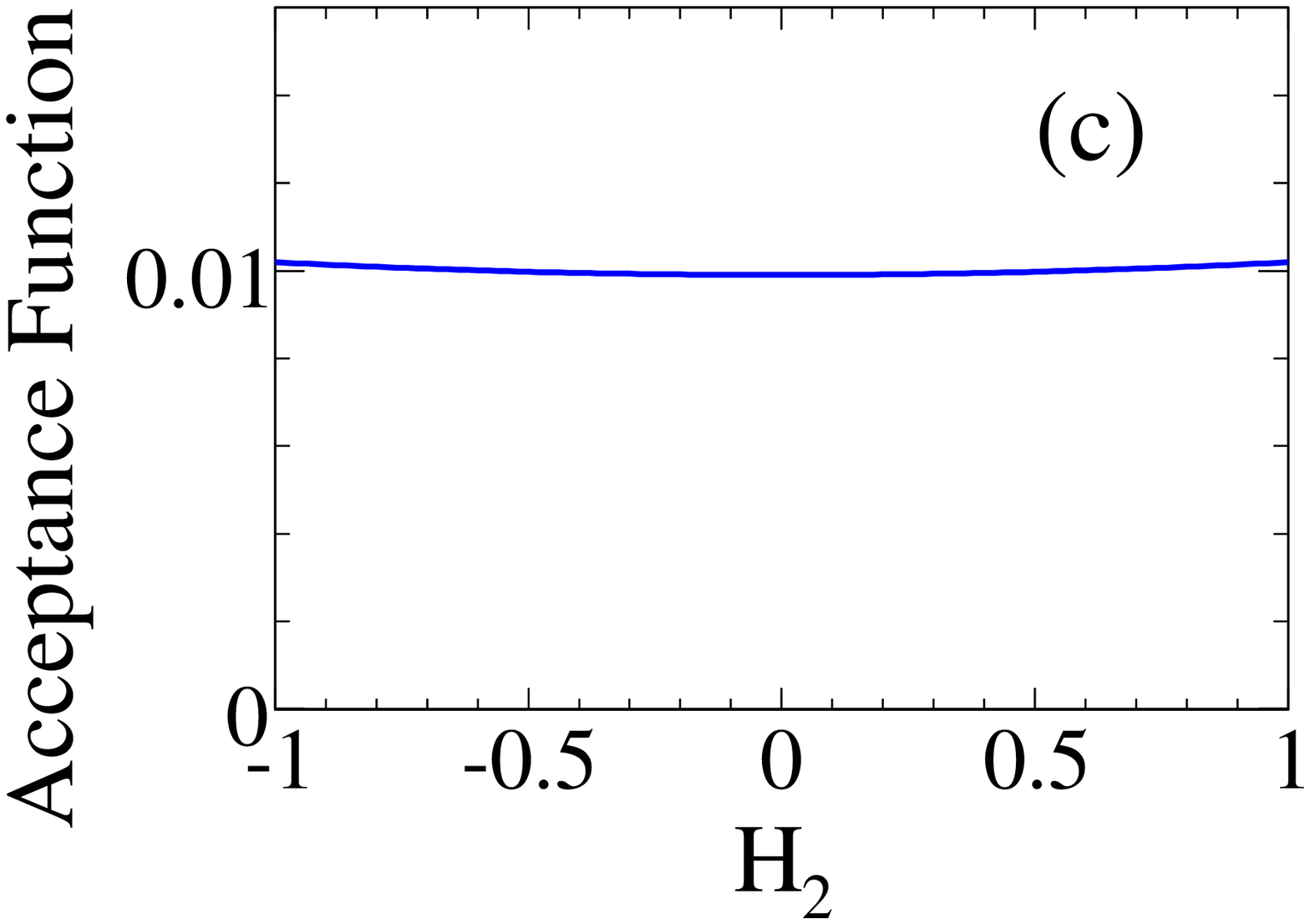}
}
\caption{\label{fig:acc-func}
Angular acceptance functions for ${\cal H}_1$ 
in $\varphi K^{\pm}\pi^{\mp}$ (a), in $\varphi K^0_S\pi^0$ (b) , and for ${\cal H}_2$ (c).
These plots show only relative 
efficiency between different helicity points with arbitrary $y$-axis units. 
The $D^{\pm}_{(s)}$-meson veto causes the sharp acceptance dips near 
${\cal H}_1 = 0.8$ seen in (a). 
}
\end{figure*}

The background parameterizations for candidate masses 
include resonant components to account for resonance production in the background.
The background shape for the helicity parameterization
is also separated into contributions
from combinatorial background and from real mesons, both
fit by low-degree polynomials.

The mass-helicity PDF is the
ideal distribution from Eqs.~(\ref{eq:inter-01}--\ref{eq:fang-interf}), 
multiplied by an empirically-determined acceptance function
${\cal{G}}({\cal H}_1,{\cal H}_2,\Phi)
\equiv{\cal{G}}_1({\cal H}_1)\times{\cal{G}}_2({\cal H}_2)$,
which is a parameterization of the relative reconstruction
efficiency as a function of helicity angles. 
It was found with detailed MC simulation that resolution effects 
in the helicity angles introduce negligible effects in the
PDF parameterization and fit performance, and they are therefore ignored.
The angles between the final state particles and their parent
resonances are related to their momenta. 
The signal acceptance effects parameterized with the function
${\cal{G}}({\cal H}_1,{\cal H}_2,\Phi)$ are due to kinematic
correlations, whereas the detector geometry correlations are negligible.
Therefore the above uncorrelated parameterization as a function
of two helicity angles was found to be appropriate
and was validated with detailed MC simulation.

Momentum in the laboratory frame is strongly correlated with detection efficiency.
Thus we have acceptance effects in the helicity
observables ${\cal H}_i$, most evident for the large values
of ${\cal H}_1$ corresponding to the slow $\pi$ from the $K^*$ meson.
However, these acceptance effects are not present for the $\Phi$ angle;
 there is no 
correlation with the actual direction with respect to the detector, 
which is random for the $B$ decays. 
The acceptance effects for the two helicity angles ${\cal H}_1$ and 
${\cal H}_2$ are shown in Fig.~\ref{fig:acc-func}. 
We obtain the acceptance functions from the fit to the signal MC
 helicity distribution, 
with the known relative components of longitudinal and transverse 
amplitudes generated with $B^0\to\varphi K^{*}(892)^0$ MC. 
The $D^{\pm}_{(s)}$-meson veto causes the sharp acceptance dips around 0.8
in the ${\cal{G}}_1({\cal H}_1)$ function in the 
$\BorBbar^0\to \varphi K^\pm\pi^\mp$ analysis.

The interference between the $J=1$ or $J=2$ and the
$S$-wave $(K\pi)$ contributions is modeled with
the term $2{\Ree}(A_{J} A^*_{0})$ in Eq.~(\ref{eq:fang-interf}) with the
four-dimensional angular and $m_{K\pi}$ dependence,
as discussed in detail in Section~\ref{sec:note-interference}.
It has been shown in the decays $B^0\to J/\psi (K\pi)^{*0}_0$
and $B^\pm\to\pi^\pm(K\pi)^{*0}_0$~\cite{jpsikpi} that the amplitude
behavior as a function of $m_{K\!\pi}$ is consistent with that observed by LASS except for
a constant phase shift. 
Integrating the probability distribution over $({\cal H}_1, {\cal H}_2, \Phi)$,
the interference term $2{\Ree}(A_{J} A^*_{0})$ should vanish.
However, as we introduce the detector acceptance effects on $({\cal H}_1, {\cal H}_2)$,
the interference contribution becomes non-zero.
The total yield of the two modes is corrected before calculating the branching fractions.
The effect can be estimated by comparing the integral of $B^0\to\varphi K_2^{*}(1430)^0$,
$B^0\to\varphi (K\pi)_0^{*0}$ and the interference probability contribution.
We find that the interference term accounts for $3.5\%$ of the total yield.
Accordingly, we scale the yields of $B^0\to\varphi K_2^{*}(1430)^0$ 
and $B^0\to\varphi (K\pi)_0^{*0}$ modes by $96.5\%$ while calculating
the branching fraction. This effect is negligible for the
$B^0\to\varphi K^{*}(892)^0$ decay due to the relatively small fraction
of the $B^0\to\varphi (K\pi)_0^{*0}$ contribution to the lower $K\!\pi$ mass range.

The parameterization of the nonresonant signal-like contribution
$B^0\to f_0K^{*0}\to (K^+K^-)K^{*0}$ is identical to the signal in
the primary kinematic observables $m_{\rm ES}$, $\Delta E$,  
${\cal F}$, and the $m_{K\!\pi}$ mass 
but is different in the angular and $m_{K\!\Kbar\!}$ distributions. 
For $B^0\to f_0K^{*0}$, the ideal angular distribution
is uniform in $\Phi$ and ${\cal H}_2$ and is proportional
to ${\cal H}^2_1$, due to angular momentum conservation.
We use a coupled-channel B-W function
to model the $K^+K^-$ mass distribution for the $f_0$~\cite{f0mass}.
The broad invariant mass distribution of $f_0$ compared to the
narrow $\varphi$ resonance was found to account well for 
any broad $m_{K\!\Kbar\!}$ contribution. This PDF parameterization is
further varied as part of the systematic uncertainty studies.
The $m_{K\!\pi}$ distribution in $B^0\to f_0K^{*0}$ is parameterized 
as a $J=1$ contribution in the lower mass range and as a $J=0$ contribution  
in the higher mass range, see Section~\ref{sec:note-mass}.

For the $B^0\to\varphi K^{0}_S\pi^{0}$ mode, the $\Phi$ angle is integrated over, 
so that no interference terms appear in the fit. An additional PDF for the
$\Delta t$ distribution is used for both the signal and background, 
which is discussed next. The treatment of other observables is similar to those of 
$\BorBbar^0\to\varphi K^{\pm}\pi^{\mp}$. 

Time-dependent \CP asymmetries are determined using the difference of
\Bz meson proper decay times $\Delta t \equiv t_{\rm sig} - t_{\rm tag}$,
where $t_{\rm sig}$ is the proper decay time of the signal $B$ ($B_{\rm sig}$)
and $t_{\rm tag}$ is that of the other $B$ ($B_{\rm tag}$). The
$\Delta t$ distribution for $B_{\rm sig}$ decaying to a \CP eigenstate 
\begin{widetext}
\begin{eqnarray}
\begin{tabular}{ll}
$f(\Delta t, Q_{\rm tag})\sim $ ${\displaystyle\frac{e^{-\left|\Delta t\right|/\tau_B}}{4\tau_B}}  
\times$&$\left\{ 1 - Q_{\rm tag} \Delta w + Q_{\rm tag}\mu(1-2\omega)\right.$\\ 
\vspace{-0.25cm}\\
&$\left. +\left(Q_{\rm tag}(1-2w)+\mu (1-Q_{\rm tag}\Delta \omega)\right)\left[
S\sin(\Delta m_B\Delta t)-C\cos(\Delta m_B\Delta t)\right]\right\}$
\end{tabular}
\label{eq:deltatpdf}
\end{eqnarray}
\end{widetext}
is convolved with a resolution function ${\cal R}$.
The parameter $Q_{\rm tag}= +1(-1)$ when the tagging meson $B^0_{\rm tag}$
is a $B^0(\Bbar^0)$, $w$ is
the average mistag probability, and $\Delta w$ and $\mu$
describe the difference in mistag probability and the tagging efficiency
asymmetry between $B^0$ and $\Bbar^0$ mesons, respectively. The time
distribution of combinatorial
background is assumed to have zero lifetime.

The $\Delta t$ resolution function ${\cal R}$ is the sum of three Gaussian functions  
representing the core, tail, and outer part of the distribution, 
weighted by the $\Delta t$ error for the core and the tail:
\begin{widetext}
\begin{eqnarray}
{\cal R}(\Delta t,\sigma_{\Delta t})=f_{\rm core}G(\Delta t,\nu_{\rm core}\sigma_{\Delta t},\sigma_{\rm core}\sigma_{\Delta t})
+f_{\rm tail}G(\Delta t,\nu_{\rm tail}\sigma_{\Delta t},\sigma_{\rm tail}\sigma_{\Delta t})
+f_{\rm out}G(\Delta t,\nu_{\rm out},\sigma_{\rm out})
\label{E_Resolution}
\end{eqnarray}
\end{widetext}
where $G(\Delta t,\sigma_{\Delta t}; \nu,\sigma)$ is a Gaussian distribution
with mean $\nu$ and standard deviation $\sigma$, and $f$ is the corresponding fraction.
We have verified in simulation that the parameters of the
resolution function for signal events are compatible with those
obtained from the $B_{\rm flav}$ sample, a data sample
of fully-reconstructed $\Bz \to D^{(*)-} \pip/ \rho^+/a^+_1$ decays. 
Therefore we use the $B_{\rm flav}$ parameters for better precision. 
The background $\Delta t$ distribution is parameterized by the 
\CP-asymmetric PDF $f(\Delta t, c_{\rm tag}) = 1\pm {\cal A}_{\rm bkgd}(c_{\rm tag})$, 
convolved with the resolution function.  
The parameters of the background $\Delta t$ PDF are determined in the fit to
data from sidebands in $m_{\rm ES}$.

The detailed description of the treatment of the $S$ and $C$ terms of different
contributing amplitudes is given in Section~\ref{sec:note-time}.
Eq.~(\ref{eq:deltatpdf}) is applicable to the time evolution of each
of the five components in the angular distribution,
three longitudinal ($S_{J0}$ and $C_{J0}$ for $J=0$, $1$, and $2$)
and two transverse ($S_{JT}$ and $C_{JT}$ for $J=1$ and $2$).
All five $S_{J0}$ and $S_{JT}$ parameters are expressed in terms of  
$\Delta\phi_{00}$ and the other polarization and $C\!P$ parameters
entering the $\BorBbar^0\to\varphi K^{\pm}\pi^{\mp}$ PDF description,
while the five  $C_{J0}$ and $C_{JT}$ parameters are expressed through
other polarization and $C\!P$ parameters only, as shown in
Eqs.~(\ref{eq:sphikstzero1430}--\ref{eq:cjt}).

For the combinatorial background we
establish the functional forms and initial parameter values of the
PDFs with data from sidebands in $m_{\rm ES}$  or $\Delta E$.  We then refine the
main background parameters (excluding the resonance-mass central values
and widths) by allowing them to vary in the final fit so that they are
determined by the full data sample.
Overall, there are 51 free background parameters in the joint fit. 

\subsection{Analysis validation}

We validate the analysis selection and fit performance 
with a number of cross-check analyses.
To test the treatment of combinatorial background in the PDF, we 
perform fits on the data collected below the
$\Upsilon(4S)$ resonance, on GEANT-based~\cite{Agostinelli:2002hh} MC simulation of about
three times the statistics of the data sample for both $\qqbar$
production (continuum) and generic $\Upsilon(4S)\to B\Bbar$ decays. 
We also test the contribution of several dozen exclusive $B$ meson decays
which could potentially mimic the signal with statistics of more than an 
order of magnitude greater than their expectation. No significant bias
in the background treatment was found. Systematic uncertainties 
associated with this treatment are discussed in the next subsection.

\begin{figure}[htbp]
\centerline{
\setlength{\epsfxsize}{0.5\linewidth}\leavevmode\epsfbox{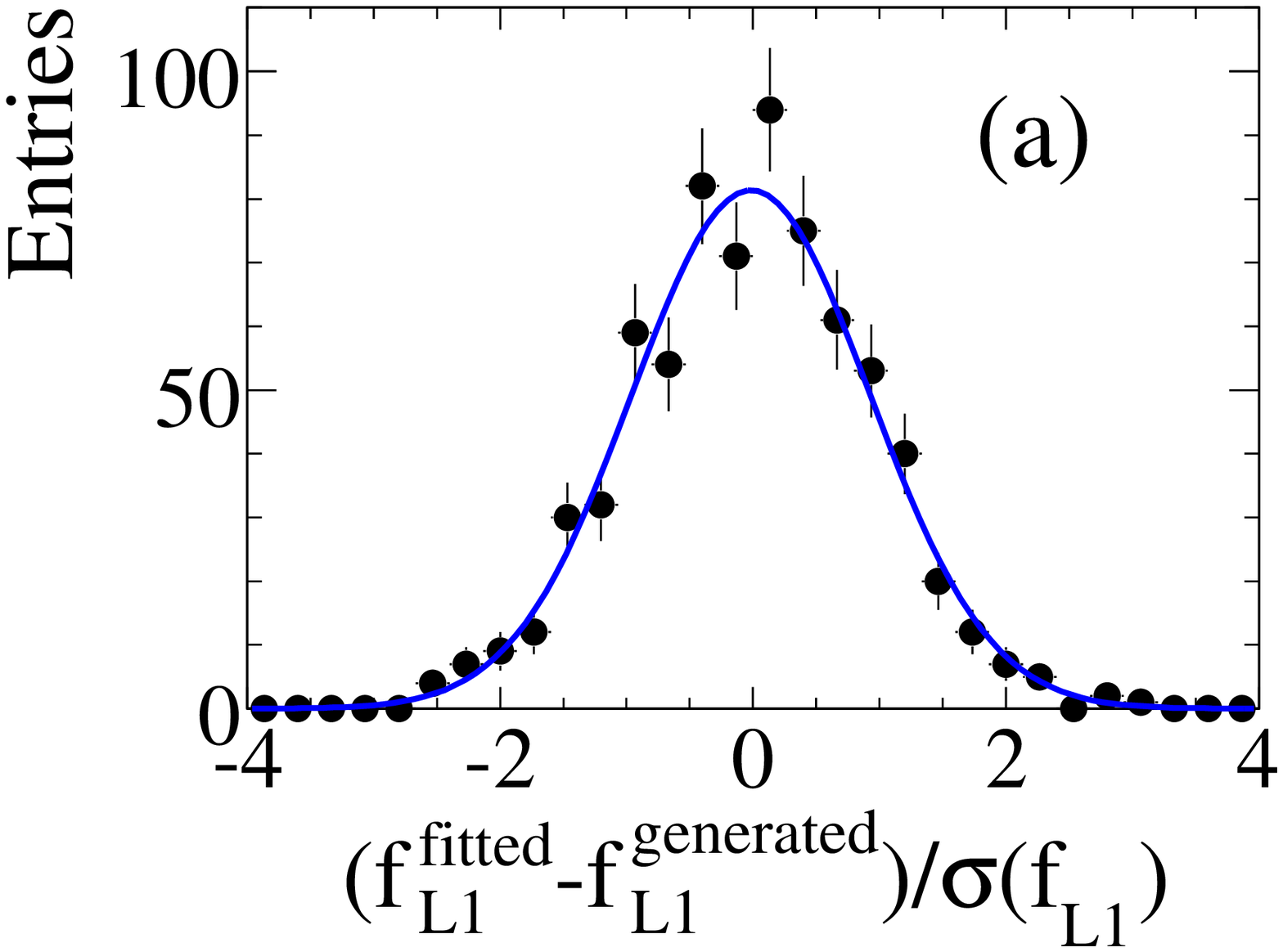}
\setlength{\epsfxsize}{0.5\linewidth}\leavevmode\epsfbox{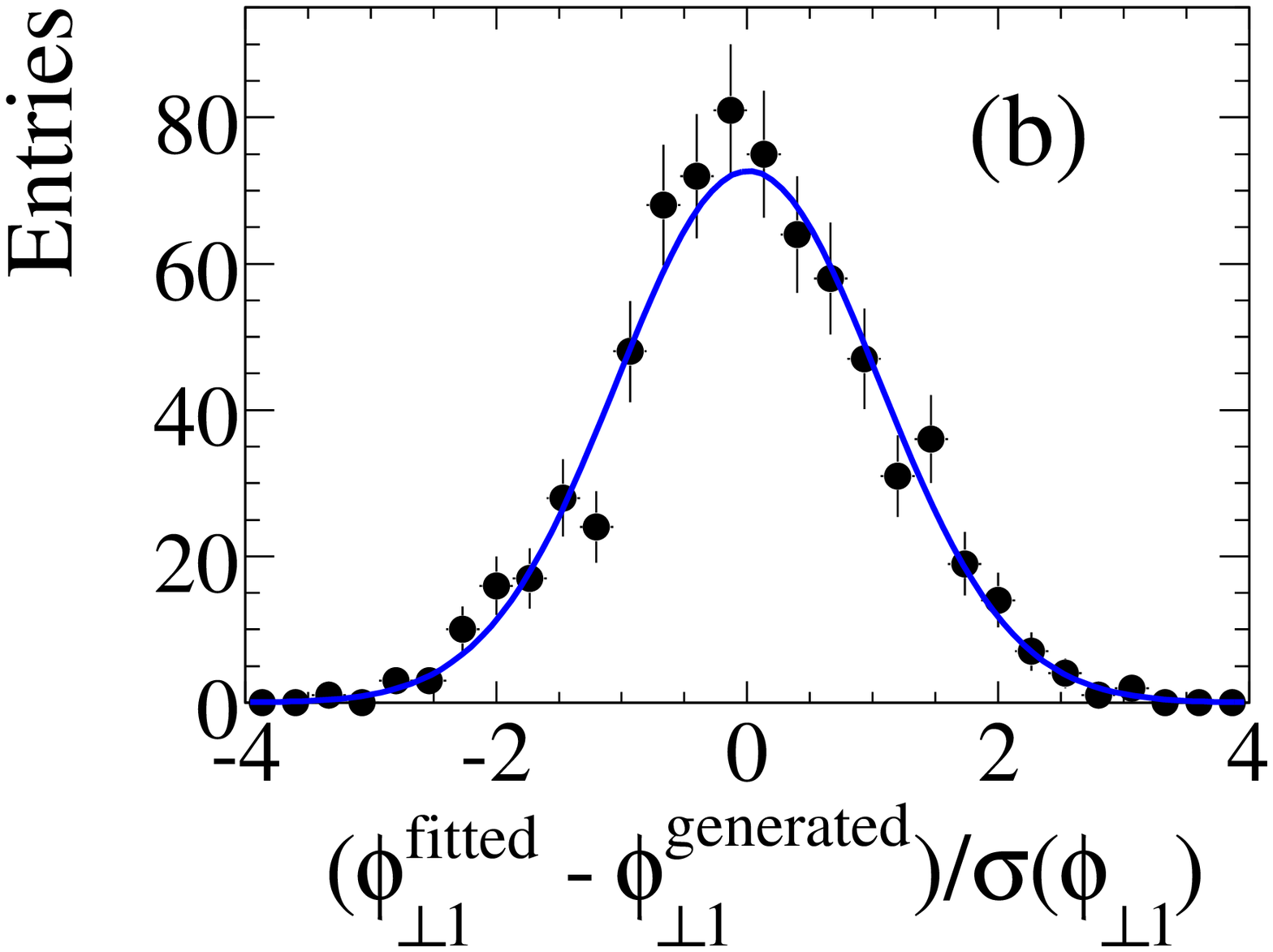}
}
\centerline{
\setlength{\epsfxsize}{0.5\linewidth}\leavevmode\epsfbox{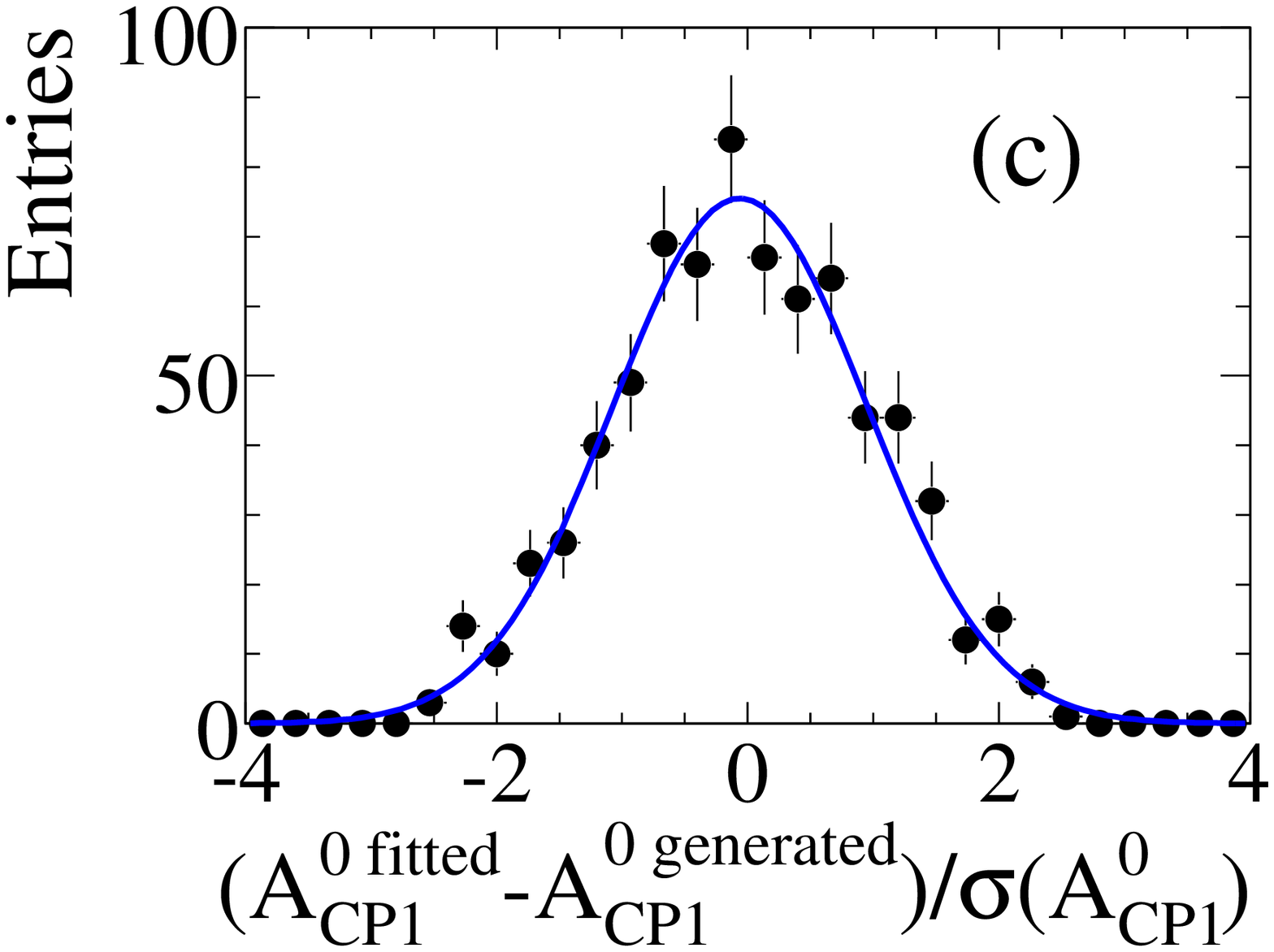}
\setlength{\epsfxsize}{0.5\linewidth}\leavevmode\epsfbox{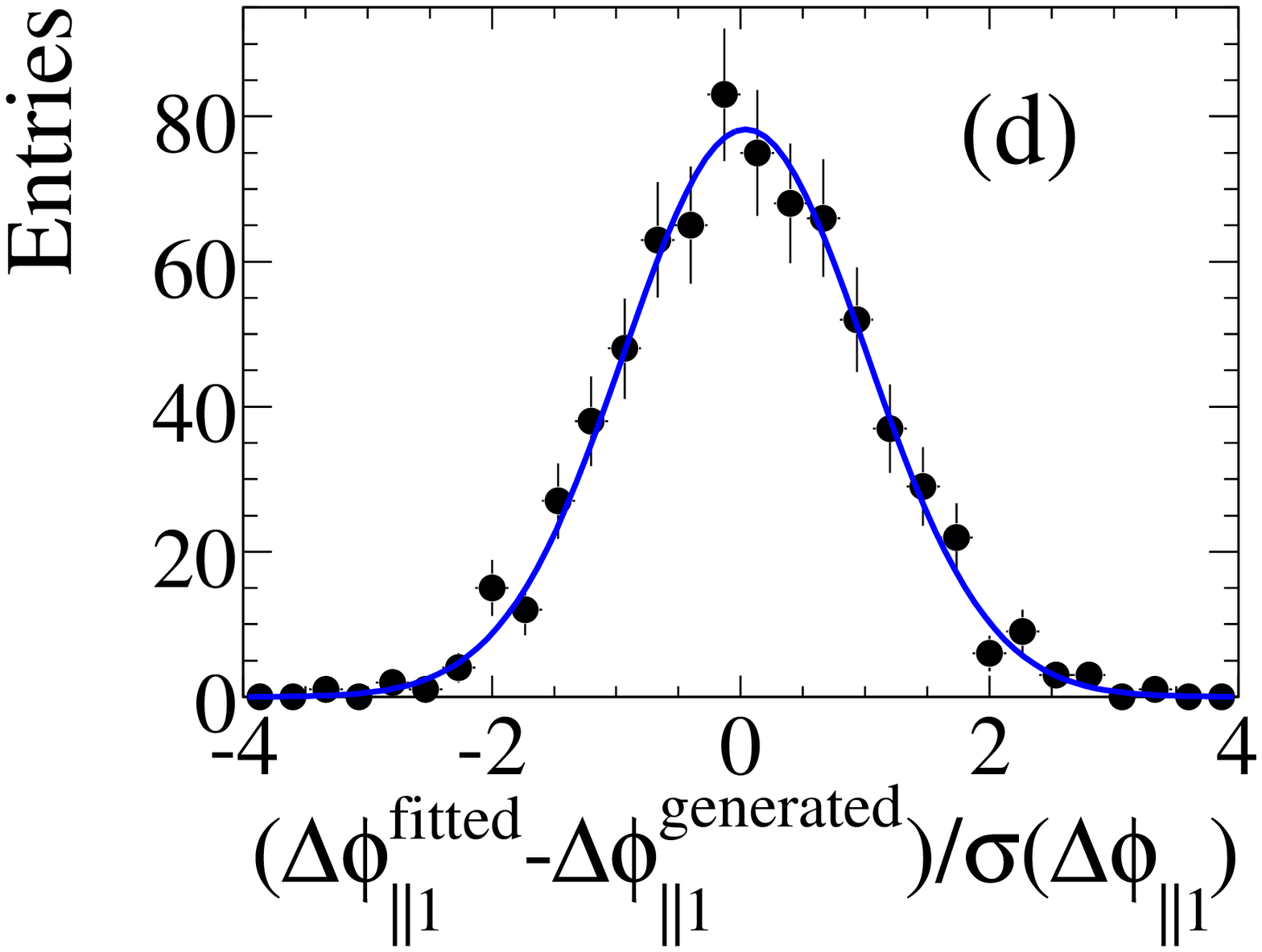}
}
\caption{\label{fig:valid-pull}
Distributions of $(x^{\rm fitted}-x^{\rm generated})/\sigma(x)$
for a large number of generated MC experiments,
where a Gaussian fit is superimposed, and $x$ denotes signal parameters: 
(a) $f_{L1}$,~(b) $\phi_{\perp 1}$,~(c) $A^{0}_{\CP 1}$, and 
(d) $\Delta\phi_{\parallel 1}$.}
\end{figure}

To test the signal PDF parameterization and the overall
fit performance, we generate a large number of MC experiments, each
one representing a statistically independent modeling of the fit to the data.
The signal events are taken from the generated MC samples, while 
background is generated from the PDF 
with the total sample size corresponding to the on-resonance data sample. 
We embed signal-like events according to expectation.
We find the results of the MC experiments to be in good agreement
with the expectations and the error estimates to be correct.
In Fig.~\ref{fig:valid-pull} we show examples of the 
$(x^{\rm fitted}-x^{\rm generated})/\sigma(x)$ distributions, 
where $x$ denotes one of the signal parameters.
The means and widths
of all these distributions are within about 5\% of the expected values
of zero and one, which results in negligibly small uncertainty in the
fit result.

At all analysis development steps we used a ``blind'' technique.
Numerical values of all physics parameters  $n_{{\rm sig}J}$, 
${\cal A}_{C\!P J}$, $\Delta\phi_{00}$, and $\vec{\zeta}$
were kept hidden until the analysis method and tools were decided,
validated, and fixed. Consistency between the blind fit results
and prediction from generated samples for the likelihood ${\cal L}$
and error values were used to judge the goodness of fit. 
We also examined the projection of data and likelihood fit PDFs onto individual 
variables with the enhancement of either signal or background to judge 
consistency. Detailed study of the systematic errors did not reveal
any uncertainties which would exceed statistical errors.

\subsection{Systematic uncertainties}

\begingroup
\begin{table*}[b]
\caption{\label{tab:systematic3} 
Systematic errors (\%) in reconstruction efficiency evaluation. 
The total errors combine the two subchannels according to
their weight and are dominated by the $K^\pm\pi^\mp$ channel.
We separate the $\varphi (K\pi)^{*0}_0$ and $\varphi K^{*}_0(1430)^0$
modes to account for different errors in the daughter branching
fractions. See text for details. 
}
\begin{center}
\begin{ruledtabular}
\setlength{\extrarowheight}{1.5pt}
\begin{tabular}{ccccccccc}
\vspace{-3mm} &  &  & &   \cr
\vspace{-3mm} &  \multicolumn{2}{c}{$\varphi K^{*}(892)^0$}  &  \multicolumn{2}{c}{$\varphi K^{*}_2(1430)^0$} & \multicolumn{2}{c}{$\varphi (K\pi)^{*0}_0$} & \multicolumn{2}{c}{$\varphi K^{*}_0(1430)^0$}  \cr
 &  &  &  &  \cr
\hline
\hline
\vspace{-3mm} &  &  &  &  \cr
&$ K^{\pm}\pi^{\mp}$&$ K^0_S\pi^0$&$ K^{\pm}\pi^{\mp}$&$ K^0_S\pi^0$&$ K^{\pm}\pi^{\mp}$&$ K^0_S\pi^0$&$ K^{\pm}\pi^{\mp}$&$ K^0_S\pi^0$\cr
\hline
\vspace{-3mm} &  &  & &   \cr
 track finding               & {2.0} & 1.0 & {2.0} & 1.0 & {2.0} & 1.0 & 2.0 & 1.0 \cr
 PID                      & {2.1} &1.1& {2.1}&1.1 & {2.1}&1.1&2.1&1.1 \cr
$K^0_S$ selection & -- & 3.5 & -- & 3.5 & -- & 3.5 & -- & 3.5 \cr
$\pi^0$ selection & -- & 3.0 & -- & 3.0 & -- & 3.0 & -- & 3.0 \cr
 MC statistics               & {0.2}&0.3 & {0.2}&0.3 & {0.3}&0.4&0.3&0.4  \cr
 polarization    & 0.2 & 1.1& 0.2 & 1.7& \multicolumn{2}{c}{--} & \multicolumn{2}{c}{--} \cr
 event selection       & \multicolumn{2}{c}{1.0} & \multicolumn{2}{c}{1.0} & \multicolumn{2}{c}{1.0}  & \multicolumn{2}{c}{1.0}  \cr
 thrust angle $\theta_T$  & \multicolumn{2}{c}{1.0} & \multicolumn{2}{c}{1.0} & \multicolumn{2}{c}{1.0} & \multicolumn{2}{c}{1.0} \cr
 vertex requirement          & \multicolumn{2}{c}{2.0} & \multicolumn{2}{c}{2.0} & \multicolumn{2}{c}{2.0}  & \multicolumn{2}{c}{2.0} \cr
 $\varphi$ branching fraction   & \multicolumn{2}{c}{1.2}  & \multicolumn{2}{c}{1.2} & \multicolumn{2}{c}{1.2}  & \multicolumn{2}{c}{1.2}  \cr
 $K^*$ branching fraction   & \multicolumn{2}{c}{0.0}  & \multicolumn{2}{c}{2.4} & \multicolumn{2}{c}{0.0}  & \multicolumn{2}{c}{10.8}  \cr
\vspace{-3mm}  $K^0_S$ branching fraction   & -- & 0.1 & -- & 0.1& -- & 0.1& -- & 0.1  \cr
 &  &  & &  \cr
\vspace{-3mm} &  &  &  &  \cr
\hline
\hline
\vspace{-3mm} &  &  &   & \cr
 total  &   \multicolumn{2}{c}{4.0} & \multicolumn{2}{c}{4.7} & \multicolumn{2}{c}{4.0} & \multicolumn{2}{c}{11.5}\cr
\end{tabular}
\end{ruledtabular}
\end{center}
\end{table*}
\endgroup

\begingroup
\begin{table*}[t]
\caption
{\label{tab:systematic1}
Systematic errors in the measurement of the three signal yields (\%) and 
other signal parameters (absolute values), excluding the $\Delta\phi_{00}$ measurement.
Uncertainties due to parameterization, 
acceptance function modeling,
$B$-background ($B$-bkgd), 
fit response,
interference of the $(K\Kbar)$ final states (interf.),
charge asymmetry in reconstruction,
assumptions about the unconstrained ${C\!P}$ asymmetries ${\cal A}^{\perp}_{\CP2}$ and $\Delta\phi_{\perp2}$ (${C\!P}$ asym.),
and the total errors are quoted.
The errors are not quoted if they are either small or not relevant
for a particular measurement.
See text for details. 
}
\begin{center}
\begin{ruledtabular}
\setlength{\extrarowheight}{1.5pt}
\begin{tabular}{ccccccccc}
 & PDF & acceptance & $B$-bkgd & fit & interf. & charge & ${C\!P}$ asym. & total\\ 
\hline
$\varphi (K\pi)_0^{*0}$ yield (\%)&7.1&--&3.4&2.7&--&--&--&8.3\\
$\varphi K^*(892)^0$ yield (\%)&2.3&--&1.9&1.3&1.8&--&--&3.7\\
$\varphi K^*_2(1430)^0$ yield (\%)&3.3&--&0.8&2.8&--&--&--&4.4\\
${\cal A}_{\CP0}$&0.036&0.002&0.048&0.001&--&0.020&0.008&0.064\\
$f_{L1}$&0.002&0.002&0.005&0.007&0.010&--&--&0.013\\
$f_{\perp1}$&0.001&0.001&0.006&0.004&0.010&--&--&0.013\\
$\phi_{\parallel1}$&0.007&0.001&0.010&0.017&0.078&--&--&0.081\\
$\phi_{\perp1}$&0.005&0.001&0.010&0.010&0.084&--&--&0.085\\
${\cal A}_{\CP1}$&0.007&0.001&0.008&0.012&0.018&0.020&--&0.031\\
${\cal A}^{0}_{\CP1}$&0.003&0.003&0.013&0.005&0.019&--&--&0.024\\
${\cal A}^{\perp}_{\CP1}$&0.007&0.004&0.022&0.024&0.052&--&--&0.062\\
$\Delta\phi_{\parallel1}$&0.008&0.001&0.009&0.010&0.078&--&--&0.080\\
$\Delta\phi_{\perp1}$&0.007&0.001&0.007&0.016&0.081&--&--&0.083\\
$\delta_{01}$&0.030&0.001&0.005&0.003&0.080&--&--&0.086\\
$\Delta\delta_{01}$&0.008&0.001&0.006&0.006&0.080&--&--&0.081\\
$f_{L2}$&0.006&0.001&0.016&0.033&--&--&0.004&0.037\\
$f_{\perp2}$&0.001&0.001&0.004&0.031&--&--&0.003&0.031\\
$\phi_{\parallel2}$&0.051&0.002&0.021&0.029&--&--&0.012&0.063\\
${\cal A}_{\CP2}$&0.037&0.002&0.029&0.001&--&0.020&0.005&0.051\\
${\cal A}^{0}_{\CP2}$&0.012&0.001&0.005&0.002&--&--&0.003&0.014\\
$\Delta\phi_{\parallel2}$&0.027&0.002&0.088&0.012&--&--&0.011&0.093\\
$\delta_{02}$&0.117&0.004&0.062&0.009&--&--&0.006&0.133\\
$\Delta\delta_{02}$&0.012&0.004&0.057&0.009&--&--&0.007&0.059\\
\end{tabular}
\end{ruledtabular}
\end{center}
\end{table*}
\endgroup

\begin{table}[htbp]
\caption{\label{tab:systematic2} 
Systematic errors (absolute values)  
in the measurement of $\Delta\phi_{00}$.
See Table~\ref{tab:systematic1} and text for details. 
}
\begin{center}
\setlength{\extrarowheight}{1.5pt}
\begin{tabular}{cc}
\hline
\hline
$\tau_B,~\Delta m_B$ & 0.001\\
sin$2\beta$ measurement & 0.015\\
signal $\Delta t$ resolution & 0.016\\
mistag differences& 0.019\\
$z$ scale + boost & 0.002\\
beam spot & 0.010\\
SVT alignment & 0.001\\
tag-side interference& 0.002\\
background resolution and asymmetry & 0.006\\
$B$-background & 0.024 \\
PDF & 0.009\\
acceptance & 0.004 \\
fit & 0.002 \\
$\CP$-asym. & 0.003 \\
\hline
total & $0.041$ \\
\hline
\hline
\end{tabular}
\end{center}
\end{table}

In Tables~\ref{tab:systematic3},~\ref{tab:systematic1}, and \ref{tab:systematic2}
we summarize the dominant sources of systematic errors in our measurements.
In the measurement of the 
branching fractions, we tabulate separately the multiplicative errors on selection
efficiency in Table~\ref{tab:systematic3}.
Measurement of all parameters suffers from uncertainties in the fit 
model which are discussed below and in Table~\ref{tab:systematic1}. 
One additional error in the
branching fraction measurement is the error on the number of $B^0\Bbar^0$ 
mesons produced and is estimated to be 1.1$\%$, where we assume equal decay rates of 
$\Upsilon(4S)\to B^0\Bbar^0$ and $B^+B^-$. While most of the errors are
dominated by uncertainties in the $\BorBbar^0\to\varphi K^\pm\pi^\mp$ decay mode, 
the measurement of $\Delta\phi_{00}$ has additional systematic errors
unique to the $B^0\to\varphi K^0_S\pi^0$ decay mode. Therefore all 
systematic errors on $\Delta\phi_{00}$ are quoted in Table~\ref{tab:systematic2}.

The systematic errors in the efficiency are typically due to
imperfect MC simulation and they are
obtained from independent studies, such as control samples.
They affect the errors in the branching fraction, 
but do not change the significance of the signal yield.
From a study of absolute tracking efficiency, we evaluate 
the corrections to the track finding efficiencies, resulting
in a systematic error of $0.5\%$ per track and the total 
error of 2.0\% for four tracks.
The error due to particle identification 
is about 2\% and is dominated by the kaon selection requirements.
Particle identification performance has been validated with 
high-statistics data and MC control samples, such as 
$\Dbar^*$-tagged $\Dbar\to K\pi$ decays.
The $K^0_S$ selection efficiency systematic uncertainty is taken from 
an inclusive  $K^0_S$ control sample study, giving a total 
error of 3.5\%.
The $\pi^0$ reconstruction efficiency error is estimated to be 3.0\% 
from a study of $\tau$ decays to modes with $\pi^0$ mesons.

The reconstruction efficiency has a weak dependence on the
fraction of longitudinal polarization due to a non-uniform 
acceptance function of the helicity angles.
Therefore, we use the measured value of the polarization when
computing the efficiency. The uncertainty in this measurement
translates into a systematic error in the branching fraction.
Several requirements on the multi-hadronic final state, minimum
number of charged tracks, event-shape, and vertex requirements
result in a few percent errors.
Other errors come from the uncertainty in the daughter branching
fractions for $\varphi\to K^+K^-$ and $K^*\to K\pi$.
All these errors are summarized in Table~\ref{tab:systematic3}.

When we perform the ML fit we make certain assumptions
about the signal and background distributions.
Most background parameters are allowed to vary in the fit, but
we constrain most $B$ decay parameters to the expectations
based on MC and control samples.
In order to account for the resulting uncertainty, we
vary the parameters within their errors, taking into
account correlations among the parameters. 
We obtain $m_{\rm ES}$, $\Delta E$,  and ${\cal F}$ 
uncertainties from the control samples discussed 
in Section~\ref{sec:note-reco}.
The invariant mass uncertainties incorporate
errors on the resonance parameters as quoted in Table~\ref{tab:mass} 
and Ref.~\cite{bib:Amsler2008}.
We take into account resolution in the $K\pi$ and
$K^+K^-$ invariant masses with the corresponding
errors on the absolute values of 1 MeV and 0.3 MeV,
respectively.

We separate a special class of PDF uncertainties
for the helicity angles, due to the acceptance
function. In addition to statistical errors
in the MC sample, we consider the momentum-dependent
uncertainty of the tracking efficiency. The main 
effect is on the curvature of the acceptance
function shown in Fig.~\ref{fig:acc-func} due to a 
strong correlation between the momentum of a track and
the value of the helicity angle.
Moreover, in order to study the effects of charge asymmetry
in angular distributions, we apply the acceptance correction
independently to only $B$ or only $\Bbar$ decay subsamples.
The largest deviation is taken as the ``acceptance'' 
systematic error quoted in Table~\ref{tab:systematic1}.

To estimate the effect of the $B$ meson decays which could mimic
signal, we study a full GEANT4-based MC simulation of the 
$\Upsilon(4S)\to B\Bbar$ events. We embed the categories of
events which may have $\DeltaE$ and $m_{\rm ES}$ distributions
similar to signal into the data sample and observe variations 
of the fit results which we take as systematic errors.
The nonresonant contribution
is taken into account naturally in the fit with both $K\Kbar$ and $K\pi$
contributions allowed to vary. 
The former is modeled as $B^0\to f_0K^{*0}$ and the latter 
is a part of the $S$-wave $K\pi$ parameterization.
Interference effects are studied separately.
We also take into account the uncertainty in the shape
of the $K^+K^-$ invariant mass distribution. The default
parameterization assumes the $B^0\to f_0K^*(892)^0$ decay and
we vary it to the phase-space $B^0\to K^+K^-K^*(892)^0$ 
distribution.
We constrain the number of $B^0\to \varphi K^*(892)^0$
events contributing to the higher $K\pi$ invariant mass
range based on the measured branching fraction, but we also
vary this number according to the branching fraction errors.
These errors are quoted as ``$B$-bkgd'' in 
Table~\ref{tab:systematic1}.

The selected signal $B^0\to\varphi K^{*0}$ events contain a 
small fraction of incorrectly reconstructed candidates.
Misreconstruction occurs when at least one candidate track
belongs to the decay products of the other $B$ from the $\FourS$ decay,
which happens in about 5$\%$ of the cases in the 
$\KorKbar^{*0}\to K^\pm\pi^\mp$ decay.
The distributions that show peaks for correctly reconstructed
events have substantial tails, with large uncertainties in MC
simulation, when misreconstructed events are included.
These tails and incorrect angular dependence
would reduce the power of the distributions
to discriminate between the background and the collection of
correctly and incorrectly reconstructed events. 
We choose,
therefore, to represent only the correctly reconstructed candidates
in the signal PDF, and to calculate the reconstruction efficiency
with both the correctly reconstructed and misreconstructed MC events. 
Fitting the generated samples to determine
the number of correctly reconstructed candidates has an efficiency
close to 100\% even though a few percent of selected candidates are
identified as background.
We account for this with a systematic error taken as half the
fraction of candidates identified as background and
quoted as the ``fit'' entry in Table~\ref{tab:systematic1}.
Similarly, we obtain systematic errors on other parameters from 
the largest deviation from the expectation.
This includes a potential bias from the finite resolution
in the helicity angle measurement and a possible dilution due to
the presence of the misreconstructed component.

As we discuss below, a substantial $B^0\to f_0K\pi$ contribution is found in
the lower $K\pi$ mass range, corresponding to either
$B^0\to f_0K^*(892)^0$ decays, or any other contribution with a broad
$K^+K^-$ invariant mass distribution, either resonant or nonresonant.
The uncertainties due to $m_{\!K\Kbar}$ interference are estimated 
with the samples generated according to the observed $K^+K^-$ intensity 
and with various interference phases analogous to $\delta_{0J}$ in $K\pi$. 
These are the dominant systematic errors for the $\vec{\zeta}$ parameters 
of the $B^0\to\varphi K^*(892)^0$ decay.
No significant $B^0\to f_0K\pi$ contribution is observed in the higher 
$K\pi$ mass range.

The charge bias uncertainty affects only the relative yields 
of $B$ and $\Bbar$ events.
We assign a systematic error of 2\%, which accounts
mostly for a possible asymmetry in the reconstruction of a charged kaon
from a $K^{*0}$~\cite{acp}.
Overall charge asymmetry has a negligible effect
on the angular asymmetry parameters, while the angular
dependence of the charge asymmetry is tested with the
flavor-dependent acceptance function discussed above.

There are still two $C\!P$ parameters, 
${\cal A}^{\perp}_{\CP2}$ and $\Delta\phi_{\perp2}$,
that are not measured in the
$\BorBbar^0\to\varphi \KorKbar^{*}_2(1430)^0$ $\to \varphi K^\pm\pi^\mp$ decay.
We assume zero asymmetry as the most likely value and vary them within
$\pm 0.2$ for the direct-$C\!P$ asymmetry and $\pm 0.5$ rad for the
phase asymmetry.
All the errors on the fit parameters are summarized
in Table~\ref{tab:systematic1}.

For the time-dependent measurement of $\Delta\phi_{00}$, 
the variations are quoted in  Table~\ref{tab:systematic2}.
We vary the $B^0$ lifetime and $\Delta m_B$ within their errors~\cite{bib:Amsler2008}.
We include the error of the $\beta$ measurement
from Eq.~(\ref{eq:sin2beta}).
We use the results of the sin$2\beta$ analysis~\cite{Aubert:2007hm} 
to estimate the systematic errors related to signal parameters,
such as $\Delta t$ signal resolution and mistag differences, 
detector effects ($z$ scale and boost, beamspot, SVT alignment
uncertainties), and the tag-side interference. 
Background $C\!P$-asymmetry and resolution parameters are 
determined from the sideband data and then constrained in the fit.
The constrained parameters are varied according to their errors 
and added in quadrature to compute the systematic errors.



\section{RESULTS}
\label{sec:note-results}

\begingroup
\begin{table*}[t]
\caption
{\label{tab:results1}
Analysis results: 
the reconstruction efficiency $\varepsilon_{\rm reco}$ obtained
from MC simulation; the total efficiency $\varepsilon$, 
including the daughter branching fractions~\cite{bib:Amsler2008}; 
the number of signal events $n_{{\rm sig}J}$; statistical 
significance (${\cal S}$) of the signal; the branching fraction ${\cal B}_J$;
and the flavor asymmetry ${\cal A}_{CPJ}$.
The branching fraction ${\cal B}(B^0\to\varphi (K\pi)^{*0}_0$) refers to the coherent 
sum $|A_\text{res}+A_\text{non-res}|^2$
of resonant and nonresonant $J^P=0^+$ $K\pi$ components and 
is quoted for $m_{K\!\pi}<1.6$~GeV, while the ${\cal B}(B^0\to\varphi K_0^{*}(1430)^0)$ 
is derived from it by integrating separately the B-W formula of the 
resonant $|A_\text{res}|^2$ $K\pi$ component without $m_{K\!\pi}$ restriction.
The systematic errors are quoted last.
}
\begin{center}
\begin{ruledtabular}
\setlength{\extrarowheight}{1.5pt}
\begin{tabular}{ccccccc}
\vspace{-3mm}&&&\\
mode  
 & $\varepsilon_{\rm reco}$ (\%) & $\varepsilon$ (\%) & $n_{{\rm sig}J}$ (events)
 & ${\cal S}$ ($\sigma$) & ${\cal B}_J$  ($10^{-6}$) & ${\cal A}_{CPJ}$ \cr
\vspace{-3mm} & & & & \\
\hline
\vspace{-3mm} & & & & \\
 $\varphi K_0^{*}(1430)^0$ 
 &    &  &  & 
 & $3.9\pm{0.5}\pm 0.6$ & \\
\vspace{-3mm} & & & & \\
 $\varphi (K\pi)^{*0}_0$ 
 &  & $8.3\pm 0.3$  & $172\pm{24}\pm 14$
  & 11 & $4.3\pm{0.6}\pm 0.4$ & $+0.20\pm{0.14}\pm 0.06$
\\
$\to K^{\pm}\pi^{\mp}$&$23.2\pm{0.9}$&$7.6\pm 0.3$&$158\pm{22}\pm 13$&&&\\
$\to K^0_S\pi^0$&$11.7\pm0.4$&$0.66\pm0.03$&$14\pm2\pm1$&&&\\
\vspace{-3mm} & & & &\\
 $\varphi K^{*}(892)^0$ 
 & & $11.9\pm 0.4$  & $535\pm 28\pm 20$
 & $24$ & $9.7\pm{0.5}\pm 0.5$ & $+0.01\pm{0.06}\pm 0.03$  \\
$\to K^{\pm}\pi^{\mp}$&$33.7\pm{1.3}$&$11.1\pm 0.4$&$500\pm 26\pm 19$&&&\\
$\to K^0_S\pi^0$&$13.8\pm{0.5}$&$0.78\pm0.03$&$35\pm2\pm1$&&&\\
\vspace{-3mm} & & & & \\
 $\varphi K_2^{*}(1430)^0$ 
 &  & $4.7\pm 0.2$  & $167\pm{21}\pm 8$
  & $11$ & $7.5\pm{0.9}\pm 0.5$ & $-0.08\pm{0.12}\pm 0.05$\\
$\to K^{\pm}\pi^{\mp}$&$26.7\pm{1.0}$&$4.4\pm 0.2$&$158\pm{20}\pm 7$&&&\\
$\to K^0_S\pi^0$&$8.7\pm0.3$&$0.25\pm0.01$&$9\pm1\pm1$&&&\\
\vspace{-3mm} & & & &\\
\end{tabular}
\end{ruledtabular}
\end{center}
\end{table*}
\endgroup

\begingroup
\begin{table*}[bth]
\caption{\label{tab:results2}
Summary of the results, see Table~\ref{tab:parameters} for definition of the parameters.
The branching fractions ${\cal B}_J$ and flavor asymmetries ${\cal A}_{C\!P J}$ 
are quoted from Table~\ref{tab:results1}.
The systematic errors are quoted last.
The dominant fit correlation coefficients (${\rho}$) 
are presented for the $\varphi K^{*}(892)^0$ and $\varphi K_2^{*}(1430)^0$ modes where 
we show correlations of ${\delta_0}$  with ${\phi_\parallel}/{\phi_\perp}$
and of ${\Delta\delta_0}$ with ${\Delta\phi_\parallel}/{\Delta\phi_\perp}$.
}
\begin{center}
{
\begin{ruledtabular}
\setlength{\extrarowheight}{1.5pt}
\begin{tabular}{cccccc}
\vspace{-3mm} & & & & & \\
   parameter
 &  $\varphi K^{*}_0(1430)^0$ & $\varphi K^{*}(892)^0$ & ${\rho}$ & $\varphi K^{*}_2(1430)^0$  & ${\rho}$ \cr
 &  $J=0$ & $J=1$ &  & $J=2$  & \cr
\vspace{-3mm} & & &  &\\
\hline
\vspace{-3mm} & & & & &\\
  ${\cal B}_J$ $(10^{-6})$ 
&  $3.9\pm{0.5}\pm 0.6$ & $9.7\pm 0.5\pm 0.5$ & &$7.5\pm{0.9}\pm 0.5$ & 
\cr
\vspace{-3mm} & & & & &\\
  ${f_{LJ}}$  
& & $0.494\pm0.034\pm0.013$ & \multirow{2}{13mm}{~{\Large\}}$-48\%$} &$0.901^{+0.046}_{-0.058}\pm 0.037$ & \multirow{2}{13mm}{~{\Large\}}$-15\%$}
\cr
\vspace{-3mm} & & & & &\\
  ${f_{\perp J}}$ 
& & $0.212\pm0.032\pm0.013$& & $0.002^{+0.018}_{-0.002}\pm 0.031$ &
\cr
\vspace{-3mm} & & & & &\\
  ${\phi_{\parallel J}}$ (rad) 
& & $2.40\pm0.13\pm0.08$ &\multirow{2}{13mm}{~{\Large\}}~$62\%$}& $3.96\pm 0.38\pm 0.06$   &  --
\cr
\vspace{-3mm} & & & & &\\
  ${\phi_{\perp J}}$  (rad) 
&  &$2.35\pm0.13\pm0.09$& & -- &
\cr
\vspace{-3mm} & & & & &\\
  ${\delta_{0J}}$  (rad) 
& & $2.82\pm0.15\pm0.09$ & $34\%/25\%$&$3.41\pm0.13\pm 0.13$  & $19\%$
\cr
\vspace{0mm} &  & &&&\\
  ${\cal A}_{C\!P J}$ 
& $+0.20\pm{0.14}\pm 0.06$ & $+0.01\pm0.06\pm0.03$& & $-0.08\pm{0.12}\pm 0.05$ & 
\cr
\vspace{-3mm} & & & & &\\
  ${\cal A}_{C\!P J}^0$ 
& &  $+0.01\pm0.07\pm0.02$ & \multirow{2}{13mm}{~{\Large\}}$-47\%$}& $-0.05\pm0.06\pm0.01$   &  {--}
\cr
\vspace{-3mm} &  & &&&\\
  ${\cal A}_{C\!P J}^{\perp}$ 
& &  $-0.04\pm0.15\pm0.06$& & -- & 
\cr
\vspace{-3mm} &  & &&&\\
  $\Delta \phi_{\parallel J}$  (rad) 
& &   $+0.22\pm0.12\pm0.08$ &  \multirow{2}{13mm}{~{\Large\}}~$62\%$}& $-1.00\pm0.38\pm0.09$  
&  -- \cr
\vspace{-3mm} & & & & & \\
  $\Delta \phi_{\perp J}$  (rad) 
& &  $+0.21\pm0.13\pm0.08$ && -- & 
\cr
\vspace{-3mm} & & & & & \\
  ${\Delta\delta_{0J}}$  (rad) 
& &  $+0.27\pm0.14\pm0.08$ & $35\%/24\%$& $+0.11\pm0.13\pm0.06$ & $16\%$
\cr
\vspace{0mm} & & & & & \\
  ${\Delta\phi_{00}}$  (rad) 
&{$0.28\pm0.42\pm0.04$}&  & & &
\cr
\end{tabular}
\end{ruledtabular}
}
\end{center}
\end{table*}
\endgroup

\begin{figure}[htbp]
\centerline{
\setlength{\epsfxsize}{0.50\linewidth}\leavevmode\epsfbox{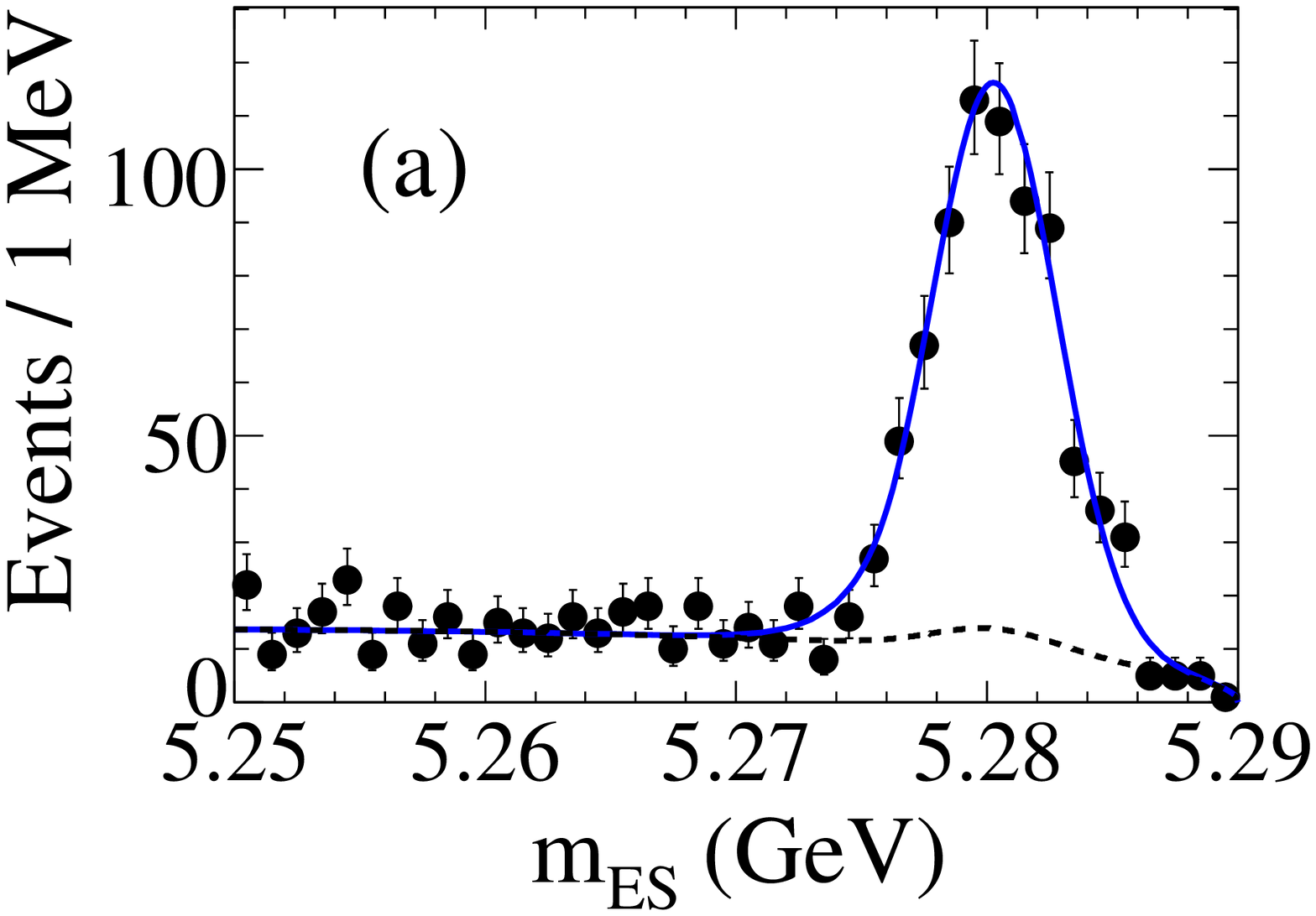}
\setlength{\epsfxsize}{0.50\linewidth}\leavevmode\epsfbox{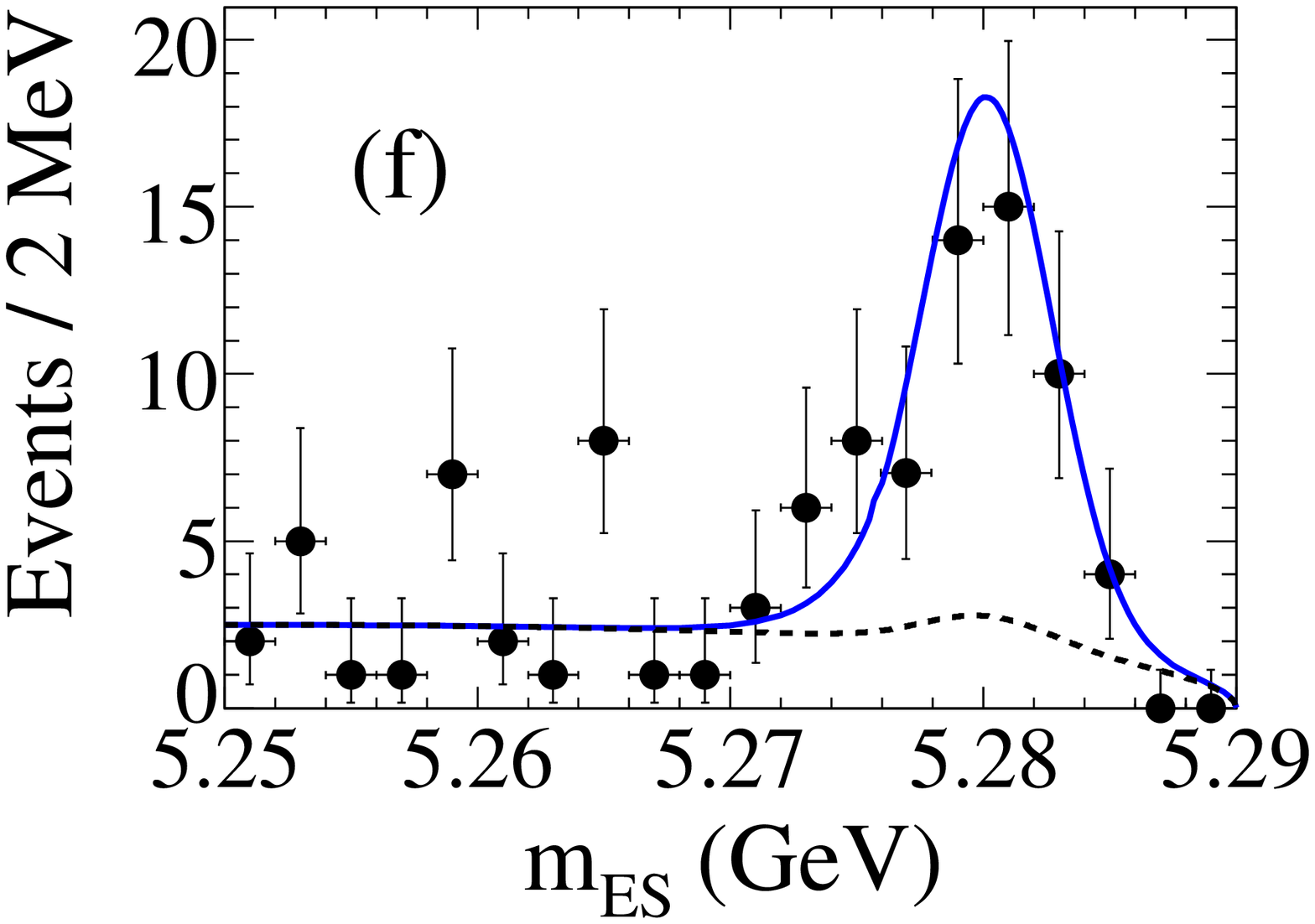}
}
\centerline{
\setlength{\epsfxsize}{0.50\linewidth}\leavevmode\epsfbox{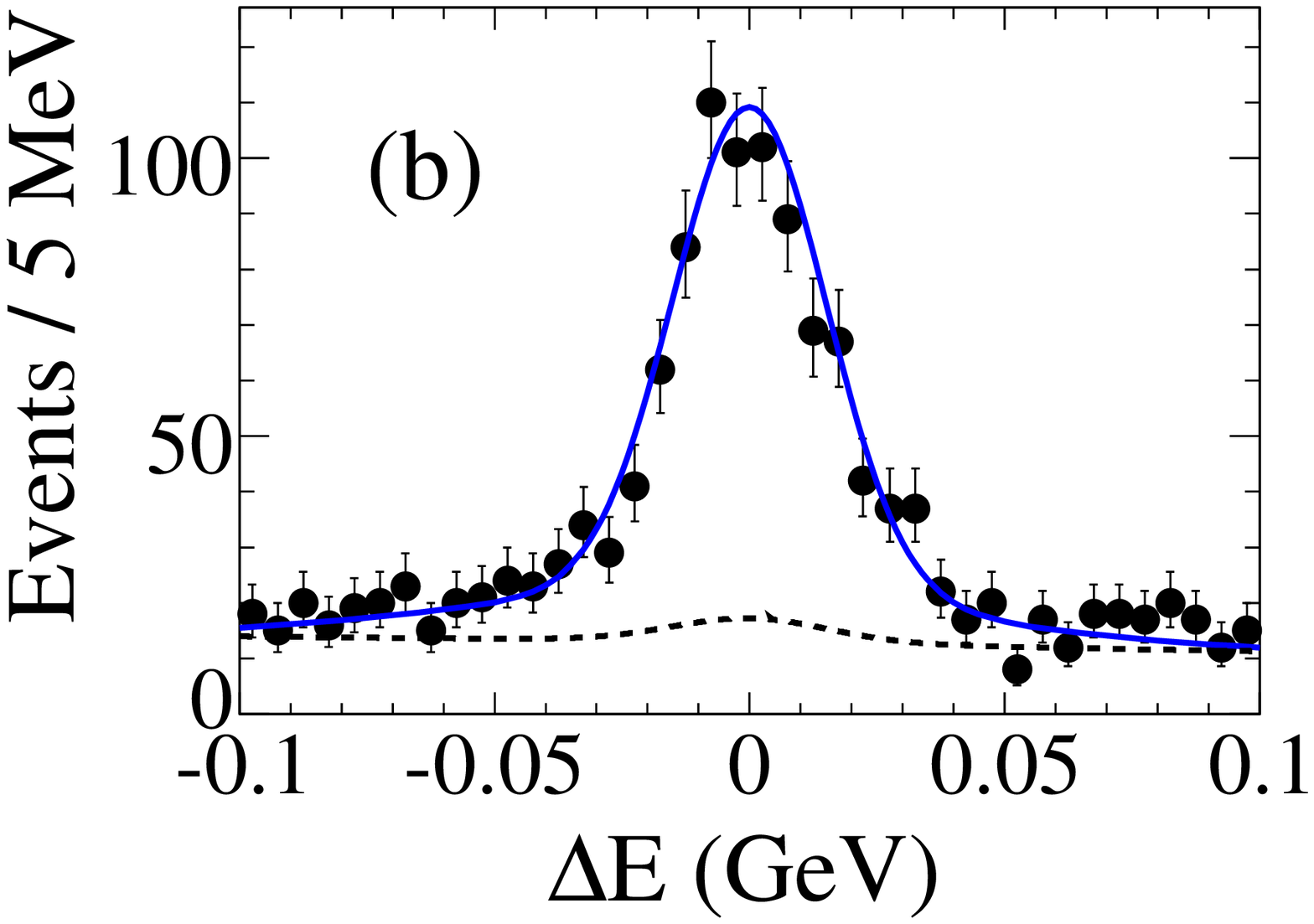}
\setlength{\epsfxsize}{0.50\linewidth}\leavevmode\epsfbox{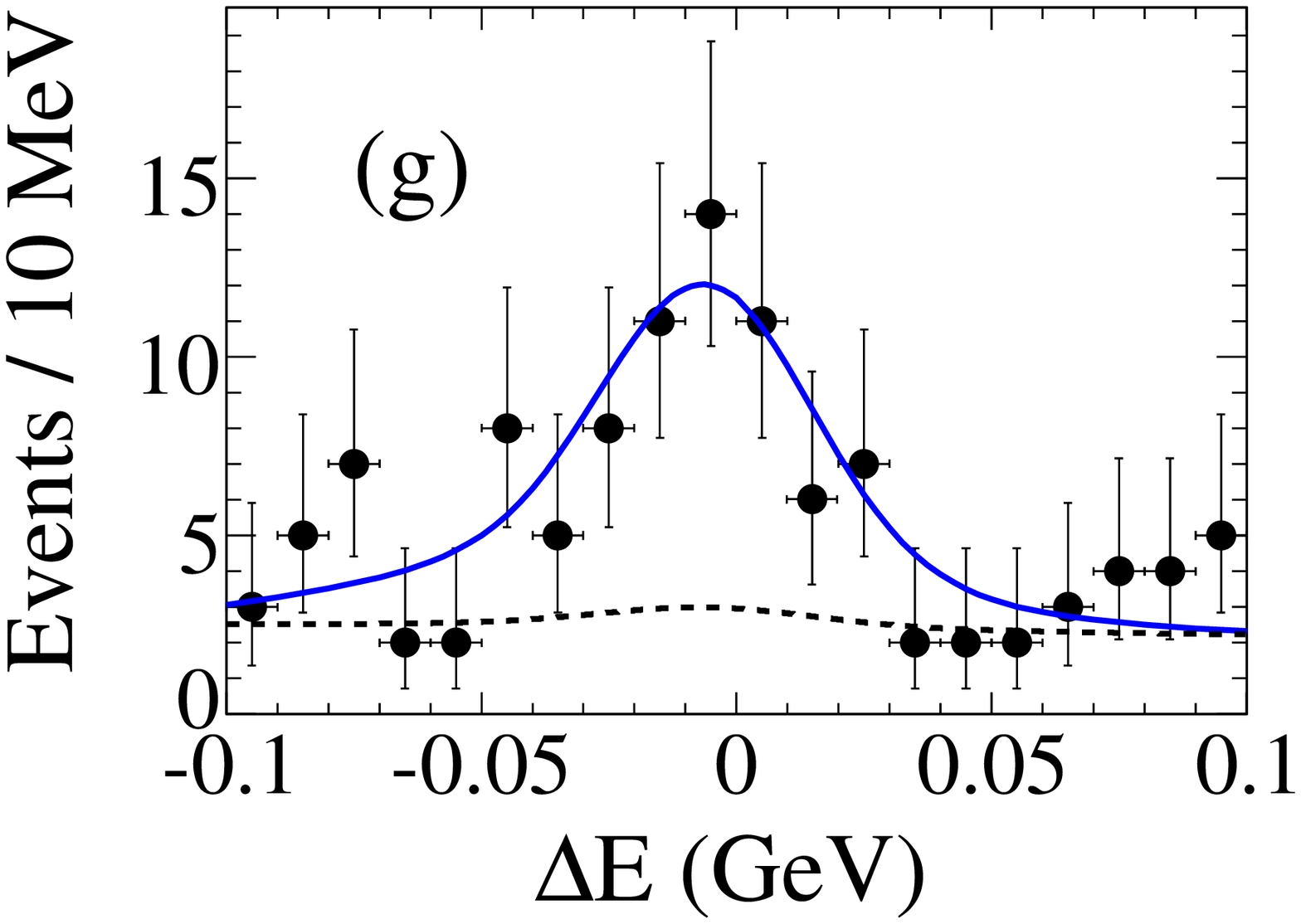}
}
\centerline{
\setlength{\epsfxsize}{0.50\linewidth}\leavevmode\epsfbox{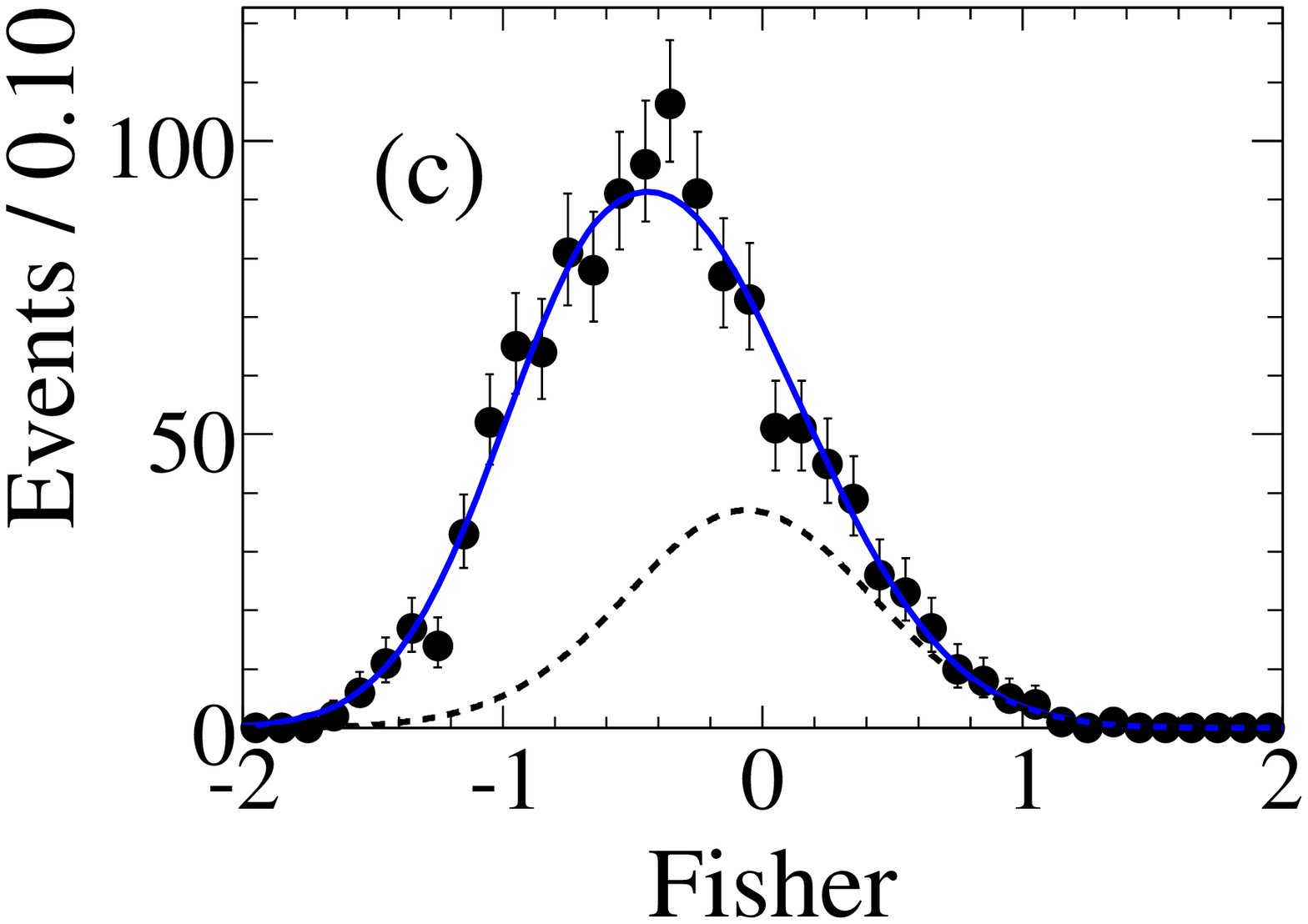}
\setlength{\epsfxsize}{0.50\linewidth}\leavevmode\epsfbox{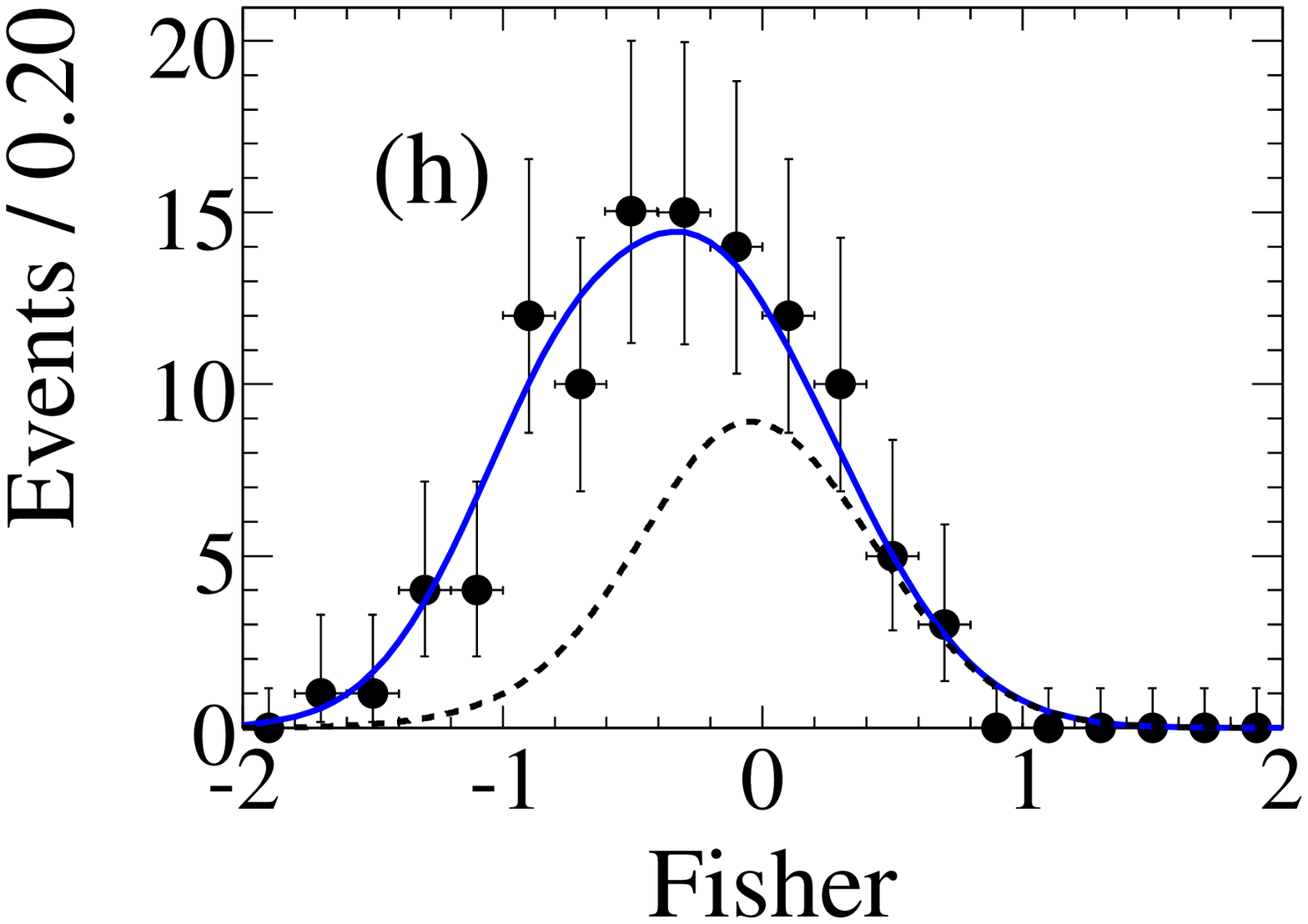}
}
\centerline{
\setlength{\epsfxsize}{0.50\linewidth}\leavevmode\epsfbox{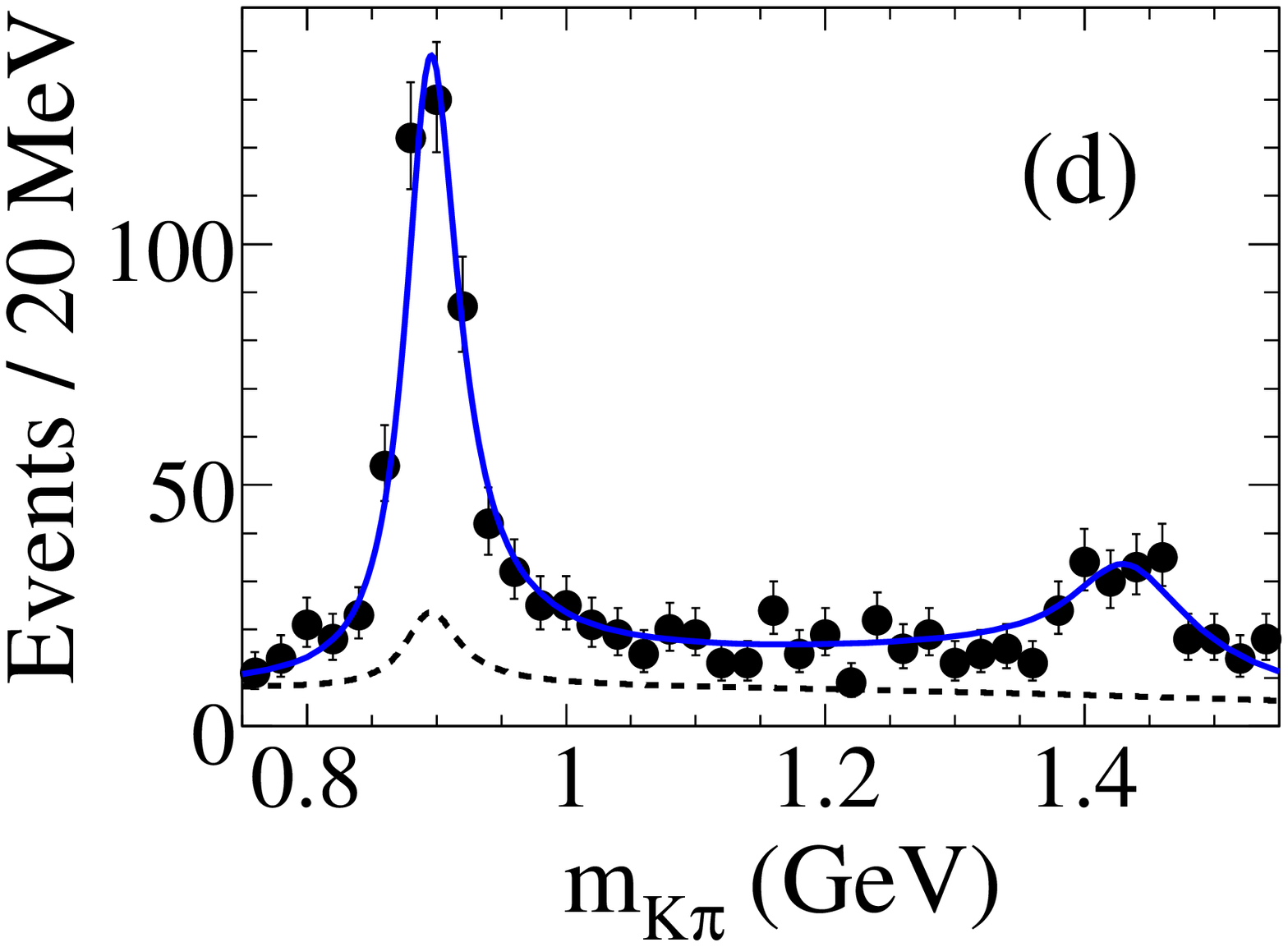}
\setlength{\epsfxsize}{0.50\linewidth}\leavevmode\epsfbox{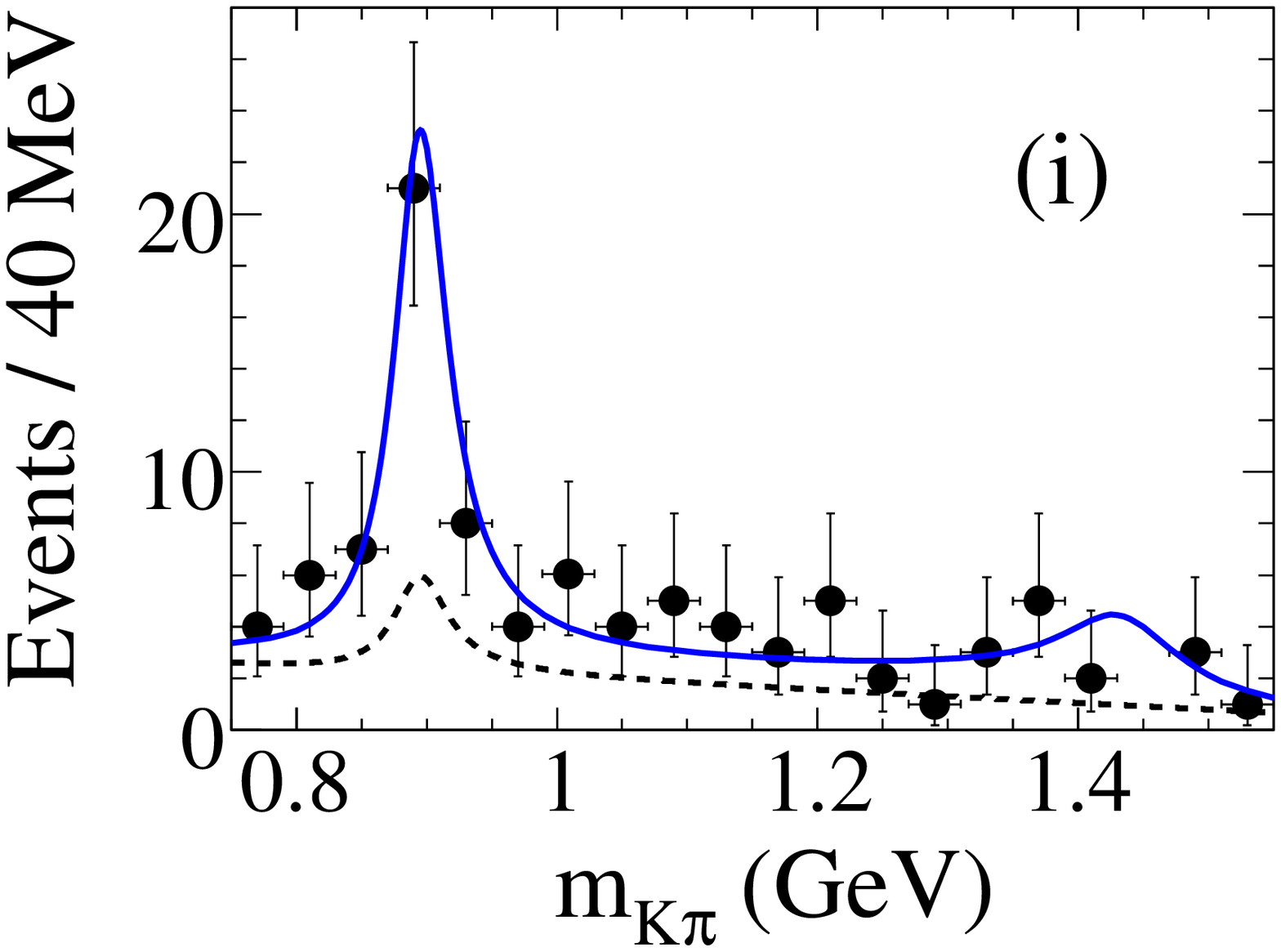}
}
\centerline{
\setlength{\epsfxsize}{0.50\linewidth}\leavevmode\epsfbox{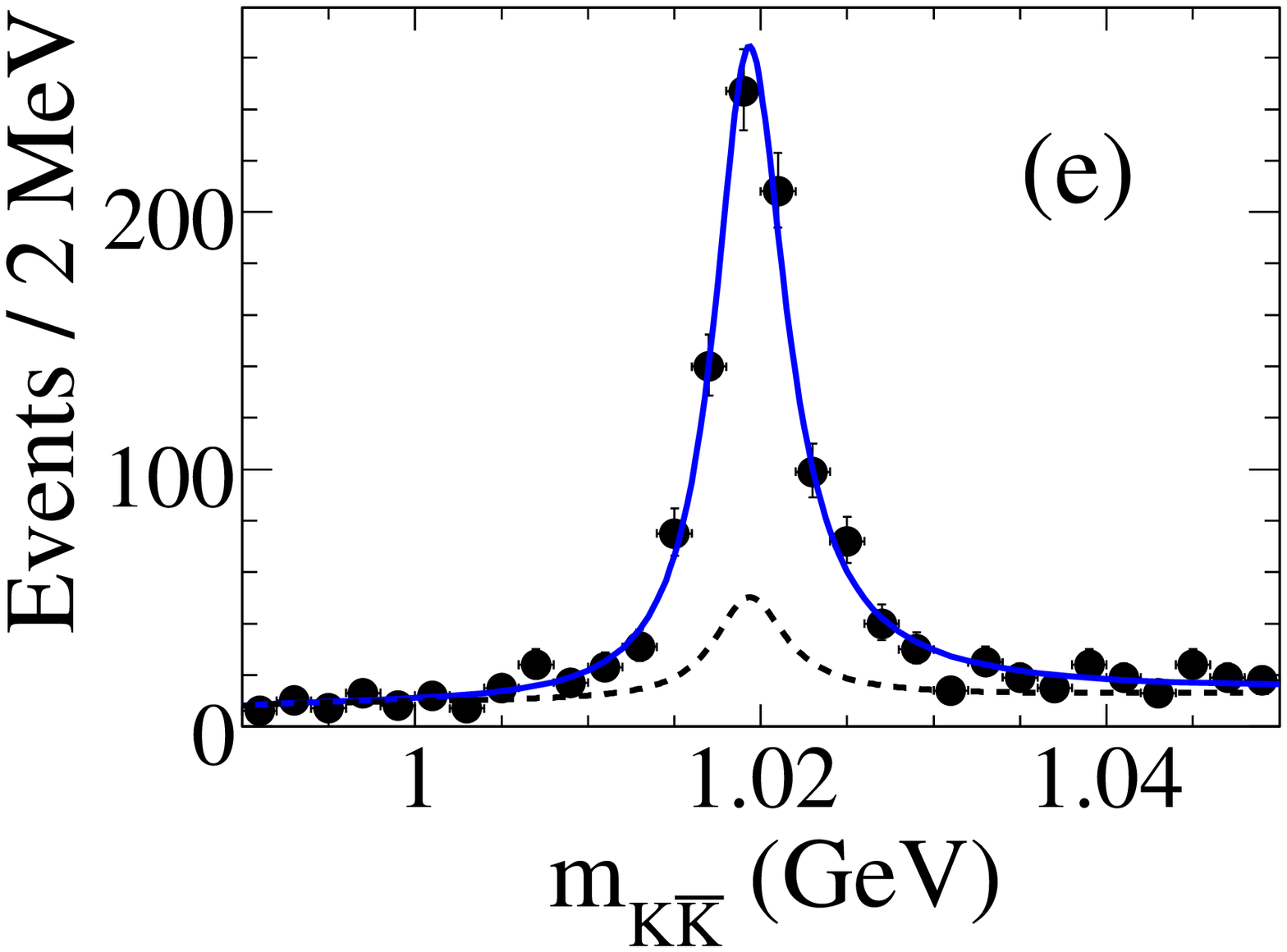}
\setlength{\epsfxsize}{0.50\linewidth}\leavevmode\epsfbox{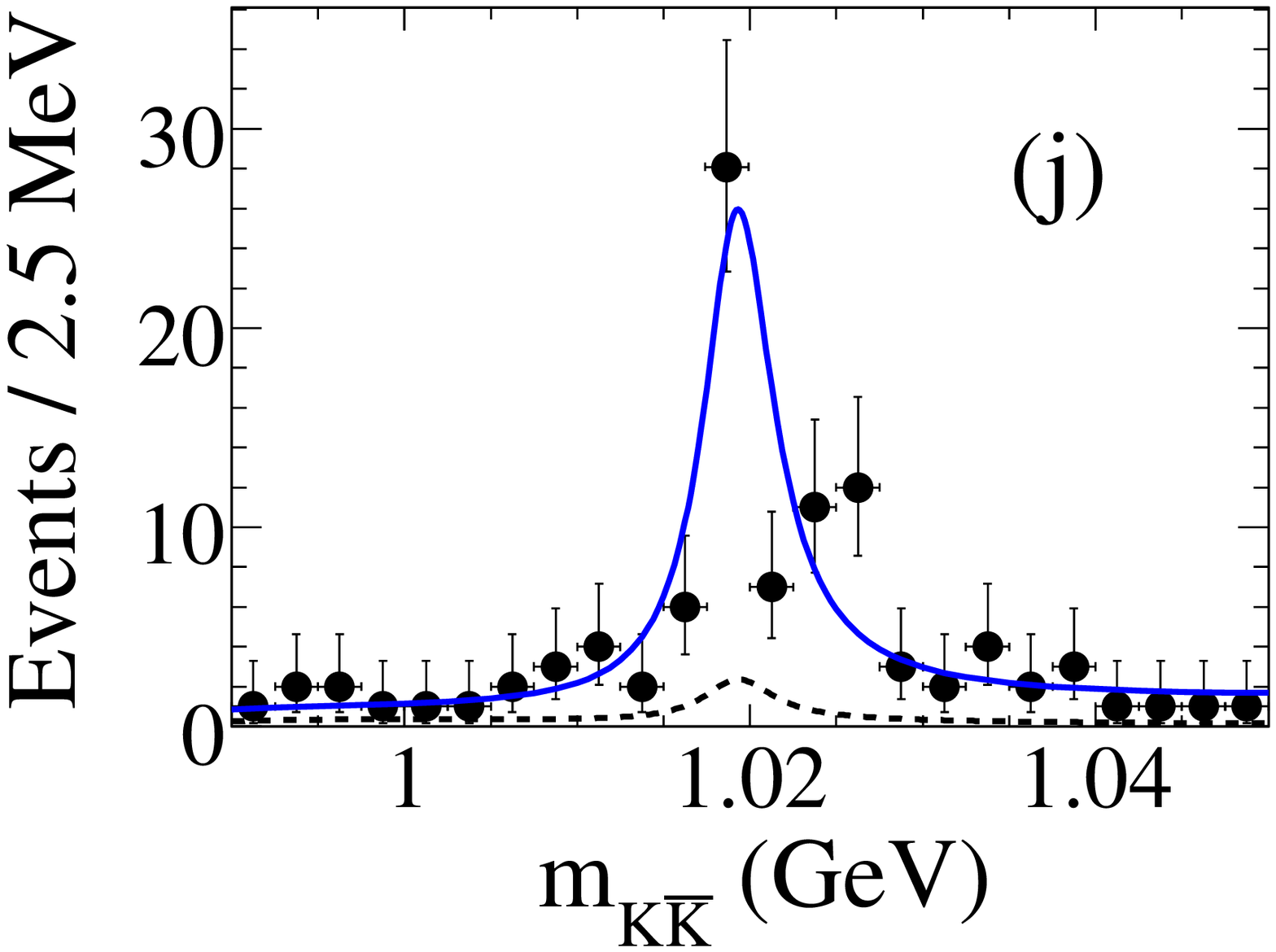}
}
\caption{\label{fig:kine-projection}
Projections onto the variables $m_{\rm ES}$, $\Delta E$, ${\cal F}$, $m_{K\!\pi}$, 
and $m_{K\!\Kbar}$ for the signal $B\to\varphi K\!^{\pm}\pi^{\mp}$ (left) and 
$B\to\varphi K\!^0_S\pi^0$ (right) candidates.
Data distributions are shown with a requirement on the signal-to-background
probability ratio calculated with the plotted variable excluded.
The solid (dashed) lines show the signal-plus-background
(background) PDF projections.
}
\end{figure}

\begin{figure}[htbp]
\centerline{
\setlength{\epsfxsize}{0.50\linewidth}\leavevmode\epsfbox{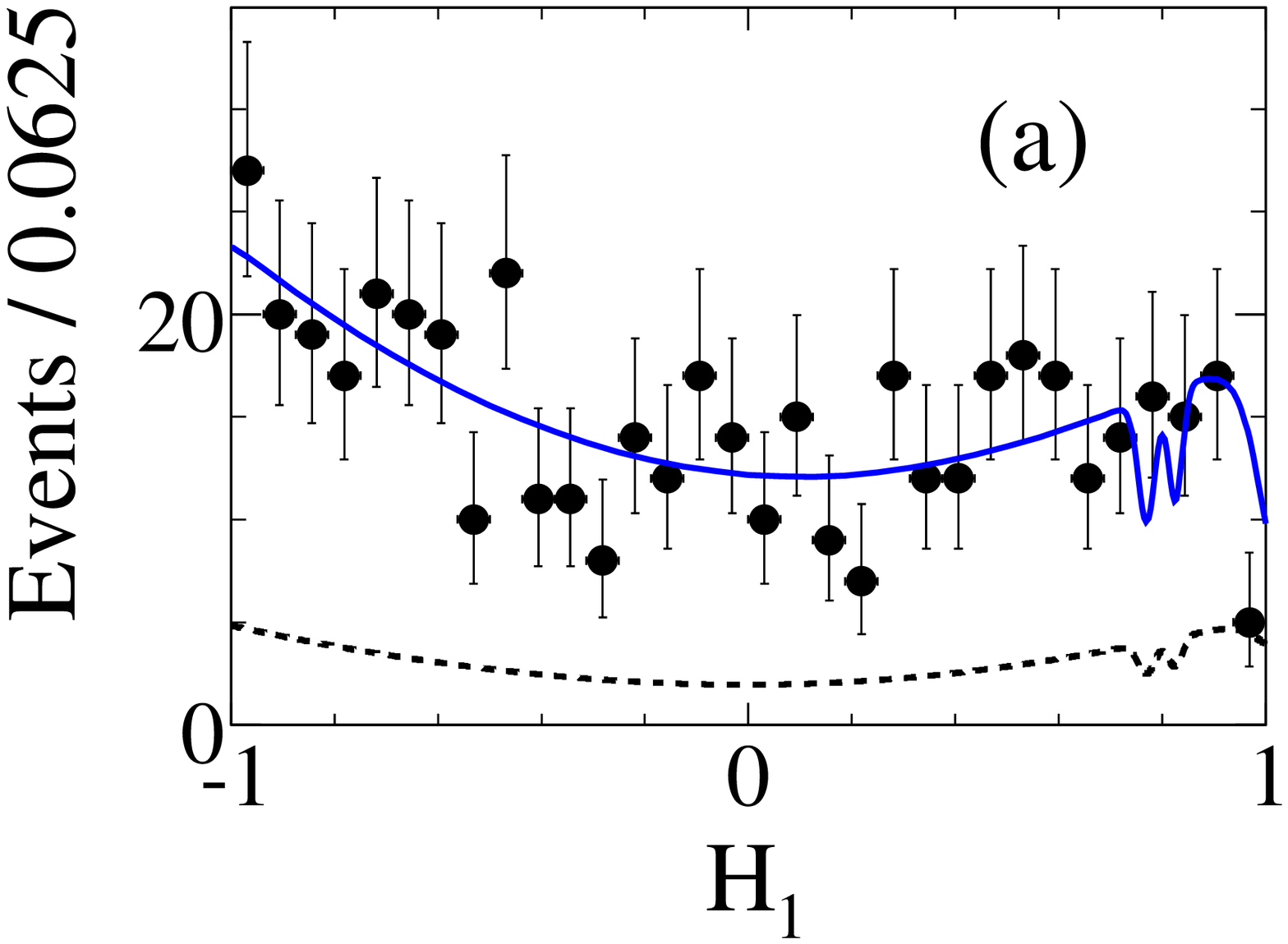}
\setlength{\epsfxsize}{0.50\linewidth}\leavevmode\epsfbox{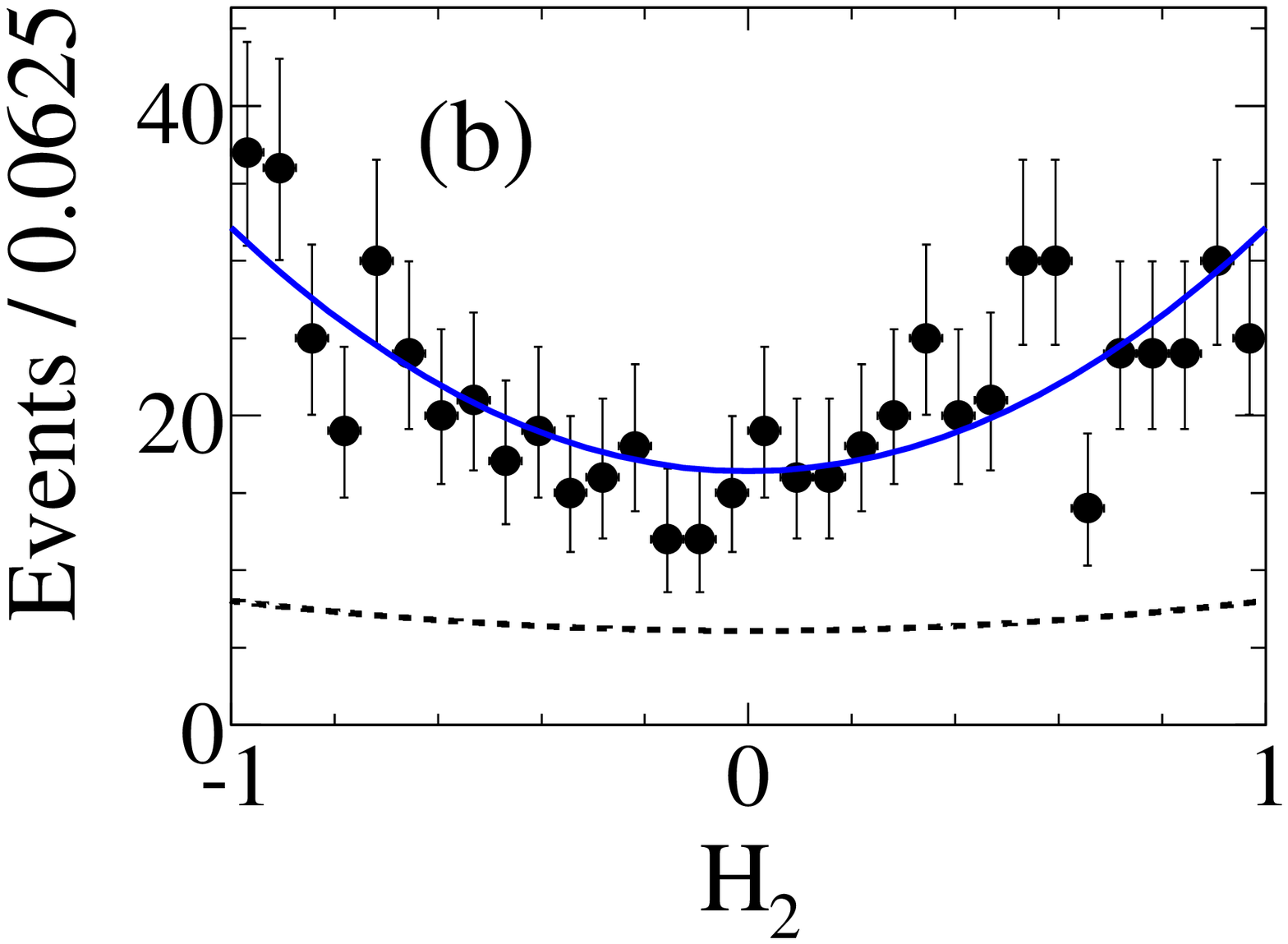}
}
\centerline{
\setlength{\epsfxsize}{0.50\linewidth}\leavevmode\epsfbox{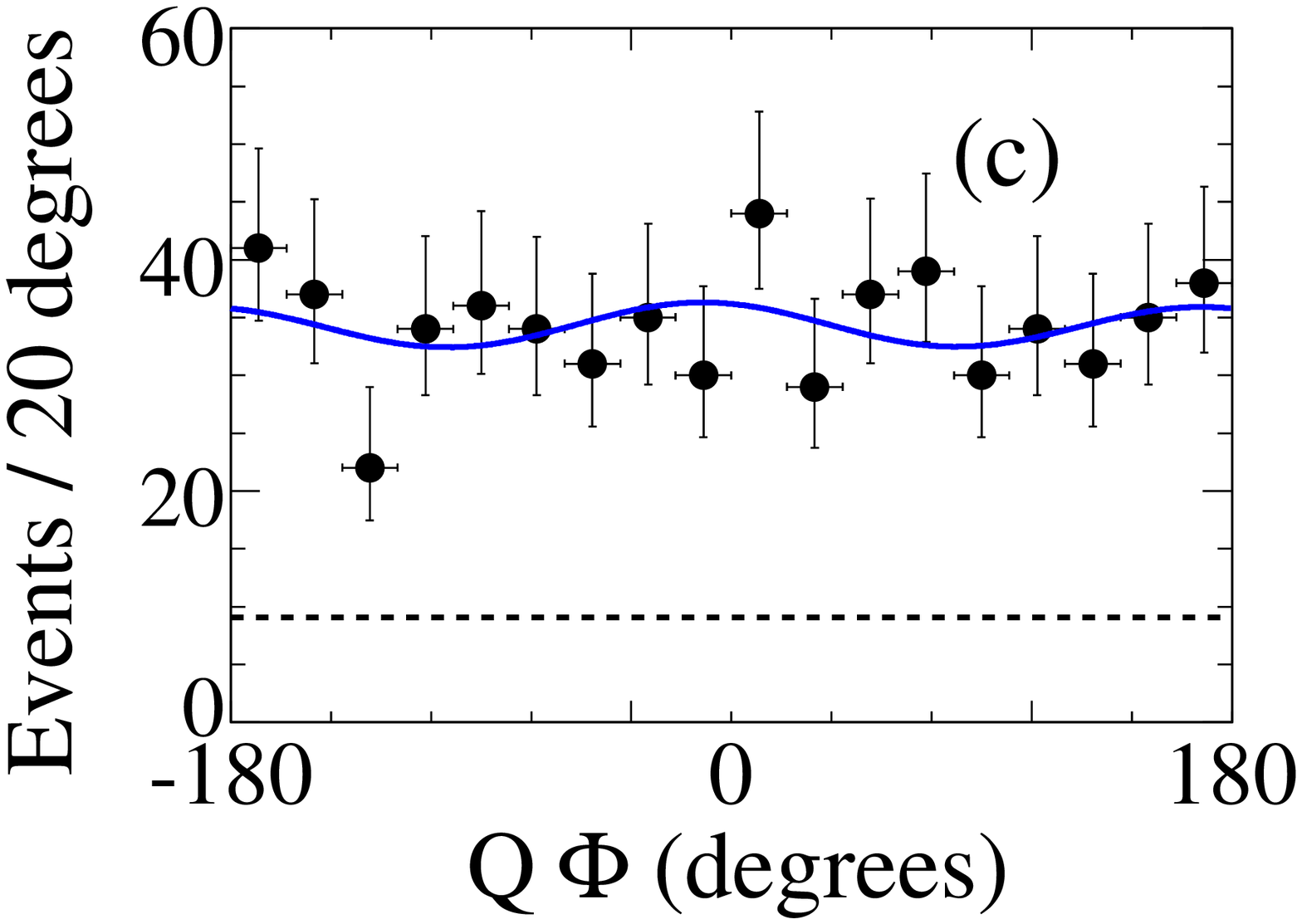}
\setlength{\epsfxsize}{0.50\linewidth}\leavevmode\epsfbox{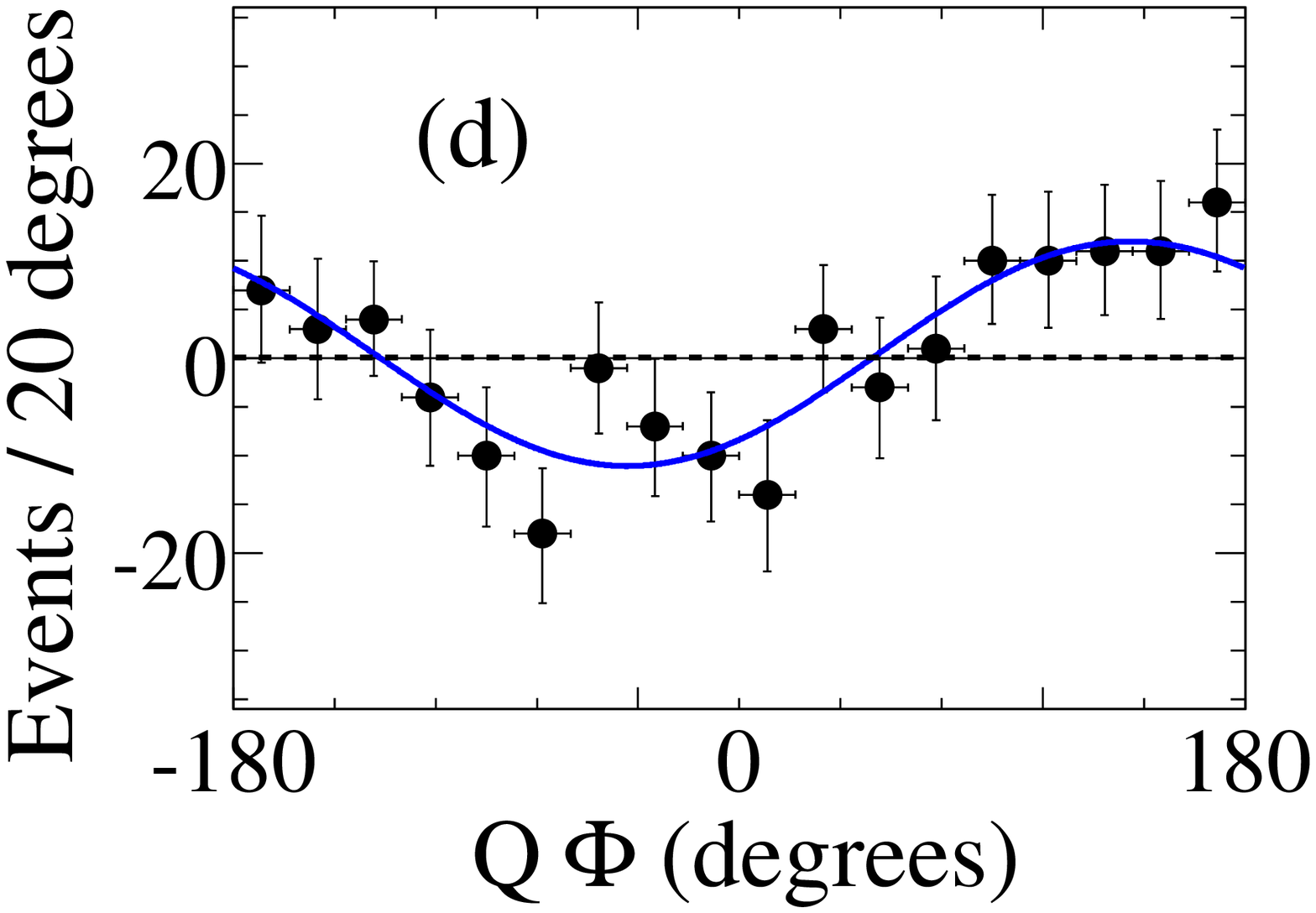}
}
\centerline{
\setlength{\epsfxsize}{0.50\linewidth}\leavevmode\epsfbox{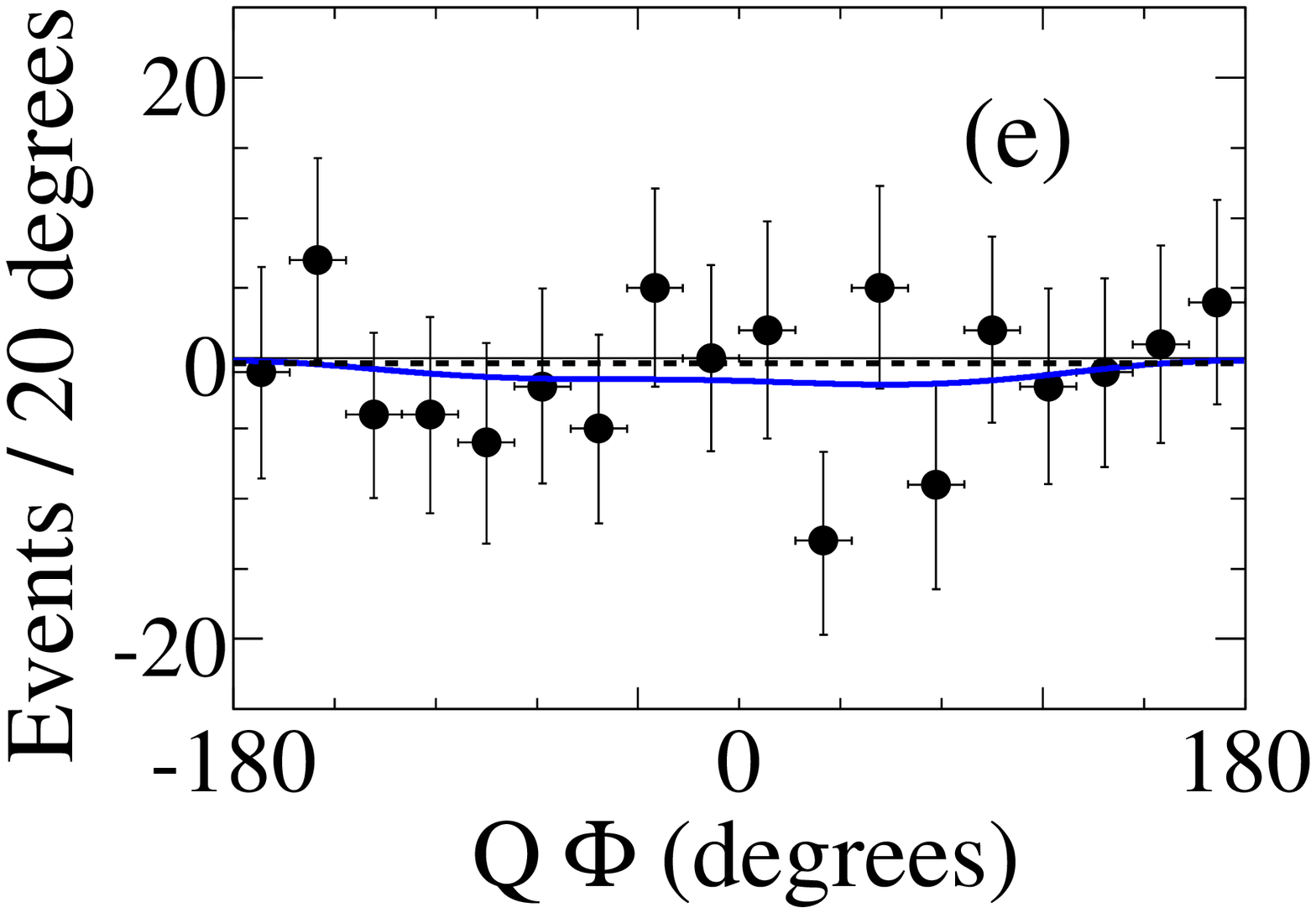}
\setlength{\epsfxsize}{0.50\linewidth}\leavevmode\epsfbox{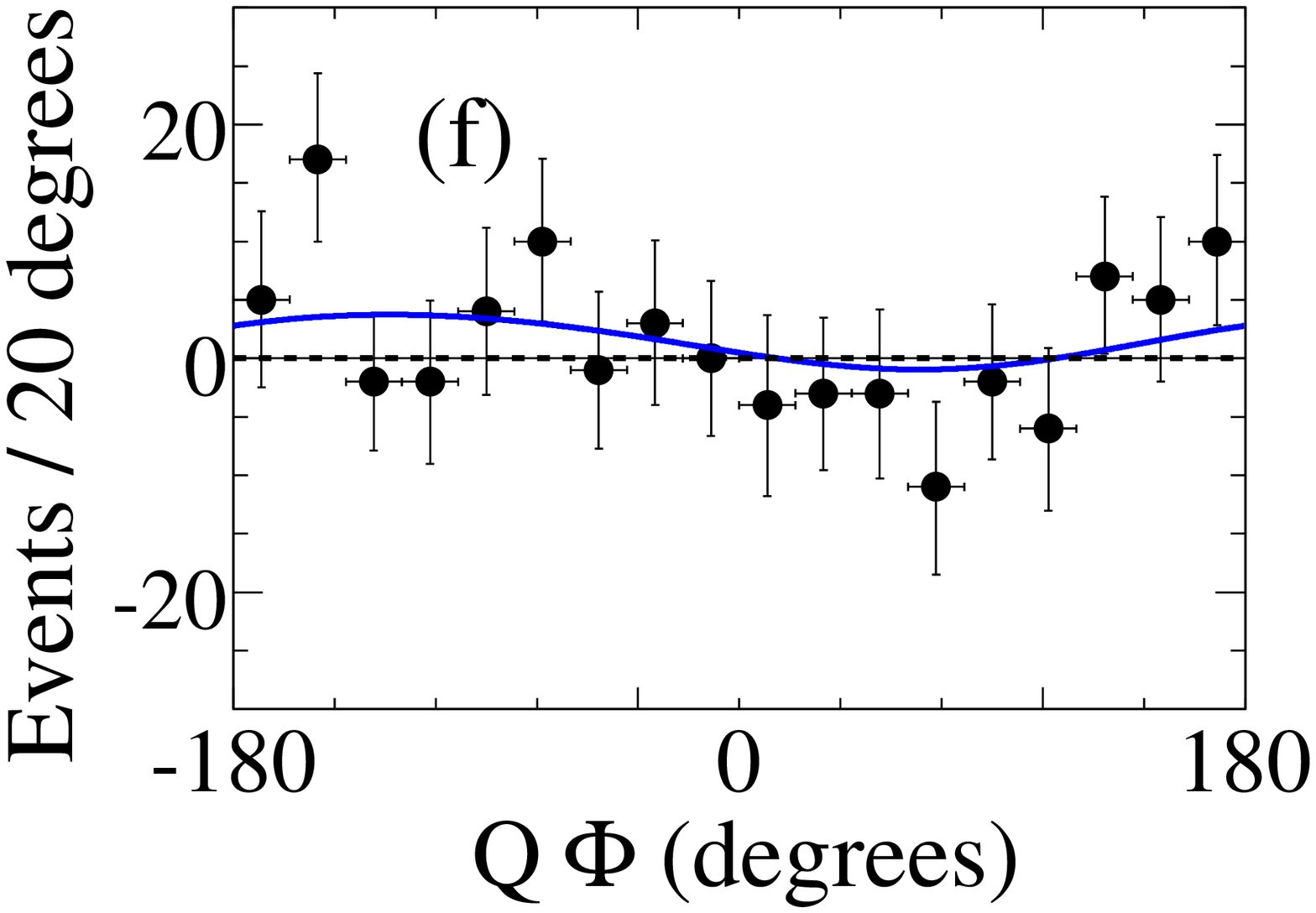}
}
\caption{\label{fig:lowmass-projection}
Projections onto the variable ${\cal H}_1$ for the lower $m_{K\!\pi}$ range in (a) and ${\cal H}_2$ in (b).
Projections onto $Q\times\Phi$ for the lower $m_{K\!\pi}$ range in (c)-(f) where
$Q$ changes sign between the $B$ decays and the $\Bbar$ decays.
The distributions are shown
for the signal $B^0\to\varphi K^{*}(892)^0$ candidates 
following the solid (dashed) line definitions in Fig.~\ref{fig:kine-projection}.
The $\Phi$ angle projections are shown for different combinations of event yields
with certain requirements on the $B$ flavor and ${\cal H}_1\times {\cal H}_2$ product signs,
as discussed in the text. The $D^\pm_{(s)}$-meson veto causes the 
sharp acceptance dips near ${\cal H}_1 = 0.8$ seen in (a).
}
\end{figure}

\begin{figure}[htbp]
\centerline{
\setlength{\epsfxsize}{0.50\linewidth}\leavevmode\epsfbox{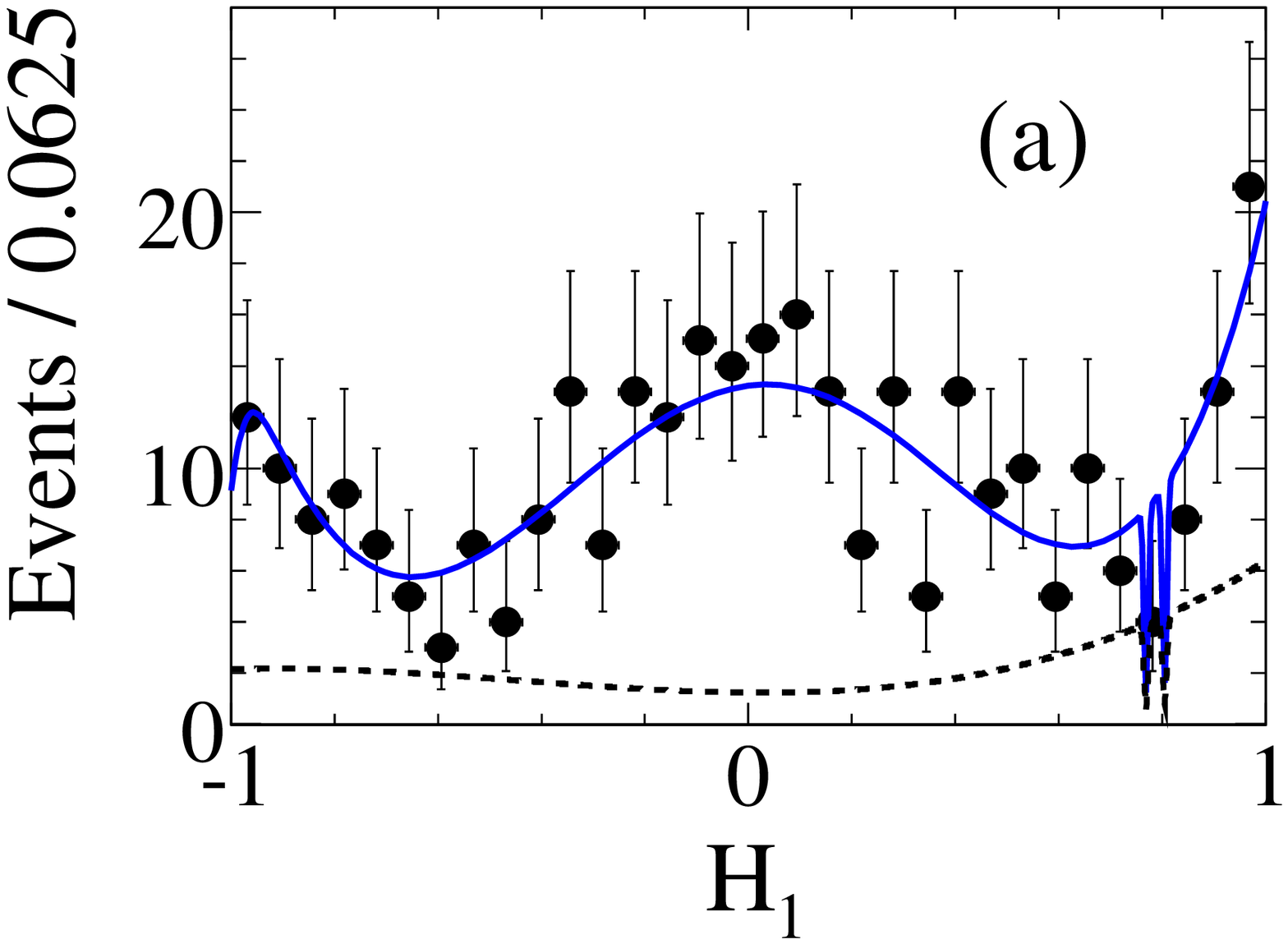}
\setlength{\epsfxsize}{0.50\linewidth}\leavevmode\epsfbox{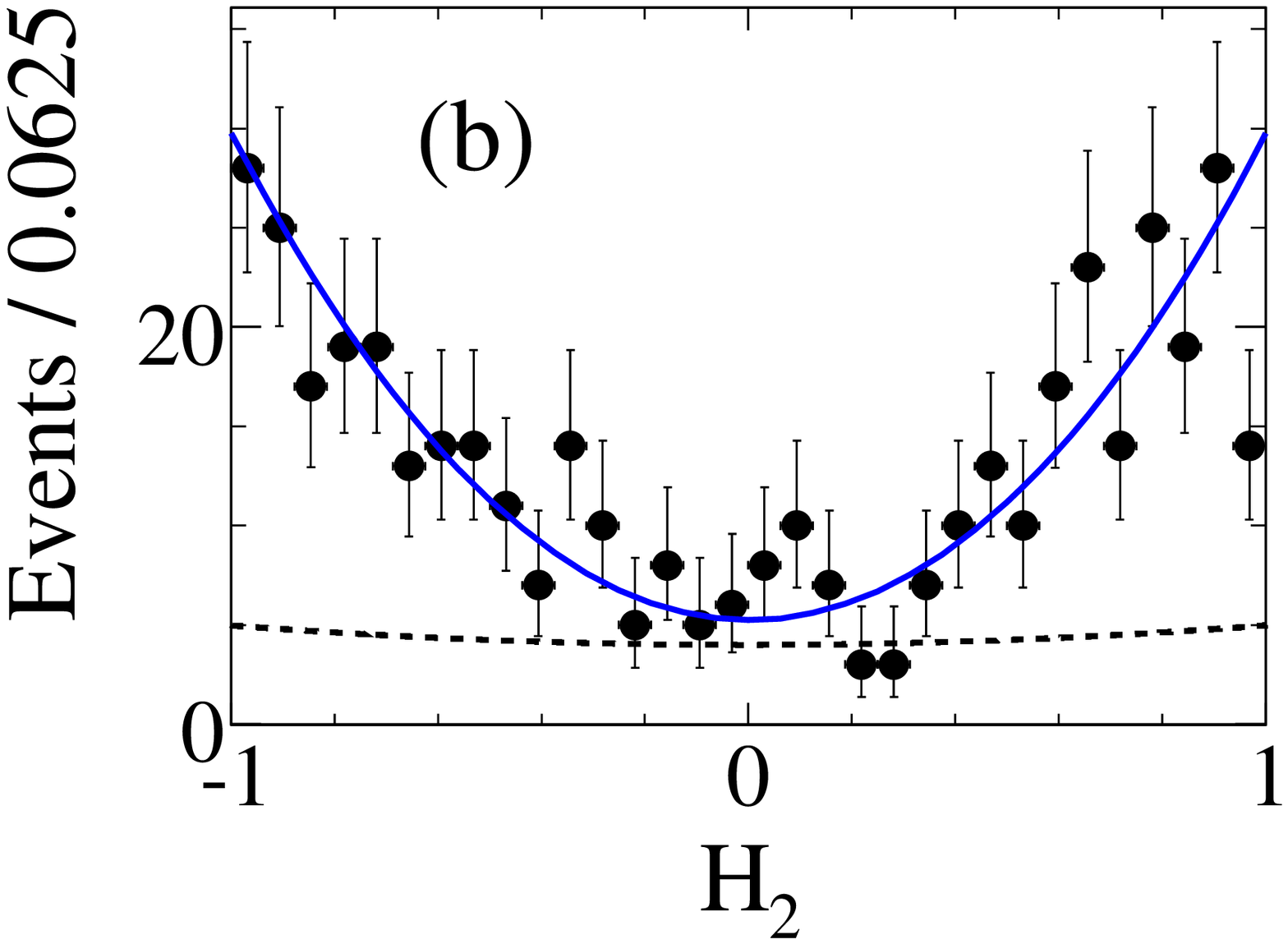}
}
\caption{\label{fig:highmass-projection}
Same as Fig.~\ref{fig:lowmass-projection} (a,b), but
for the signal $B^0\to\varphi K_2^{*}(1430)^0$ and
$\varphi (K\pi)_0^{*0}$ candidates combined. The $D^{\pm}_{(s)}$-meson veto 
causes the sharp acceptance dips near ${\cal H}_1 = 0.8$ seen in (a).
}
\end{figure}

\begin{figure}[t]
\centerline{
\setlength{\epsfxsize}{0.50\linewidth}\leavevmode\epsfbox{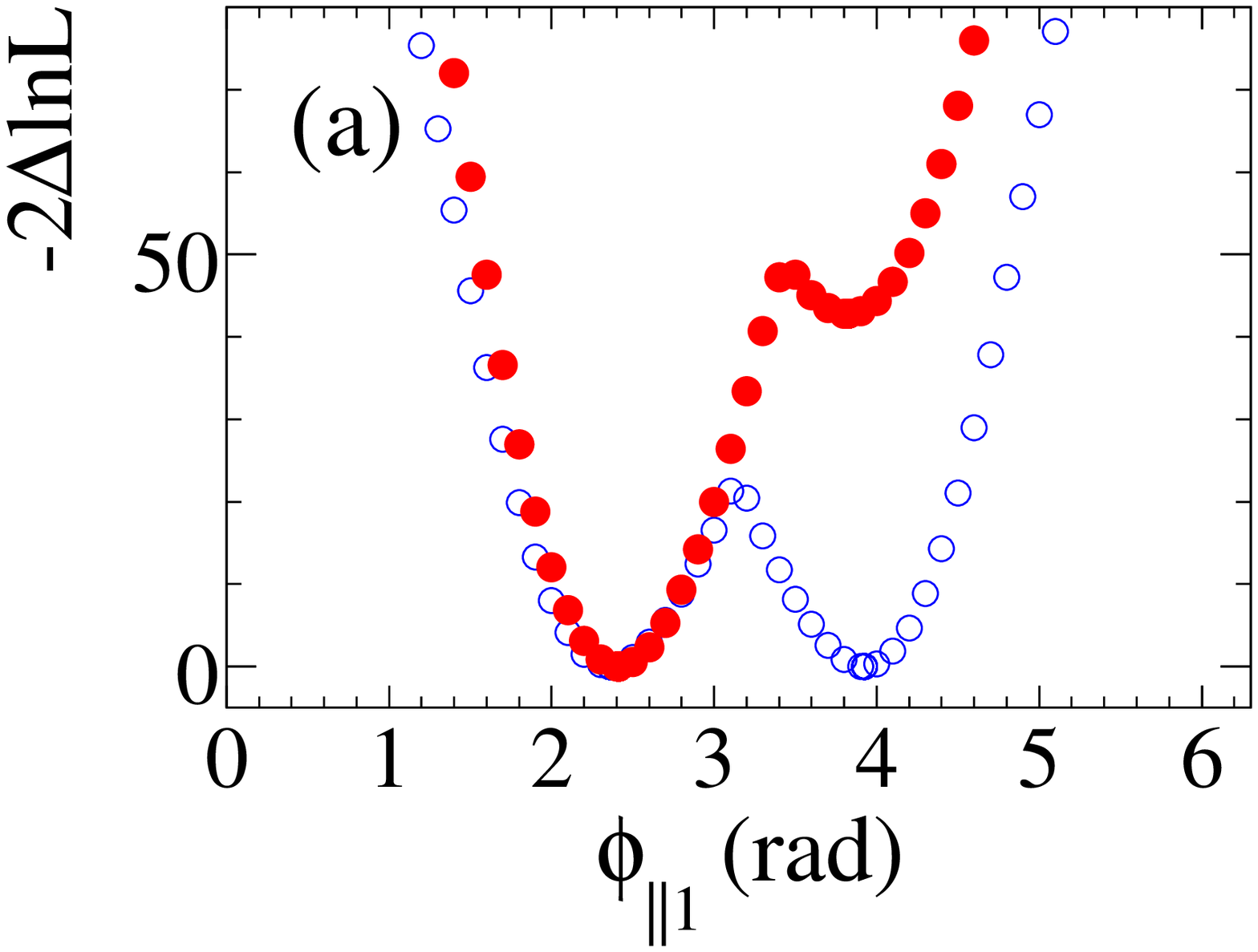}
\setlength{\epsfxsize}{0.50\linewidth}\leavevmode\epsfbox{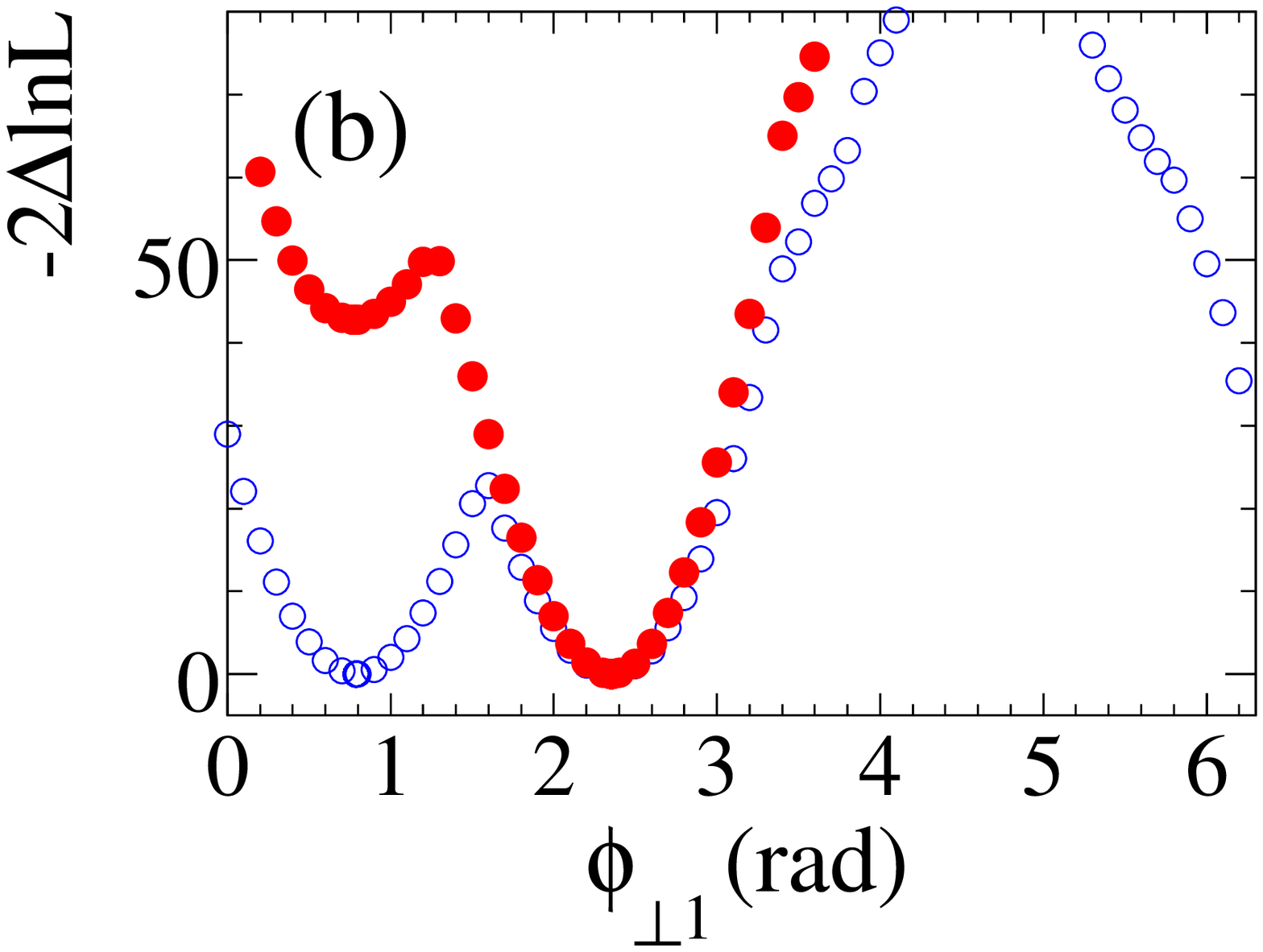}
}
\caption{\label{fig:scan-phild} Scan of $\chi^2=-2\ln({\cal L/L_{\rm max}})$ as a function
of $\phi_{\parallel1}$ and $\phi_{\perp1}$ for the $\varphi K^\pm\pi^\mp$ decays 
in the lower $m_{K\!\pi}$ range, 
where the filled circles are the results with the interference term, and the open circles without the interference term.
Two discrete solutions are visible in the case without interference, while this ambiguity
is resolved with the interference term.
The values of $\Delta\phi_{\parallel1}$ and $\Delta\phi_{\perp1}$
have been constrained in the range $(-0.5,0.5)$ in order to reject
ambiguities with larger values of
$\Delta\phi_{\parallel1}$ and $\Delta\phi_{\perp1}$.
}
\end{figure}

\begin{figure*}[t]
\centerline{
\setlength{\epsfxsize}{0.50\linewidth}\leavevmode\epsfbox{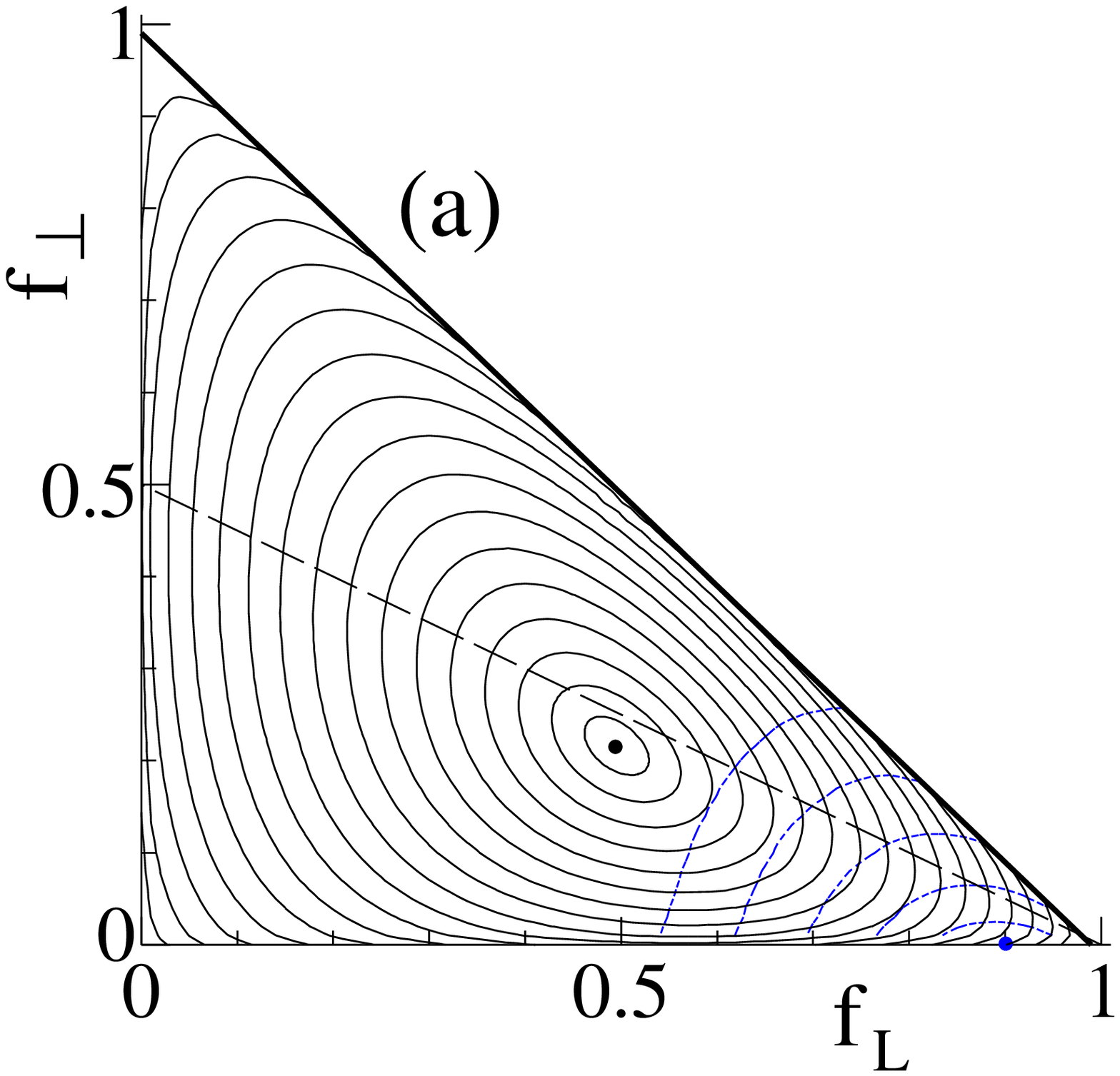}
\setlength{\epsfxsize}{0.45\linewidth}\leavevmode\epsfbox{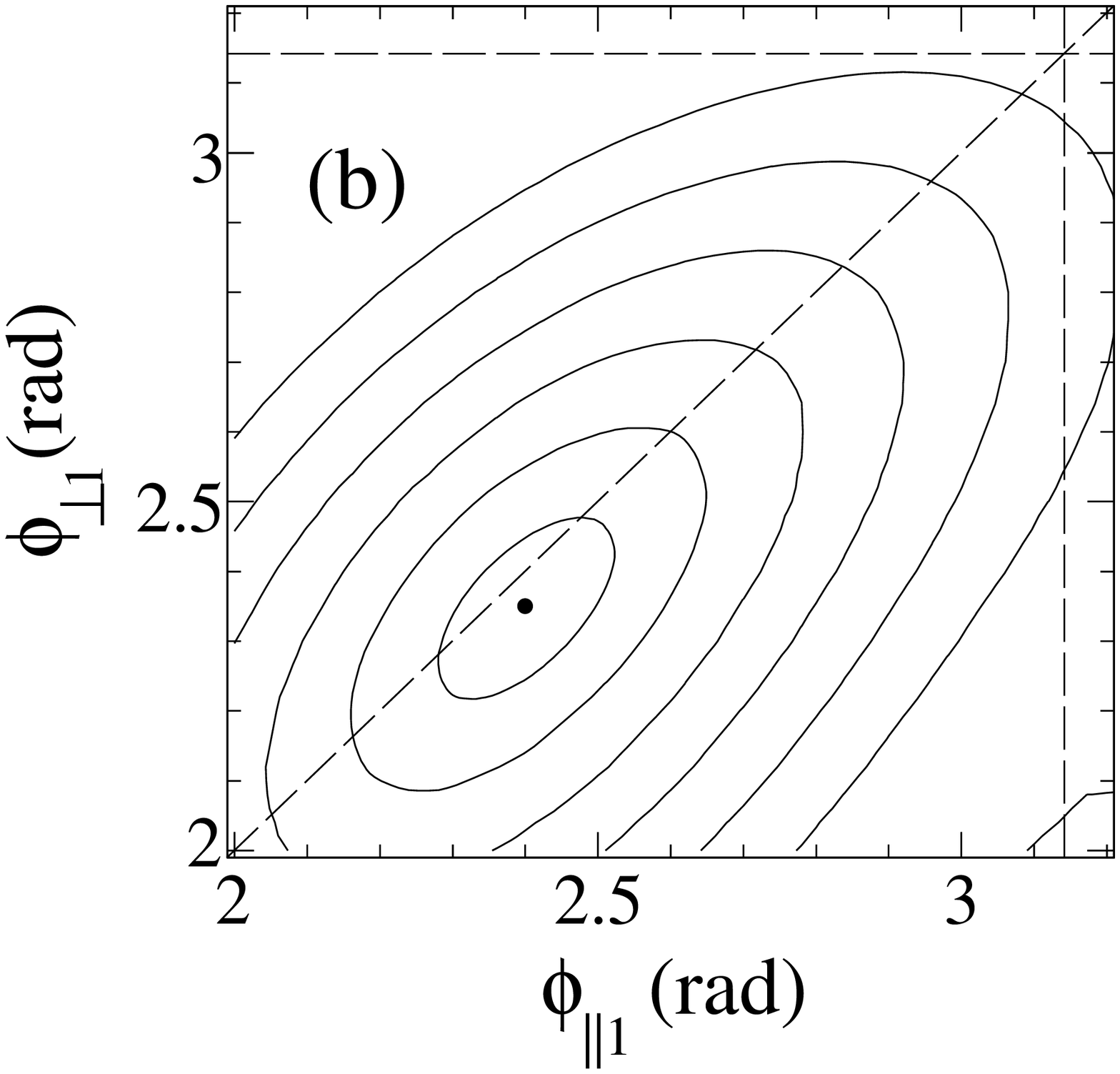}
}
\caption{\label{fig:contourPlot} 
Contours corresponding to the unit intervals of
$\sqrt{-2\Delta\ln{\cal L}}$ 
for polarization $f_{\perp J}$ and $f_{LJ}$ (a) 
and phase $\phi_{\perp 1}$  and $\phi_{\parallel1}$  (b) measurements. 
Diagonal dashed lines $f_{\perp J} = (1-f_{LJ})/2$ and $\phi_{\perp 1} = \phi_{\parallel1}$ 
correspond to $|A_{J+1}| \gg |A_{J-1}|$. In (a), the solid (dashed) contours 
are the results for $J=1$ ($J=2$).
In (b) the $(\pi,\pi)$ point is indicated by the crossed dashed lines.}
\end{figure*}

\begin{figure}[htbp]
\centerline{
\setlength{\epsfxsize}{0.5\linewidth}\leavevmode\epsfbox{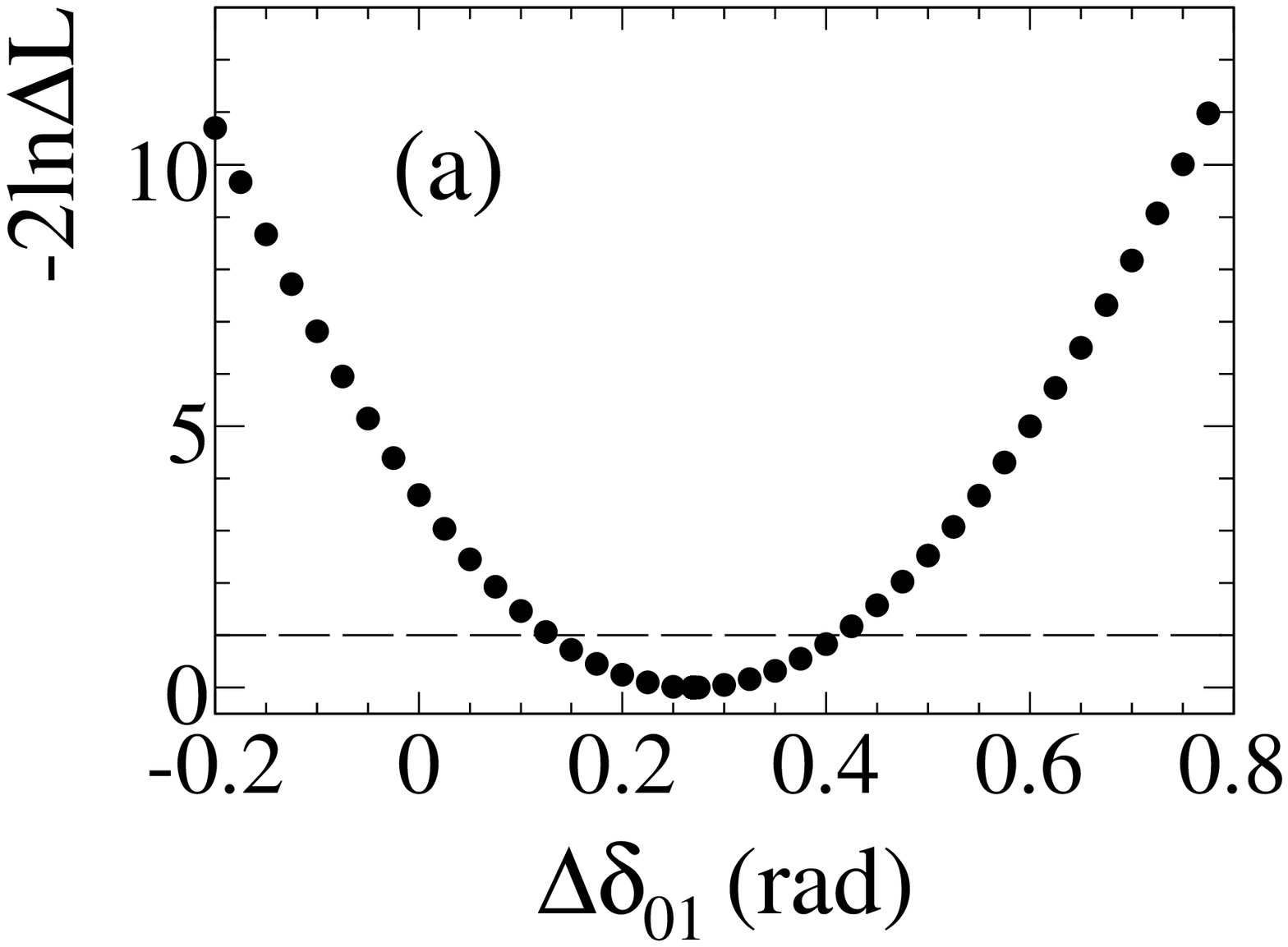}
\setlength{\epsfxsize}{0.5\linewidth}\leavevmode\epsfbox{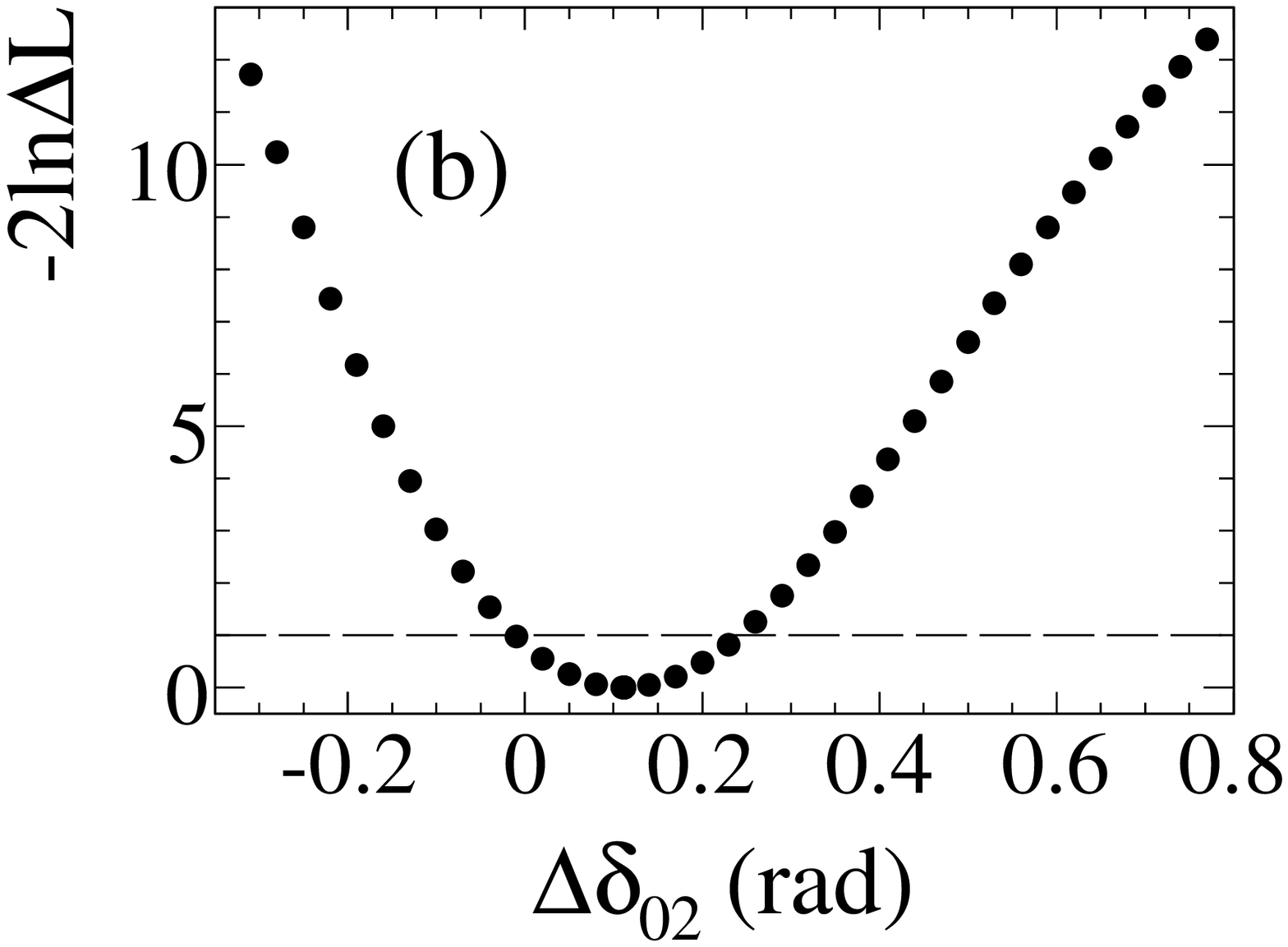}
}
\centerline{
\setlength{\epsfxsize}{0.5\linewidth}\leavevmode\epsfbox{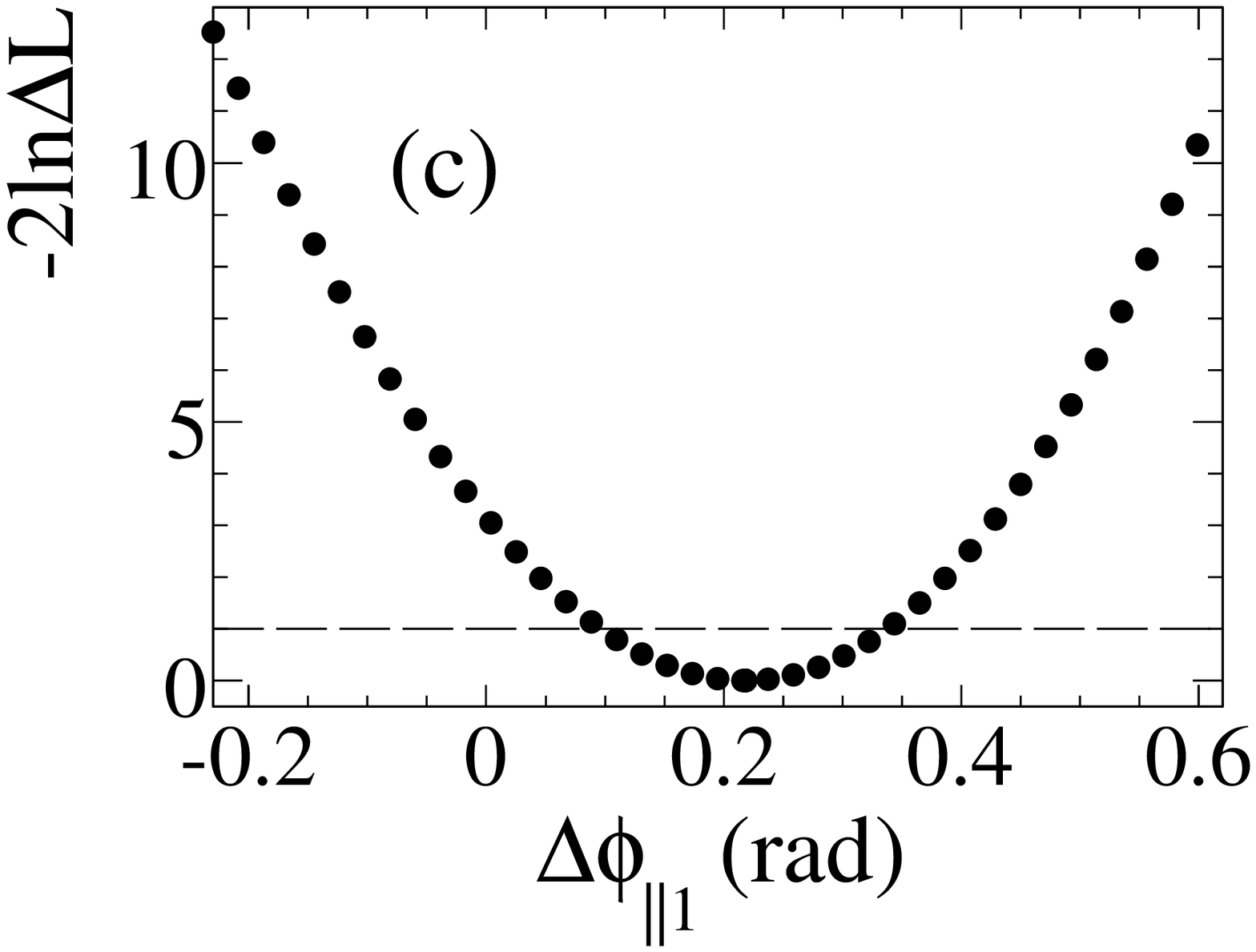}
\setlength{\epsfxsize}{0.5\linewidth}\leavevmode\epsfbox{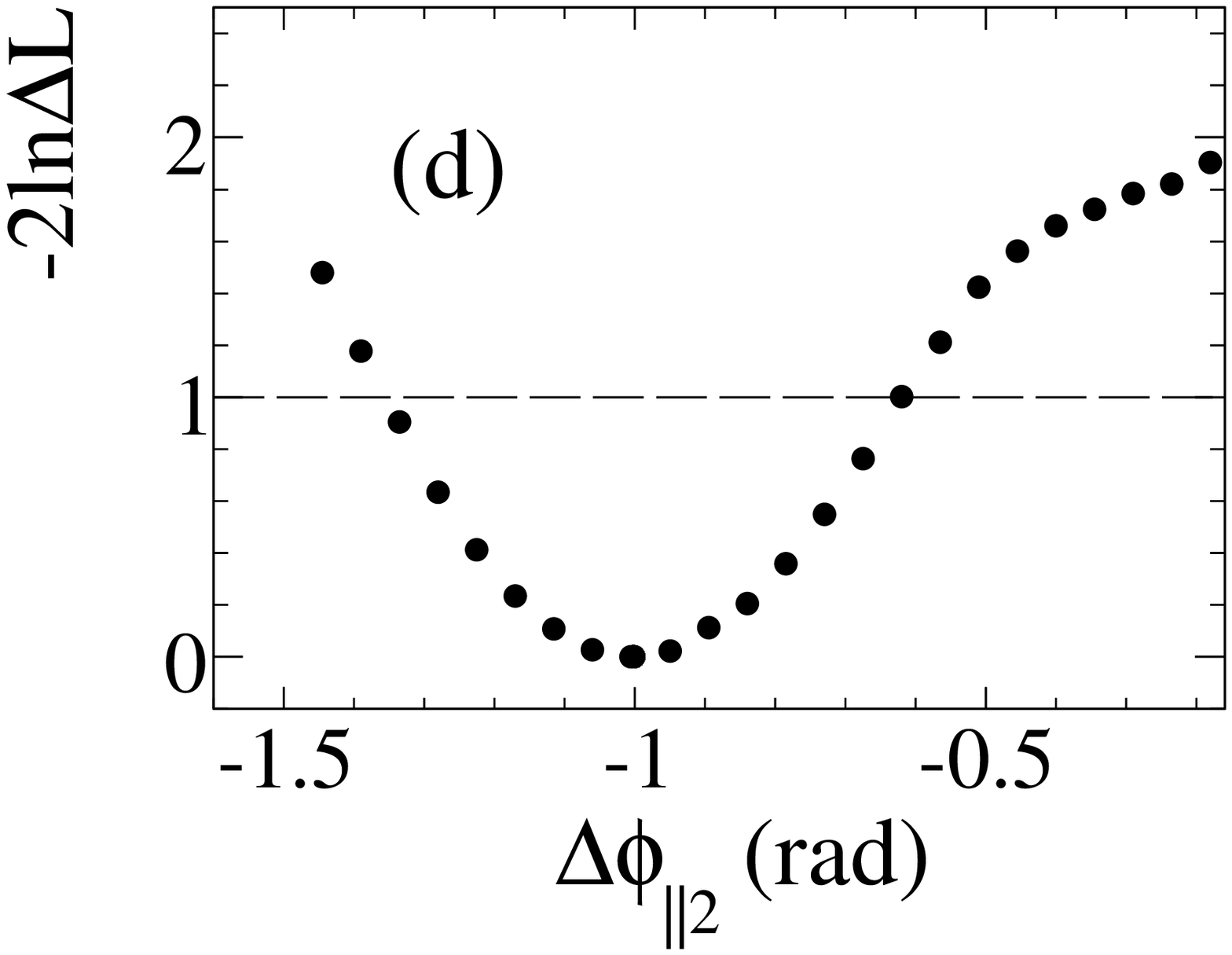}
}
\centerline{
\setlength{\epsfxsize}{0.5\linewidth}\leavevmode\epsfbox{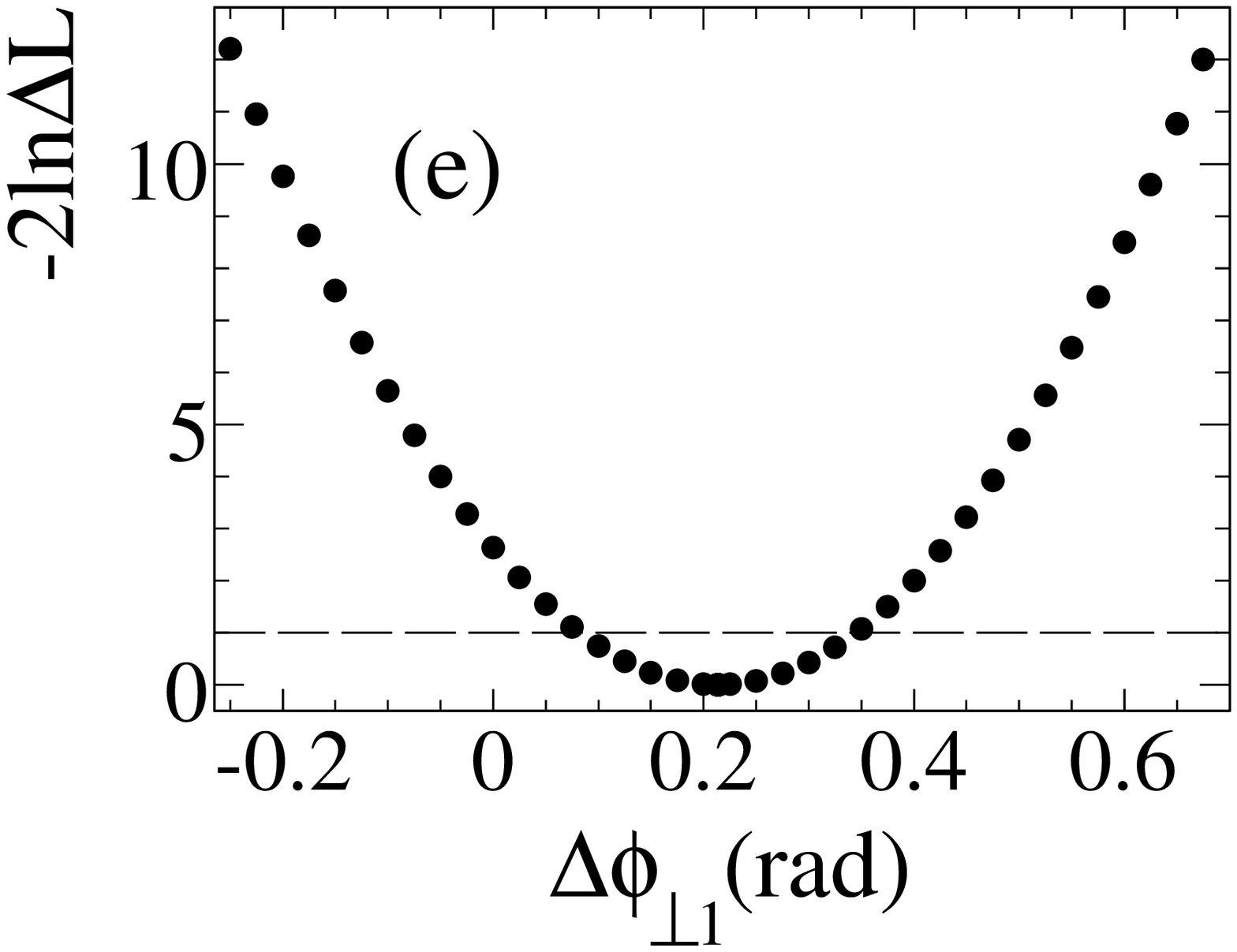}
\setlength{\epsfxsize}{0.5\linewidth}\leavevmode\epsfbox{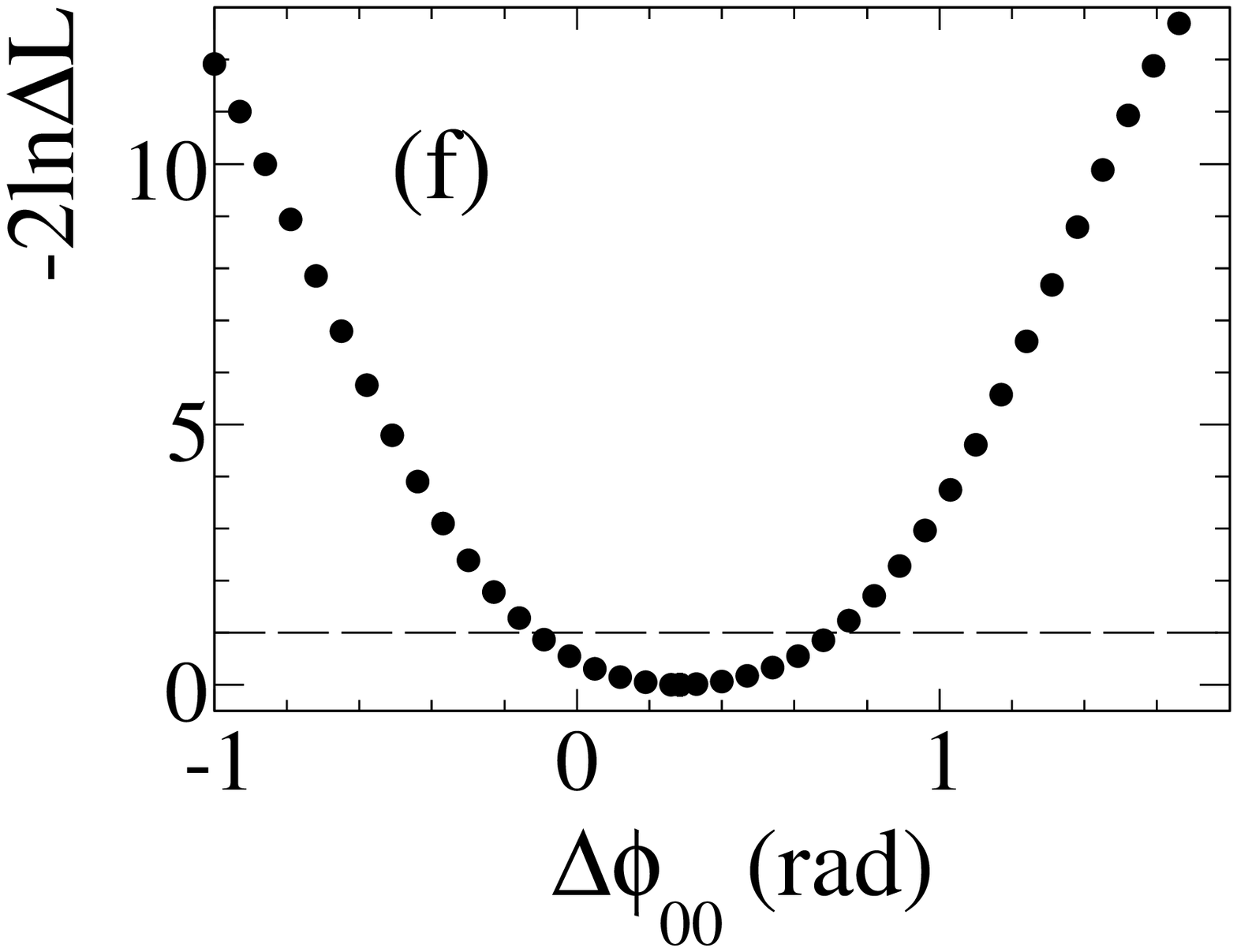}
}
\caption{\label{fig:scanPlot} 
Scan of $\chi^2=-2{\rm ln}({\cal L}/{\cal L}_{\rm max})$ as a function of (a) 
$\Delta\delta_{01}$, (b) $\Delta\delta_{02}$, (c) $\Delta\phi_{\parallel 1}$, (d) $\Delta\phi_{\parallel 2}$, (e) $\Delta\phi_{\perp 1}$, and (f) $\Delta\phi_{00}$.}
\end{figure}

\begin{figure}[h]
\centerline{
\setlength{\epsfxsize}{0.5\linewidth}\leavevmode\epsfbox{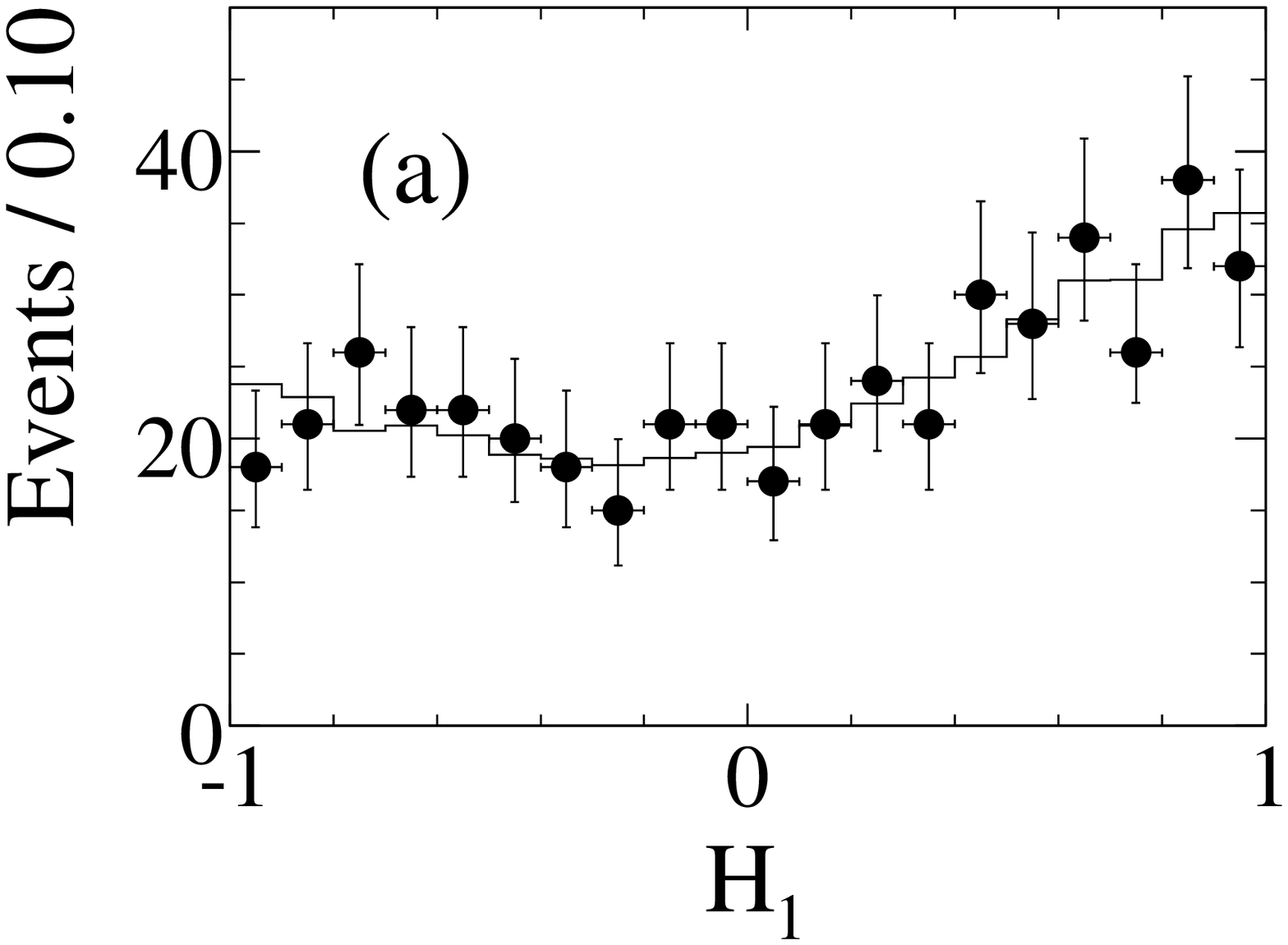}
\setlength{\epsfxsize}{0.5\linewidth}\leavevmode\epsfbox{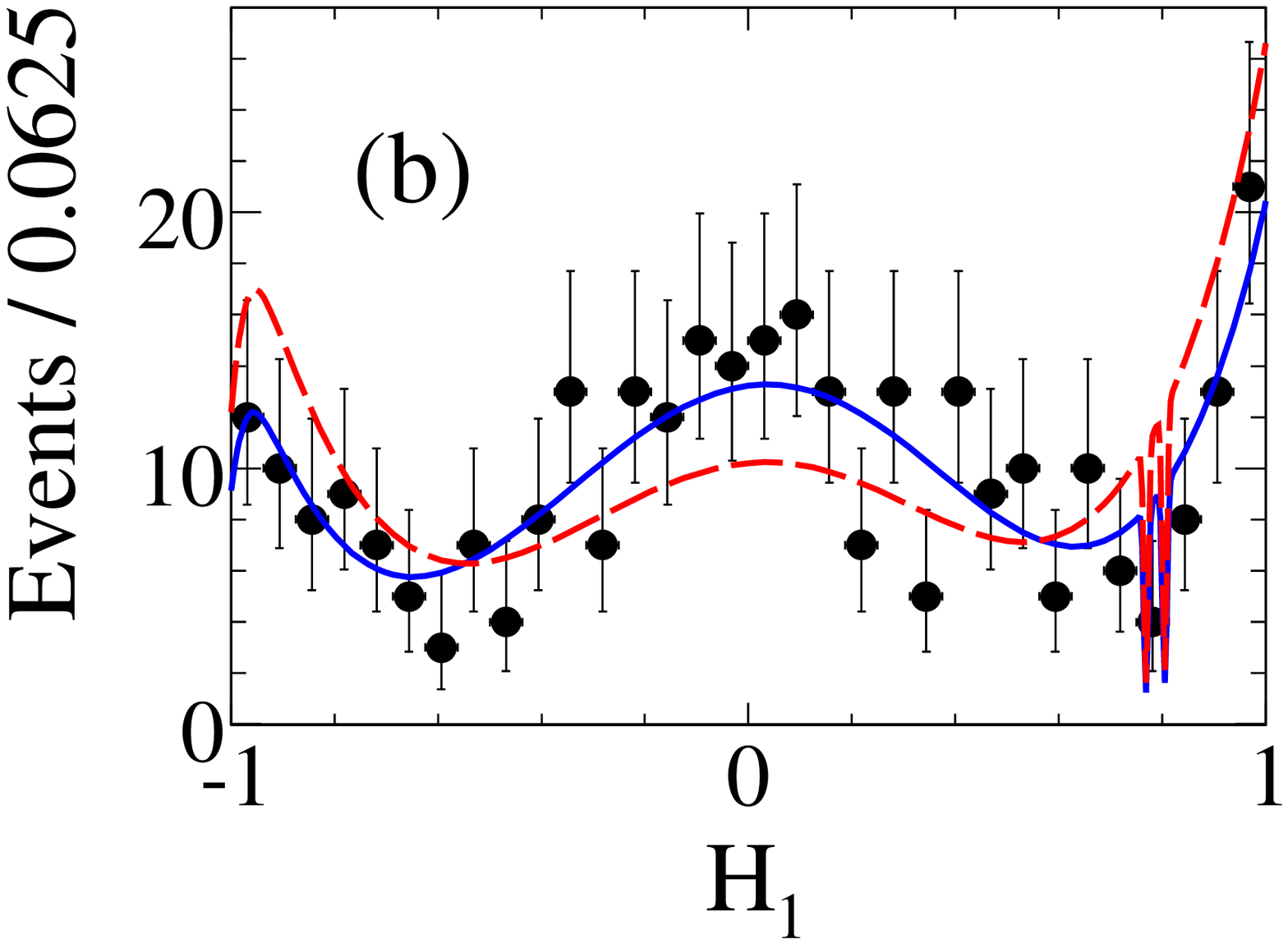}
}
\caption{\label{fig:asym}
Projections onto the variable $\pm {\cal H}_{1}$.
(a) low $m_{K\pi}$ mass range, where we project the data onto 
${\cal H}_{1}$ for $m_{K\pi} > $ 0.896 GeV and
onto $-{\cal H}_{1}$ for $m_{K\pi} < $ 0.896 GeV.
The points with error bars show data and the histogram corresponds to the
results of the MC generated with the observed polarization
parameters. 
(b) high $m_{K\pi}$ mass range, where we project the data onto ${\cal H}_{1}$.
Points with error bars represent the data, while the solid line 
represents the PDF projection with the interference term and the dashed 
line without the interference term included.
The $D^{\pm}_{(s)}$-meson veto causes the sharp acceptance dips near ${\cal H}_1 = 0.8$.
}
\end{figure}

\begin{figure}[htbp]
\centerline{
\setlength{\epsfxsize}{0.85\linewidth}\leavevmode\epsfbox{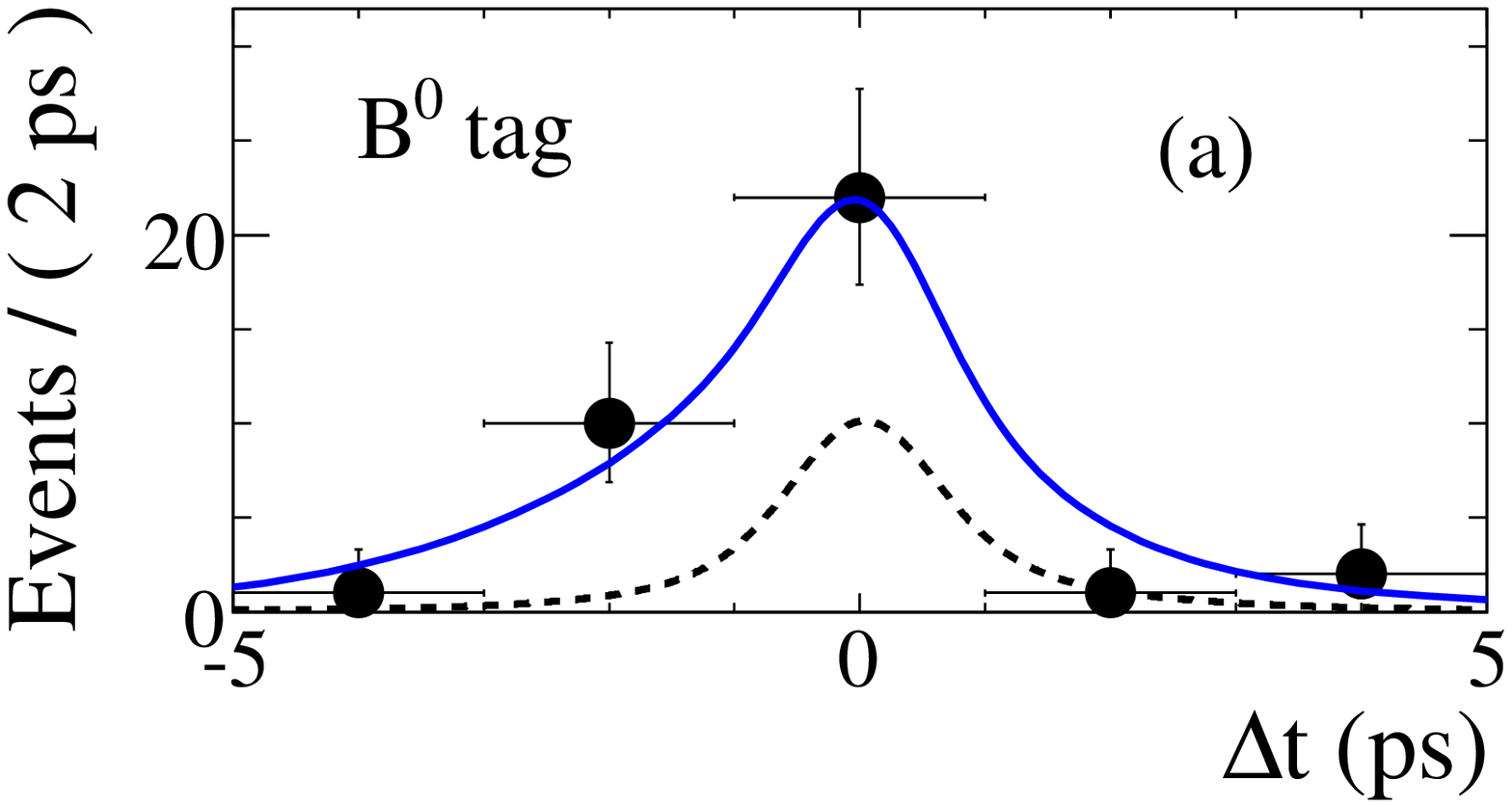}
}
\vspace*{-6pt}
\centerline{
\setlength{\epsfxsize}{0.85\linewidth}\leavevmode\epsfbox{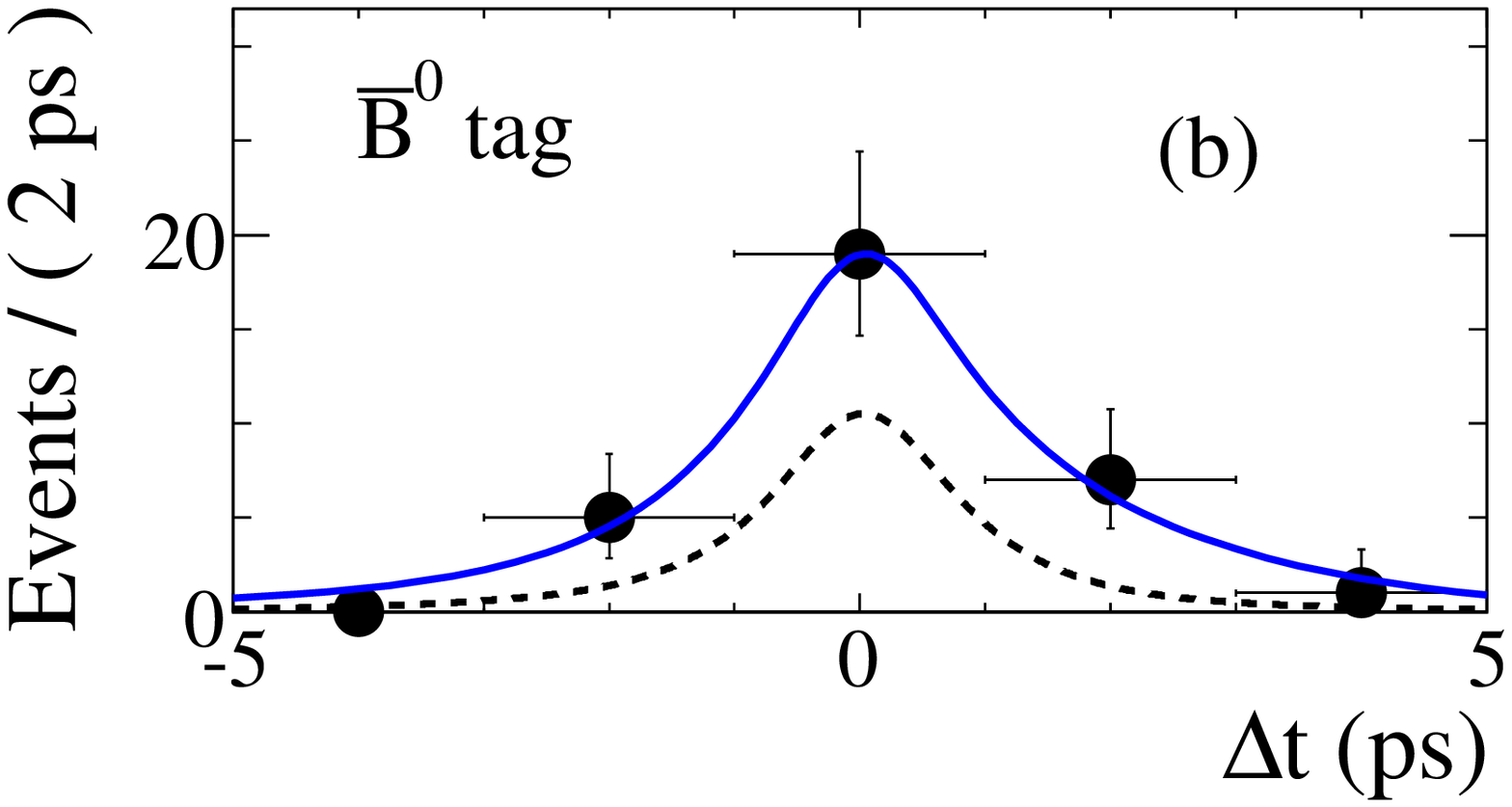}
}
\vspace*{-6pt}
\centerline{
\setlength{\epsfxsize}{0.85\linewidth}\leavevmode\epsfbox{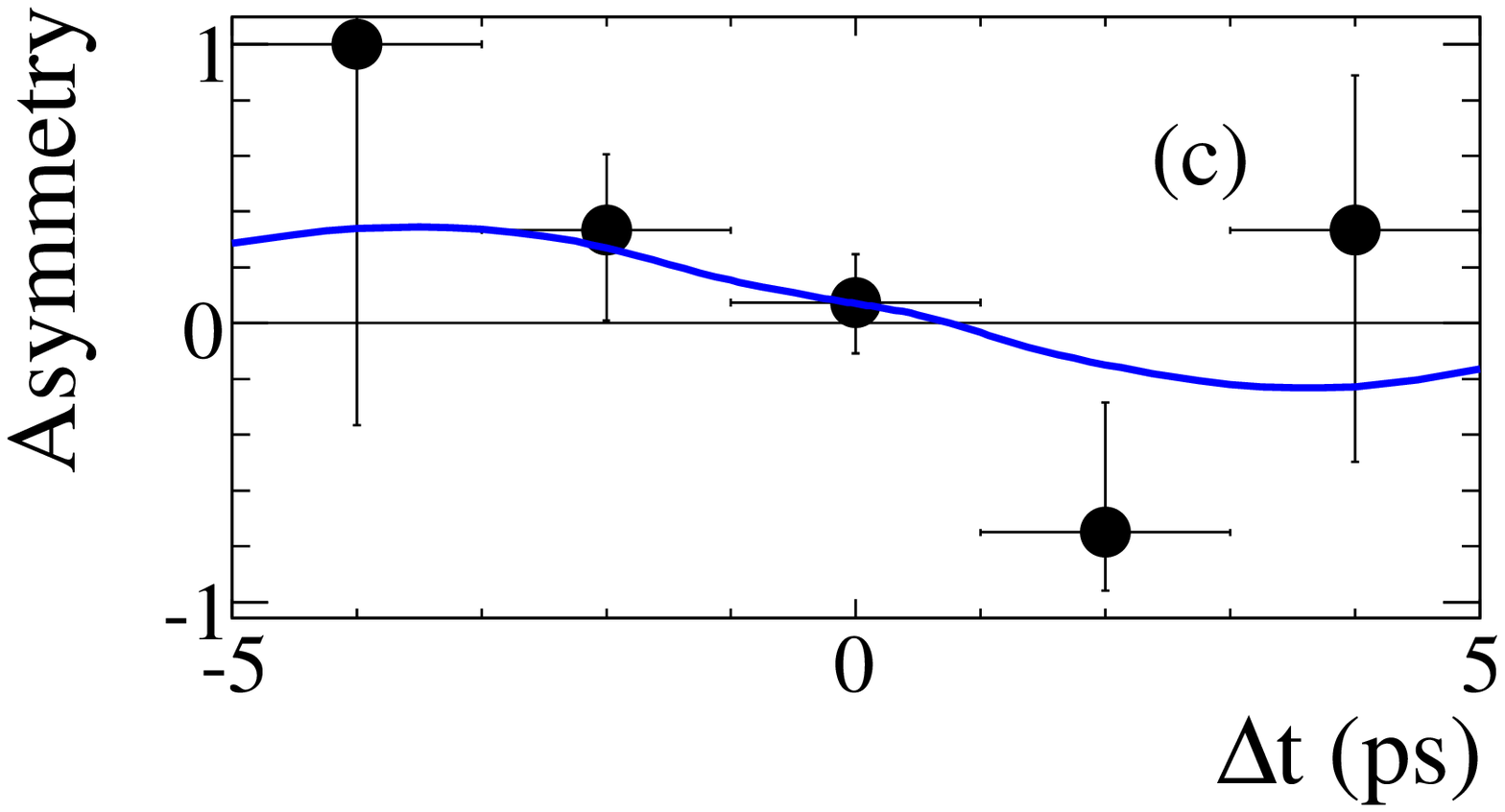}
}
\vspace*{-10pt}
\caption{\label{fig:dT-projection}
The distribution of $\Delta t$ for events in the signal region, for $B^0_{\rm tag}$ (a) and 
$\Bbar^0_{\rm tag}$ (b) events with the fit result overlaid. 
The solid (dashed) lines show the signal-plus-background
(background) PDF projections.
The asymmetry ${\cal A}(\Delta t)$ shown in (c) is defined in Eq.~(\ref{eq:adeltat}). 
}
\end{figure}

We observe a non-zero yield with more than 10\,$\sigma$ significance,
including systematic uncertainties, in each of the three
$B^0\to\varphi K^{*0}$ decay modes.
In Figs.~\ref{fig:kine-projection}--\ref{fig:highmass-projection}
we show projections onto the discriminating variables.
For illustration, the signal fraction is enhanced with a requirement on the
signal-to-background probability ratio
to be greater than a value within the range (0.85 -- 0.95), 
calculated with the plotted variable
excluded. This requirement is at least 50\% efficient for the signal events.

In Tables~\ref{tab:results1} and~\ref{tab:results2}
the $n_{{\rm sig}J}$, ${\cal A}_{C\!P J}$, $\Delta\phi_{00}$, and
{$\vec{\zeta}$}~$\equiv\{f_{LJ}$, $f_{\perp J}$, $\phi_{\parallel J}$,
$\phi_{\perp J}$, $\delta_{0J}$, ${\cal A}_{C\!PJ}^0$, ${\cal A}_{C\!PJ}^{\perp}$,
$\Delta \phi_{\parallel J}$, $\Delta \phi_{\perp J}$, $\Delta\delta_{0J}$\}
parameters of the $B^0\to\varphi K^{*}(892)^0$,
$\varphi K_2^{*}(1430)^0$, and $\varphi(K\pi)_0^{*0}$ decays are shown.
The three quantities 
$\phi_{\perp 2}$, ${\cal A}_{C\!P2}^{\perp}$, and $\Delta \phi_{\perp 2}$, 
which characterize parity-odd transverse 
amplitude in the vector-tensor decay, are not measured
because $f_{\perp 2}$ is found to be consistent with zero.

The computed significance of the yield is 
more than 24\,$\sigma$ for $B^0\to\varphi K^{*}(892)^0$ and 
11\,$\sigma$ for $B^0\to\varphi (K\pi)_0^{*0}$.
Given the convincing presence of the $S$-wave $(K\pi)_0^{*0}$ contribution, 
we rely on the interference terms to resolve the phase ambiguities.
In the lower $m_{K\!\pi}$ range the yield of the
$\varphi({K\pi})^{*0}_0$ contribution is $75^{+20}_{-17}$ events
with a statistical significance of 9\,$\sigma$,
including the interference term.
From the measurements of the higher $m_{K\!\pi}$ range we calculate the 
contribution of $\varphi({K\pi})^{*0}_0$ to the lower mass range to be about 
61 events. This is consistent with the above result within 1\,$\sigma$.
The dependence of the interference on the $K\pi$ invariant
mass~\cite{Aston:1987ir, jpsikpi} allows us to reject the other
solution near ($2\pi-\phi_{\parallel1},\pi-\phi_{\perp1}$)
relative to that in Table~\ref{tab:results2}
for the $B^0\to\varphi K^{*}(892)^0$ decay
with a statistical significance of 6.5\,$\sigma$ 
(which becomes 5.4\,$\sigma$ when systematic uncertainties are included).
We also resolve this ambiguity with a statistical significance
of more than 4\,$\sigma$ with the $\Bbar^0$ or $B^0$ decays independently.
Fig.~\ref{fig:scan-phild} shows the $\chi^2=-2\ln({\cal L/L_{\rm max}})$
scan plots for $\phi_{\parallel}$ and $\phi_{\perp}$,
where we illustrate how the phase ambiguity is resolved.
For comparison, we show the result of the fit where interference is not
taken into account and no sensitivity to resolve the ambiguity 
is present.
The significance of the deviations of $\phi_{\parallel1}$ and 
$\phi_{\perp1}$ from $\pi$ is 5.4\,$\sigma$ 
(4.5\,$\sigma$) and 6.1\,$\sigma$ (5.0\,$\sigma$), respectively 
(including systematics in parentheses). 

Projections onto ${\cal H}_1$ and ${\cal H}_2$ 
in the lower $m_{K\!\pi}$ range in Figs.~\ref{fig:lowmass-projection} 
(a) and (b) show sizable contributions of both $\cos^2\theta$ (longitudinal)
and $\sin^2\theta$ (transverse) components.
These two plots emphasize 
${f}_{11}\,({\cal H}_1,\,{\cal H}_2,\,\Phi)$
and 
${f}_{21}\,({\cal H}_1,\,{\cal H}_2,\,\Phi)$
angular terms in Eq.~(\ref{eq:fangular}).
Similarly, projections onto ${\cal H}_1$ and ${\cal H}_2$ 
in the higher $m_{K\!\pi}$ range in Fig.~\ref{fig:highmass-projection}
show predominant longitudinal polarization 
with sizable ${f}_{12}\,({\cal H}_1,\,{\cal H}_2,\,\Phi)$
and ${f}_{10}\,({\cal H}_1,\,{\cal H}_2,\,\Phi)$ contributions, while the 
${f}_{22}\,({\cal H}_1,\,{\cal H}_2,\,\Phi)$ contribution is small.

In order to illustrate 
${f}_{31}\,({\cal H}_1,\,{\cal H}_2,\,\Phi)$
and 
${f}_{41}\,({\cal H}_1,\,{\cal H}_2,\,\Phi)$
angular distributions,
we project onto the angle $-\Phi$ for the $B$ decays and $\Phi$ for $\Bbar$ decays.
This procedure takes into account the change of sign
for the odd components with $P$-wave amplitude ($A_{1\perp}$).
This $\Phi$ angle projection is sensitive only to the
constant term and to $\cos(2\Phi)$ and $\sin(2\Phi)$ terms. 
The fact that the double
sine and cosine contributions are small
in Fig.~\ref{fig:lowmass-projection} (c)
tells us that both $(|A_{1\parallel}|^2-|A_{1\perp}|^2)$ and
$\Imm(A_{1\perp} A^*_{1\parallel})$ are relatively small, 
in agreement with the fit results. That is, 
the values of
$(1-f_{L1}-2f_{\perp1})$ and $(\phi_{\parallel1}-\phi_{\perp1})$ are small; 
see $\alpha^-_{3J}$ and $\alpha^-_{4J}$ 
in Eqs.~(\ref{eq:alpha3}) and (\ref{eq:alpha4}).

In order to emphasize 
${f}_{51}\,({\cal H}_1,\,{\cal H}_2,\,\Phi)$
and 
${f}_{61}\,({\cal H}_1,\,{\cal H}_2,\,\Phi)$
angular terms in Eq.~(\ref{eq:fangular}), or
$\cos$$\Phi$ and $\sin$$\Phi$ distributions,
we show the difference between
the above $\Phi$ angle projections for events with
${\cal H}_1\times{\cal H}_2>0$ and with
${\cal H}_1\times{\cal H}_2<0$.
This gives us contributions to 
$\alpha^-_{5J}$ and $\alpha^-_{6J}$ in Eqs.~(\ref{eq:alpha5}) and (\ref{eq:alpha6}),
while background and all other signal contributions cancel.
This projection is shown in Fig.~\ref{fig:lowmass-projection} (d), where we 
see good agreement between the data
and the fit results. This plot indicates a sizable $\cos\Phi$ component 
due to $\Ree(A_{1\parallel} A^*_{10})$ and an 
asymmetric $\sin\Phi$ component due to
$\Imm(A_{1\perp} A^*_{10})$. The latter asymmetry is visible and
reflects the presence of a strong phase with a statistical significance
of 6.4\,$\sigma$.

In order to emphasize $C\!P$ asymmetries in the angular
distributions we make two other similar projections onto the
$\Phi$ angle, but now we plot the difference between the
$B$ and $\Bbar$ decays. Figs.~\ref{fig:lowmass-projection} (e) and (f) show distributions of
(N$_{B>}$$-$N$_{\Bbar>}$+N$_{B<}$$-$N$_{\Bbar<}$)
and
(N$_{B>}$$-$N$_{\Bbar>}$$-$N$_{B<}$+N$_{\Bbar<}$),
where N$_{B>}$ denotes the number of $B$ decays with
${\cal H}_1{\cal H}_2>0$,
N$_{\Bbar>}$ is the number of $\Bbar$ decays with
${\cal H}_1{\cal H}_2>0$,
N$_{B<}$ is the number of $B$ decays with
${\cal H}_1{\cal H}_2<0$, and
N$_{\Bbar<}$ is the number of $\Bbar$ decays with
${\cal H}_1{\cal H}_2<0$.
In all cases we project onto $-\Phi$ for the $B$ decays
and onto $\Phi$ for the $\Bbar$ decays.
Fig.~\ref{fig:lowmass-projection} (e) is sensitive to $C\!P$ asymmetries in
the $\cos(2\Phi)$ and $\sin(2\Phi)$ terms
in Eqs.~(\ref{eq:alpha3}) and (\ref{eq:alpha4}), 
while Fig.~\ref{fig:lowmass-projection} (f) is sensitive to $C\!P$ asymmetries in
the $\cos\Phi$ and $\sin\Phi$ terms
in Eqs.~(\ref{eq:alpha5}) and (\ref{eq:alpha6}).
In particular, there is a hint of a sine wave contribution in 
Fig.~\ref{fig:lowmass-projection} (f), corresponding to
the non-zero measurement of $\Delta\phi_{\perp1}=+0.21\pm0.13\pm0.08$,
though not significant enough to constitute evidence for $C\!P$ violation.

In Table~\ref{tab:results2} we summarize the correlation among the primary 
fit parameters. There are two large correlation effects of approximately 50\% evident 
for the $B^0\to\varphi K^*(892)^0$ decay in the fit, that is between the 
longitudinal and transverse fractions ($f_{L1}$ and $f_{\perp 1}$) and between 
the two phases ($\phi_{\parallel1}$ and $\phi_{\perp1}$). 
In Fig.~\ref{fig:contourPlot} we show the likelihood function contour plots for 
the above pairs of correlated observables as well as for $f_{L2}$ and $f_{\perp 2}$. 
Fig.~\ref{fig:scanPlot} shows the $\chi^2=-2\ln({\cal L/L_{\rm max}})$
distributions for the $C\!P$ violation phase parameters $\Delta\delta_{xJ}$ and 
$\Delta\phi_{xJ}$, where $x$ stands for either $\perp$, $\parallel$, or 0.

The $B^0\to f_0 K^{*0}$ category accounts for final states
with $K^+K^-$ from either $f_0$, $a_0$, or any other broad $K^+K^-$
contribution under the $\varphi$. Its yield is consistent with zero 
in the higher $m_{K\!\pi}$ range and is $84\pm19$ events in the 
lower $m_{K\!\pi}$ range.
Due to uncertainties in the nature of this contribution, we do not
calculate its branching fraction, but include it in the evaluation
of systematic uncertainties in other parameters as discussed above.

In Fig.~\ref{fig:asym} we illustrate the effect of interference
in the $K\pi$ invariant mass.
In the vector-scalar $K\pi$ interference (lower mass range)
the interference term is linear in ${\cal H}_{1}$,  
creating a forward-backward asymmetry.
However, due to variation of the B-W phase this
effect cancels when integrated over the $m_{K\!\pi}$ range.
For $\delta_0\simeq\pi$ we expect
the coefficient in front of ${\cal H}_{1}$ to be positive above
$m_{K\!\pi}\simeq 0.896$ GeV and negative below this value.
Thus, we create Fig.~\ref{fig:asym} (a) to emphasize this effect.
If $\delta_{01}\simeq 0$, the sign of the forward-backward asymmetry
would be reversed and we would see more events on the left
as opposed to the right. 
Thus $\delta_{01}\simeq \pi$ is preferred.
We note that this plot has only
partial information about the interference while the
multi-dimensional fit uses the full information to extract
the result.

Figure~\ref{fig:asym} (b) shows a  similar effect in the 
tensor-scalar $K\pi$ interference (higher mass range).
In this case interference could either enhance events 
in the middle of the ${\cal H}_{1}$ distribution and 
deplete them at the edges, or the other way around.
Fig.~\ref{fig:asym} (b) indeed shows significant 
improvement in the ${\cal H}_{1}$ parameterization
with the inclusion of interference. 
It also corresponds to the observed value $\delta_{02}\simeq \pi$.

Because of the low significance of our measurements of 
$f_{\parallel2}=(1-f_{L2}-f_{\perp2})$ (1.9\,$\sigma$)
and $f_{\perp2}$ (0\,$\sigma$)
in the $B^0\to\varphi K^{*}_2(1430)^0$ decay
we have insufficient information to constrain $\phi_{\parallel2}$
at higher significance or to measure $\phi_{\perp2}$,
$A^{\perp}_{\CP2}$, or $\Delta\phi_{\perp2}$, which we constrain to zero
in the fit.

Finally, we fit a single parameter $\Delta\phi_{00}$ in the time evolution after 
combining all available
$B^0\to\varphi K^0_S\pi^0$ charmless final states, which are
dominated by spin-0, 1, and 2 $(K\pi)$ combinations.
The distribution of the time difference $\Delta t$ and
the time-dependent asymmetry are shown in Fig.~\ref{fig:dT-projection}.
The parameter $\Delta\phi_{00}$ is measured to be
$0.28\pm0.42\pm0.04$, as shown in Table~\ref{tab:results2}.


\section{CONCLUSION}
\label{sec:note-consclusion}

In conclusion, we have performed an amplitude analysis and searched
for $C\!P$ violation in the angular distribution of 
$B^0\to\varphi K^{*0}$ decays with tensor, vector, and scalar $K^{*0}$ mesons.
Our results are summarized in Tables~\ref{tab:results1}
and~\ref{tab:results2} and supersede corresponding measurements
in Ref.~\cite{babar:phikst}.
In this analysis we employ several novel techniques
for $C\!P$-violation and polarization measurements 
in the study of a single $B$-decay topology $B^0\to\varphi(K\pi)$. 
We use the time-evolution of the $B^0\to\varphi K^0_S\pi^0$ channel to extract
the $C\!P$-violating phase difference $\Delta\phi_{00}=0.28\pm0.42\pm 0.04$
between the $B$ and $\Bbar$ decay amplitudes. 
We use the dependence on the $K\pi$ invariant mass of the interference
between the scalar and vector, or scalar and tensor components to resolve discrete 
ambiguities of both the strong and weak phases.
Twelve parameters are measured
for the vector-vector decay, nine parameters for the vector-tensor decay,
and three parameters for the vector-scalar decay,
including the branching fractions, $C\!P$-violation parameters,
and parameters sensitive to final-state interactions.

The $(V$-$A)$ structure of the weak interaction and the
$s$-quark spin-flip suppression in the process shown in Fig.~\ref{fig:Diagram_phiKst}
suggest $|A_{J0}|\gg|A_{J+1}|\gg|A_{J-1}|$~\cite{bvv1, newtheory}.
The relatively small value of ${f_{L1}}=0.494\pm{0.034}\pm 0.013$
and the relatively large value of ${f_{\perp1}}=0.212\pm{0.032}\pm 0.013$
in the vector-vector decay remain a puzzle. 
The naive expectation is that the $(V$-$A)$ nature of the weak
decays requires that an anti-quark originating from the
$\bar{b}\to\bar{q}W^+$ decay be produced in helicity state
$+{1\over 2}$.
This argument applies to the penguin loop
which is a purely weak transition, $\bar{b}\to\bar{s}$, with
the double-$W$ coupling in the Standard Model.
The $\bar{s}$ anti-quark can couple
to the $s$ quark (see Fig.~\ref{fig:Diagram_phiKst})
to produce the $\varphi$ state with
helicity of either $\lambda=0$ or $\lambda=+1$, but not
$\lambda=-1$. However, the $K^*$ state should have
the same helicity as the $\varphi$ due to angular momentum
conservation. The $\lambda=+1$ state is not allowed in this
case because both $s$ and $\bar{s}$ quarks would have
helicity $+{1\over 2}$, in violation 
of helicity conservation in the vector
coupling, ${\rm g}\to{s}\bar{s}$.

The spin-flip can alter both of the above requirements, but its
suppression factor is of order $\sim{m_V}/m_B$ for each flip,
where ${m_V}$ is the mass of the $\varphi$ or $K^{*0}$ mesons.
Thus, we arrive at the expectation
$|A_{J0}|\gg|A_{J+1}|\gg|A_{J-1}|$, or
$|A_{J0}|\gg|A_{J\perp}|$ and $A_{J\perp}\simeq A_{J\parallel}$,
where $A_{J+1}$ is suppressed by one spin flip, while
$A_{J-1}$ is suppressed by two spin flips.
New physics could have different interactions,
alter the spin-helicity expectations and result
in a large fraction of transverse polarization.
Alternatively, strong interaction effects might change 
this expectation~\cite{newtheory}.
The value of $f_{L2}=0.901^{+0.046}_{-0.058}\pm 0.037$
in vector-tensor decays is not compatible with that 
measured in vector-vector decays, while it is compatible
with the expectation from the spin-flip analysis above.
This points to a unique role for the spin-1 particle
recoiling against the $\varphi$ in
the $B\to\varphi K^*$ polarization puzzle.

In the $B^0\to\varphi K^{*}(892)^0$ decay we obtain the
solution ${\phi_{\parallel1}}\simeq{\phi_{\perp1}}$ without
discrete ambiguities. Combined with the approximate solution
$f_{L1}\simeq1/2$ and $f_{\perp1}\simeq(1-f_{L1})/2$, this results
in the approximate decay amplitude hierarchy
$|A_{10}|\simeq|A_{1+1}|\gg|A_{1-1}|$
(and $|\Abar_{10}|\simeq|\Abar_{1-1}|\gg|\Abar_{1+1}|$).
We find more than 5\,$\sigma$ (4\,$\sigma$) deviation, including
systematic uncertainties, of ${\phi_\perp} ({\phi_\parallel})$
from either $\pi$ or zero in the
$B^0\to\varphi K^{*}(892)^0$ decay, indicating the presence of
final-state interactions (FSI) not accounted for in the naive factorization.
The effect of FSI is evident in the phase shift of the cosine
distribution in Fig.~\ref{fig:lowmass-projection} (d).

From the definition in Eq. (\ref{eq:sphikstzero1430}) 
and the measurement of $\Delta\phi_{00}$, we
determine the parameter
\begin{eqnarray}
\sin(2\beta_{\rm eff})=\sin(2\beta+2\Delta\phi_{00})=0.97^{+0.03}_{-0.52} \, ,
\label{eq:ress2b}
\end{eqnarray}
as measured with the $B^0\to\varphi K^{*}_0(1430)^0$ decay.
Our measurements of eleven $C\!P$-violation 
parameters rule out a significant part of the
physical region and are consistent with no
$C\!P$ violation in the direct decay, but are consistent 
with the measurement of $\sin(2\beta_{\rm eff})$ in Eq.~(\ref{eq:ress2b}).
The current precision on $\sin(2\beta_{\rm eff})$ is still statistics-limited, 
although we constrain $\sin(2\beta_{\rm eff})>0.15$ at the 90\% confidence level,
which is consistent with the Standard Model $C\!P$ violation due 
to $B^0$-$\Bbar^0$ mixing. 
This analysis provides techniques for future experiments 
to extract this parameter from the $B^0\to\varphi K\pi$ decays.

Other significant non-zero $C\!P$-violation parameters
would indicate the presence of new amplitudes
with different weak phases.
The parameters $\Delta{\phi_{\perp J}}$ and $\Delta{\phi_{\parallel J}}$ are particularly
interesting due to their sensitivity to the weak phases of the
amplitudes without hadronic uncertainties~\cite{bvvreview2006},
such as the relative weak phases of $A_{J+1}$ and $A_{J0}$,
while the $C\!P$-violation parameter $\Delta\delta_{0J}$ represents
potential differences of weak phases among decay modes.

We note that the measurement of $\sin(2\beta_{\rm eff})$ 
in Eq.~(\ref{eq:ress2b}) is not the primary result of this analysis, 
but only an interpretation of the $\Delta\phi_{00}$ measurement.
Equivalently, there could be six other effective $\sin(2\beta)$ 
measurements as shown in Eqs.~(\ref{eq:sphikst892a}--\ref{eq:sphikst892c}).
However, all of them would be highly correlated due to the same
dominant uncertainty coming from $\Delta\phi_{00}$.
Rather than give them all here, we provide an illustration of our measurements 
with the following differences using the results 
in Table~\ref{tab:results2} as input:
\begin{eqnarray}
{\sin}(2\beta-2\Delta\delta_{01}) - {\rm sin}(2\beta) = -0.42^{+0.26}_{-0.34} \, ,~~
  \label{eq:ress2b1} \\
{\sin}(2\beta-2\Delta\phi_{\parallel 1}) - {\rm sin}(2\beta) = -0.32^{+0.22}_{-0.30} \, ,~~
  \label{eq:ress2b2} \\
{\sin}(2\beta-2\Delta\phi_{\perp 1}) - {\rm sin}(2\beta) = -0.30^{+0.23}_{-0.32} \, ,~~
  \label{eq:ress2b3} \\
{\sin}(2\beta-2\Delta\phi_{\perp 1}) - {\rm sin}(2\beta-2\Delta\phi_{\parallel 1})  = 0.02\pm0.23 \, ,~~
  \label{eq:ress2b4} \\
{\rm sin}(2\beta-2\Delta\delta_{02}) - {\rm sin}(2\beta) = -0.10^{+0.18}_{-0.29} \, .~~
  \label{eq:ress2b5} 
\end{eqnarray}
Systematic uncertainties are included in the errors quoted 
in Eqs.~(\ref{eq:ress2b}--\ref{eq:ress2b5}).

Taking the example in Eq.~(\ref{eq:ress2b4}), we see that because of the
positive correlation between $\Delta\phi_{\perp 1}$ and
$\Delta\phi_{\parallel 1}$, we achieve a precision of $\pm0.23$ on the
measurement of the difference between values of $\sin(2\beta_{\rm eff})$
from the parity-odd ($A_{1\perp}$) and parity-even ($A_{1\parallel}$)
decay amplitudes. This precision is significantly better than that of
the measurement of $\sin(2\beta_{\rm eff})$ itself because of the
cancellation of common uncertainties.
A significant deviation from zero would indicate a $C\!P$-violating
contribution to either the parity-odd or
parity-even amplitude but not the other.  A similar comparison would be
the values of $\sin(2\beta_{\rm eff})$ measured in $B\to\eta^\prime{K}$
and $B\to\varphi{K}$ decays. 
This measurement of the $\sin(2\beta_{\rm eff})$ difference with 
the parity-odd and parity-even amplitudes is possible with the 
angular analysis alone without any time-dependent measurement.

Among other results in this analysis, we note the significant yield of events
(more than 5$\,\sigma$ statistical significance)
in the category $B^0\to (K^+K^-)K^*(892)^0$, where $(K^+K^-)$ reflects
an $S$-wave contribution, which could be an $f_0$ or $a_0$ meson, or any other
scalar component. This decay is of the type scalar-vector.
We have already observed such decays with $B^0\to\varphi K^*_0(1430)^0$,
as discussed in this paper. Therefore it is plausible that the two decays are
related by SU$(3)$ symmetry. However, we do not report the branching fraction of 
$B^0\to (K^+K^-)K^*(892)^0$ because the exact nature of the process is
not known, and a more detailed study together with $B^0\to (\pi^+\pi^-)K^*(892)^0$
is required. Nonetheless, interference between the
$B^0\to (K^+K^-)K^*(892)^0$ and $B^0\to\varphi K^*(892)^0$ decays provides a 
further path for relating strong and weak phases in the two processes,
similar to the interference studies presented in this analysis. 
At present this interference is considered in the study of systematic
uncertainties with good prospects for phase measurements from higher
statistics experiments.


\section{ACKNOWLEDGEMENT}
We are grateful for the 
extraordinary contributions of our \pep2\ colleagues in
achieving the excellent luminosity and machine conditions
that have made this work possible.
The success of this project also relies critically on the 
expertise and dedication of the computing organizations that 
support \babar.
The collaborating institutions wish to thank 
SLAC for its support and the kind hospitality extended to them. 
This work is supported by the
US Department of Energy
and National Science Foundation, the
Natural Sciences and Engineering Research Council (Canada),
the Commissariat \`a l'Energie Atomique and
Institut National de Physique Nucl\'eaire et de Physique des Particules
(France), the
Bundesministerium f\"ur Bildung und Forschung and
Deutsche Forschungsgemeinschaft
(Germany), the
Istituto Nazionale di Fisica Nucleare (Italy),
the Foundation for Fundamental Research on Matter (The Netherlands),
the Research Council of Norway, the
Ministry of Education and Science of the Russian Federation, 
Ministerio de Educaci\'on y Ciencia (Spain), and the
Science and Technology Facilities Council (United Kingdom).
Individuals have received support from 
the Marie-Curie IEF program (European Union) and
the A. P. Sloan Foundation.


\end{document}